\documentclass{article}
\usepackage{graphicx} 
\usepackage{amsmath,geometry,xcolor,hyperref,physics,bigfoot,amssymb,lscape,mathtools,amstext,array,tabularx,amsfonts}
\usepackage{cite}

\newcommand{\nn}{\nonumber \\}
\newcommand{\beq}{\begin{equation}}
\newcommand{\eeq}{\end{equation}}
\newcommand{\bea}{\begin{eqnarray}}
\newcommand{\eea}{\end{eqnarray}}

\def\b{\beta}

\def\k{\kappa}                    

\def\m{\mu}
\def\n{\nu}
  
\def\r{\rho}                                     
\def\s{\sigma}                                   

\def\6{\partial}

\def\ft#1#2{{\textstyle{{\scriptstyle #1}\over {\scriptstyle #2}}}}
\def\fft#1#2{{#1 \over #2}}
\def\del{\partial}
\def\sst#1{{\scriptscriptstyle #1}}
\def\dalemb#1#2{{\vbox{\hrule height .#2pt
        \hbox{\vrule width.#2pt height#1pt \kern#1pt
                \vrule width.#2pt}
        \hrule height.#2pt}}}

\def\0{{\sst{(0)}}}
\def\1{{\sst{(1)}}}
\def\2{{\sst{(2)}}}
\def\3{{\sst{(3)}}}
\def\4{{\sst{(4)}}}
\def\5{{\sst{(5)}}}
\def\6{{\sst{(6)}}}
\def\7{{\sst{(7)}}}
\def\8{{\sst{(8)}}}
\def\n{{\sst{(n)}}}

 \let\b=\beta   
    \let\k=\kappa
 \let\m=\mu \let\n=\nu   \let\r=\rho
\let\s=\sigma

\def\bd{\begin{document}} \def\ed{\end{document}}
\def\ds{\documentstyle} \let\fr=\frac \let\bl=\bigl \let\br=\bigr
\let\Br=\Bigr \let\Bl=\Bigl 
\let\bm=\bibitem
\let\na=\nabla
\let\pa=\partial \let\ov=\overline 
\def\ba{\begin{eqnarray}}
\def\ea{\end{eqnarray}}
\def\ft#1#2{{\textstyle{{\scriptstyle #1}\over {\scriptstyle #2}}}}
\def\fft#1#2{{#1 \over #2}}
\def\del{\partial}
\def\sst#1{{\scriptscriptstyle #1}}
\def\oneone{\rlap 1\mkern4mu{\rm l}}
\def\ie{{\it i.e.\ }}
\def\via{{\it via}}
\def\semi{{\ltimes}}
\def\str{{\rm str}}
\def\jm{{\rm j}}
\def\im{{\rm i}}
\def\mapright#1{\smash{\mathop{-\!\!\!-\!\!\!-\!\!\!-\!\!\!-\!\!\!
             \longrightarrow}\limits^{#1}}}
\def\maprightt#1#2{\smash{\mathop{-\!\!\!-\!\!\!-\!\!\!-\!\!\!-\!\!\!
             \longrightarrow}\limits^{#1}_{#2}}}

\newcommand{\ho}[1]{$\, ^{#1}$}
\newcommand{\hoch}[1]{$\, ^{#1}$}
\newcommand{\ra}{\rightarrow}
\newcommand{\lra}{\longrightarrow}
\newcommand{\Lra}{\Leftrightarrow}
\newcommand{\bp}{\tilde \beta^\prime}
\def\rme{{\rm e}}

\numberwithin{equation}{section}

\begin{document}

\begin{titlepage}
\begin{center}
\vskip .5in 
\noindent

{\Large \bf{SCFT deformations via uplifted solitons} }
\bigskip\medskip

Dimitrios Chatzis\footnote{d.chatzis.2322097@swansea.ac.uk}, Ali Fatemiabhari\footnote{a.fatemiabhari.2127756@swansea.ac.uk}, Carlos Nunez\footnote{c.nunez@swansea.ac.uk} and Peter Weck\footnote{p.j.weck@swansea.ac.uk} \\

\bigskip\medskip
{\small 
Department of Physics, Swansea University, Swansea SA2 8PP, United Kingdom}

\vskip .5cm 
\vskip .9cm 
     	{\bf Abstract }\vskip .1in
\end{center}

\noindent
A holographic method for implementing a particular supersymmetry-preserving deformation to 4d SCFTs is presented. At the heart of  the procedure is a soliton solution of minimal $d=5$ gauged supergravity. Embedding this solution into ten- and eleven-dimensional string theory backgrounds of the form AdS$_5 \times M$, we systematically construct a range of new solutions. Each  holographically realizes a twisted compactification of the SCFT$_4$ dual to the original background. In the IR, the resulting SQFTs flow to gapped three-dimensional systems. Using a variety of holographic observables, we give evidence for this interpretation and for confinement in the deformed SQFTs. Our method applies to any holographic solutions admitting a consistent truncation to minimal $d=5$ gauged supergravity, and can likely be generalized to solutions with other AdS$_d$ factors.
 \noindent
\vskip .5cm
\vskip .5cm
\vfill
\eject

\end{titlepage}


\tableofcontents

\newpage


\section{Introduction}
\label{1-intro}
In this rather long and dense paper we study the holographic description of UV CFTs that are deformed by VEVs, flowing to confining and gapped systems. The topic is motivated from
both the holographic and field-theoretic viewpoints. 
We start the introduction with our general motivation (in a historical framework), describe the more particular deformation we aim to apply, and then discuss the systems in which we study it, ending with a summary of the contents of this work. The introduction is lengthy so that the reader interested in just the basic idea can focus on section \ref{1b}, while the level of detail in the rest of the paper is intended for colleagues who may want to work in this area. A shorter companion paper \cite{Chatzis:2024top} presents the highlights.

\subsection{Background and motivation}\label{1a}
After the formulation of the AdS/CFT correspondence\cite{Maldacena:1997re} and its initial refinements  \cite{Gubser:1998bc,Witten:1998qj}, extending the idea to Quantum Field Theories (QFTs) with less symmetries than 
${\cal N}=4$ Super Yang-Mills became a natural goal. 
The extension to either gapped theories or conformal theories with less symmetries allowed for the study of phenomena like confinement, symmetry breaking, and the presence of condensates from the new holographic point of view.

Papers like \cite{Itzhaki:1998dd} and \cite{Boonstra:1998mp} dealing with non-conformal field theories in different dimensions initiated this line of
research. These were followed by \cite{Witten:1998zw,Girardello:1999hj,Polchinski:2000uf}, which studied QFTs with minimal or no-SUSY in four dimensions using their geometric realisations.
Soon after that, the works
\cite{Klebanov:1998hh,Gubser:1998fp,Klebanov:2000nc,Klebanov:2000hb,Gubser:2004qj,Butti:2004pk,Dymarsky:2005xt},
produced a very clear geometric picture of non-perturbative aspects of a two node quiver field theory with ${\cal N}=1$ SUSY (a quasi-marginal deformation of the Klebanov-Witten ${\cal N}=1$ conformal field theory, flowing to a confining IR ${\cal N}=1$ QFT). For a summary of these developments, see \cite{Gubser:2004tf}.

A second line of work based on wrapped branes was pursued in
\cite{Witten:1998zw,Maldacena:2000yy,Acharya:2000mu,Nieder:2000kc,Naka:2002jz,Gauntlett:2000ng,Edelstein:2001pu,Nunez:2001pt,Gomis:2001vk,Karndumri:2012vh,Gomis:2001aa,Schvellinger:2001ib,Maldacena:2001pb}, extending the duality between gauge fields and strings to 
non-conformal QFTs in diverse dimensions. See  \cite{Aharony:2002up,Bertolini:2003iv,Bigazzi:2002gyi} for pedagogical reviews of this work, which gave a geometric interpretation to various non-perturbative aspects of QFT.
%
These two lines of research were joined beautifully in the works \cite{Maldacena:2009mw,Gaillard:2010qg,Caceres:2011zn,Elander:2011mh}. 

The addition of fields transforming in the fundamental representation of the gauge group was discussed first in the quenched/probe approximation. See \cite{Karch:2002sh,Kruczenski:2003be,Kruczenski:2003uq,Babington:2003vm,Nunez:2003cf,Sakai:2004cn,Erdmenger:2007cm} for some representative papers and a review. This was later improved by including the backreaction of these sources. 
The flavour branes were `smeared' so that one can work with BPS ODEs rather than BPS PDEs \cite{Casero:2006pt,Casero:2007jj,Hoyos-Badajoz:2008znk,Nunez:2010sf,Benini:2006hh,Benini:2007gx,Bigazzi:2014qsa,Bigazzi:2011db,Bea:2013jxa}.

However, some unwelcome features of these models are the following:
\begin{itemize}
    \item {When the high-energy behaviour of the QFT is field theoretical and represented by a deformed 4d UV-CFT,  the IR  part of the holographic dual background is singular. This is the case e.g. for the models in \cite{Girardello:1999hj,Polchinski:2000uf,Bobev:2018eer,Petrini:2018pjk}.}
    \item{On the other hand, for the wrapped brane models \cite{Maldacena:2000yy,Acharya:2000mu,Nieder:2000kc,Naka:2002jz,Gauntlett:2000ng,Edelstein:2001pu,Nunez:2001pt,Gomis:2001vk,Karndumri:2012vh,Gomis:2001aa,Schvellinger:2001ib,Maldacena:2001pb} or for the two node quiver system described above \cite{Klebanov:1998hh,Gubser:1998fp,Klebanov:2000nc,Klebanov:2000hb,Gubser:2004qj,Butti:2004pk,Dymarsky:2005xt}, the high energy behaviour of the QFT does not {\it strictly} reach a conformal point in the UV. This is reflected in the fact that the holographic dual is not {\it strictly} asymptotically AdS, making the application of holographic renormalisation technically challenging.}
     \item{The addition of a large number of flavours, as explained above, is technically cumbersome, requiring the smearing of branes. As a consequence, it is hard to study CFTs with flavours in four or more dimensions.}
 \end{itemize}
Our motivation is to
remedy some of these unwelcome features. To do this we consider four-dimensional super-conformal field theories (SCFTs), whose dual backgrounds have an AdS$_5$ factor in the geometry. The SCFTs include a large number of (localised) flavour branes. These SCFTs are deformed by VEVs and a RG-flow is triggered, ending in a  confining and gapped system. The holographic description mirrors this, featuring an asymptotically AdS$_5$ space that ends smoothly at a fixed value of the radial coordinate. Non-perturbative aspects of the RG-flow in the QFT can be consistently studied. To implement these ideas, we consider a deformation on the holographic side, which we describe below.

\subsection{The deformation}\label{1b}
The goal of this work is to establish a general holographic mechanism to obtain confining, (2+1)-dimensional SQFTs from known (3+1)-dimensional SCFTs. Our inspiration is \cite{Anabalon:2021tua} (see also \cite{Nunez:2023nnl, Anabalon:2024che, Anabalon:2024qhf, Fatemiabhari:2024aua,  
Nunez:2023xgl, Anabalon:2023lnk, Anabalon:2022aig} for related work), in which a supersymmetry-preserving AdS$_5$ soliton solution is identified. From the perspective of AdS$_5 / $CFT$_4$, the solution of \cite{Anabalon:2021tua} realizes a compactification of $\mathcal{N}=4$ SYM, leading to an $\mathcal{N}=2$ SYM theory (plus massive multiplets) in three dimensions. Crucially, some supersymmetry remains unbroken, thanks to a mixing of the R-symmetry of $\mathcal{N}=4$ SYM with the isometry for the compact circle. We refer to this as a `twisted' compactification throughout, although the compact manifold is just a circle (and thus flat). The field theory story for $\mathcal{N}=4$ SYM is discussed in detail in \cite{Kumar:2024pcz}.

We have generalized this mechanism to a variety of SCFTs with holographic duals of the form
\begin{equation} \label{undeformed_backgrounds}
    \text{AdS}_5 \times M_n,
\end{equation}
where $n=5$ for Type II and $n=6$ for 11d supergravity examples. In particular, using known embeddings of $d=5$ minimal gauged supergravity solutions into backgrounds the form \eqref{undeformed_backgrounds}, we have uplifted the soliton solution of \cite{Anabalon:2021tua}. The result is new families of smooth string theoretic backgrounds of the form
\begin{equation}\label{deformed_backgrounds}
    \widehat{\text{AdS}_5} \times \widehat{M_n}
\end{equation}
where the hats denote a deformation which can be summarized as follows:
\begin{itemize}
    \item Compactify one of the AdS$_5$ directions to a circle $S_\phi^1$ of size $L_\phi/2\pi$.
    \item Deform the AdS$_5$ geometry by introducing a warping function $f(r)$ which smoothly caps off the $S^1_\phi$ circle at a finite radius $r=r_\star$, at which  $f(r_\star)=0$,
\begin{equation}
    \frac{r^2}{l^2}(-dt^2+dx_1^2+dx_2^2+ d\phi^2)+\frac{l^2}{r^2} dr^2 \quad \rightarrow \quad 
    \frac{r^2}{l^2} (- dt^2+ dx_1^2 +  dx_2^2 +  f(r) d\phi^2) + \frac{l^2}{r^2} ~\frac{ dr^2}{f(r)} ,\label{ARmetric}
\end{equation}
with the solution parameters constrained by $L_\phi=\frac{l^2}{r_\star^2} \frac{4\pi}{f'(r_\star)}$ to avoid conical singularities.
\item Identify an appropriate $U(1)$ inside the isometry group of the internal manifold $M_n$, and gauge it by
\begin{equation}
    \mathcal{A} =q \left( \frac{1}{r^2}- \frac{1}{r _{\star}^2}\right) d\phi \,, \label{ARgaugefield}
\end{equation}
suitably modifying all the fluxes to ensure the full ten- or eleven-dimensional supergravity equations of motion are satisfied. 
\end{itemize}
Requiring $r_\star=(q l)^{1/3}$ ensures the preservation of four supercharges for the $d=5$ gauged supergravity solution we are uplifting \cite{Anabalon:2021tua}. Thanks to the embedding frameworks we are relying on \cite{Buchel_2007,Gauntlett:2007sm,plethora}, this ensures the preservation of higher-dimensional supersymmetry. As a result, all of the examples we will present preserve ten- or eleven-dimensional supersymmetry for this choice of solution parameters.

Let us return to the field-theoretic interpretation of these new solutions. The internal manifold $M_n$ of the undeformed backgrounds \eqref{undeformed_backgrounds} encodes various data of the dual SCFT$_4$. In our examples, we will encounter not only the case of $M_5=S^5$ and the $\mathcal{N}=4$ SYM dual, but also $M_5=Y^{p,q}$ describing $\mathcal{N}=1$ toric quiver field theories, internal manifolds describing $\mathcal{N}=2$ linear quivers, as well as some for $\mathcal{N}=1$ non-Lagrangian SCFTs. When we deform each background in the sense described above, we are implementing a particular twisted compactification in the dual field theory.

When compactifying on $S_\phi^1$ one needs to specify boundary conditions for the various fields in the SCFT. This generically breaks SUSY, as the scalars and gauge fields are assigned periodic boundary conditions while the fermions are anti-periodic. As alluded to earlier, SUSY can be preserved by turning on a background gauge field $\mathcal{A}=\mathcal{A}_\phi \, d\phi$ which mixes a $U(1)$ inside the R-symmetry with the $U(1)_\phi$ isometry of the compact circle.\footnote{While the background gauge field is constant in the boundary field theory, it has a non-trivial holonomy, and thus cannot be absorbed by a gauge transformation.} The reason is that the background gauge field modifies the covariant derivative. If its charge is tuned appropriately, massless fermions can exist and enter into supermultiplets.

The introduction of the background gauge field and the scale $L_\phi$ into the field theory break conformality while preserving some supersymmetry (again, for an appropriate choice of parameters). At low energies, our holographic construction of this twisted compactification takes the original SCFT$_4$ to a strongly coupled SQFT$_3$, thanks to the closing of $S_\phi^1$ by $f(r)$. Employing a variety of holographic observables, we find clear indications of confinement in these (2+1)-dimensional IR theories.

\subsection{Summary of contents}
We will begin by reviewing $d=5$ minimal gauged supergravity and the soliton solution of \cite{Anabalon:2021tua} in section \ref{2-5d}. The remainder of the paper will concern uplifts of this solution to ten- and eleven-dimensional supergravity and how holography can tell us about the resulting QFT duals. 

In section \ref{3-IIB}, we use the embedding prescriptions of \cite{Cvetic:1999xp,Buchel_2007} to lift into backgrounds in Type IIB. Specifically we review how to obtain the deformation of AdS$_5\times S^5$ first noted in \cite{Anabalon:2021tua}, and introduce new deformed AdS$_5\times Y^{p,q}$ backgrounds. To uplift into massive Type IIA, in section \ref{4-IIA} we use the embedding scheme developed in \cite{plethora}. This allows us to write down our deformation for any AdS$_5$ solution in massive IIA dual to an $\mathcal{N}=1$ SCFT. We focus on two particular families, based on the work of \cite{afprt2015,afpt2015,bpt2017}. Section \ref{5-11d} utilizes the fact that minimal $d=5$ gauged supergravity 
can be treated as a truncation of $d=5$ Romans' $SU(2)\times U(1)$ supergravity, allowing us to lift from there to eleven-dimensional supergravity using \cite{Gauntlett:2007sm}. Backgrounds based on the LLM family of solutions \cite{Lin:2004nb} are presented. Restricting to the ``electrostatic case'' allows us to make contact with the holographic solutions of Gaiotto-Maldacena \cite{Gaiotto:2009gz}.

Having presented a wide range of deformed backgrounds, in section \ref{6-observables} we study the features of their dual QFTs. Wilson, 't Hooft loop, and Entanglement Entropy calculations reveal signatures of confinement in the IR three-dimensional systems, as presented in sections \ref{6a}, \ref{6b}, and \ref{6c}. Probing the (deformed) AdS directions, these observables show universalities across all the examples studied.
In section \ref{6d}, we present an observable capturing the number of degrees of freedom along the full flow from UV SCFT$_4$ to IR (S)QFT$_3$, known as the flow central charge. Holographic complexity is also studied.
Other probes engaging the internal manifold are presented to study particularities of the SQFTs in section \ref{6f}. The spontaneous symmetry breaking of $U(1)_\phi$ is considered in \ref{6g}, appearing as a massive vector mode in the bulk.

We conclude with a discussion section and several appendices collecting more technical details.

\section{Review of the supersymmetric AdS$_5$ soliton}
\label{2-5d}
In this section, we study the original 5d minimal gauged supergravity solution of \cite{Anabalon:2021tua}, which is used as a seed to obtain different backgrounds investigated in this paper. The supersymmetry of the solution is discussed in appendix \ref{A2-susy}

Consider the Einstein-Maxwell-AdS system in five dimensions, developed in different works \cite{Schwarz:1983qr, Howe:1983sra, Gunaydin:1984fk,
Pernici:1985ju, Gunaydin:1984qu, 
Cvetic:1999xp}. 
The bosonic part of the action is
\begin{equation} 
S\left(  g,A\right)  =\frac{1}{16\pi G}\int d^{5}x \sqrt{-g}\left(
R+\frac{12}{l^{2}}-\frac{3}{4}\mathcal{F}_{\mu\nu}\mathcal{F}^{\mu\nu}\right)+\frac{1}{16\pi G} \int \mathcal{F}\wedge \mathcal{F}\wedge \mathcal{A}\text{ }, \label{eq:Lag5d}%
\end{equation}
where $l$ is the AdS radius. The equations of motion are%
\begin{align}
&  
d \star \mathcal{F} + \mathcal{F}\wedge \mathcal{F}=0\,\,,\nonumber\\
&  R_{\mu\nu}-\tfrac{1}{2}g_{\mu\nu}R-\tfrac{3}{2}\big[\mathcal{F}_{\mu\rho}\,\mathcal{F}_{\nu}%
{}^{\rho}-\tfrac{1}{6}g_{\mu\nu}\mathcal{F}_{\rho\sigma}\mathcal{F}^{\rho\sigma}\big]-\frac
{6}{l^{2}}\,g_{\mu\nu}=0\,. \label{eq:5dEin}
\end{align}
We restrict to solutions that satisfy $\mathcal{F}\wedge \mathcal{F}=0$, so the Chern-Simons term will not play a role.

One can obtain the solution of \cite{Anabalon:2021tua} from a double Wick rotation of an electrically charged black hole with a flat boundary. The result reads
\begin{align} \label{soliton}
ds_5^{2}=\frac{r^{2}}{l^{2}}(-dt^{2}+dx_1^{2}+dx_2^{2})&+\frac{l^2 dr^{2}}%
{r^{2} f(r)}+\frac{r^{2}}{l^{2}} f(r)d\phi^{2}, \qquad f(r)=1-\frac{\mu l^2}{r^{4}}-\frac{q^{2} l^2}{r^{6}},
\nonumber \\
\mathcal{A}=q\left(  \frac{1}{r^{2}}-\frac{1}{r_{\star}^{2}}\right) & d\phi, \qquad \mathcal{F}=d\mathcal{A}=-\frac{2q}{r^3} dr \wedge d\phi,
\end{align}
where $r_{\star}$ is the largest positive root of $f(r)$, $f(r_{\star})=0$. One can write
\begin{equation}\label{mu_eq}
\mu=\frac{(r_{\star}^{6}-q^{2}l^{2})}{l^{2}r_{\star}^{2}}.
\end{equation}
Note that the $\phi$-coordinate parameterizes a compact circle which has a finite size at $r \rightarrow \infty$, but shrinks at $r=r_{\star}$.
In order to have a smooth solution there, the periodicity of $\phi$ is fixed to
\begin{equation}
L_\phi=\frac{4\pi l^2}{r_\star^2f'(r_{\star})}.
\end{equation}
The magnetic flux through the $\phi$ direction at $r\rightarrow\infty$ is 
$\Phi=-\oint \mathcal{A}=\frac{q}{r_{\star}^{2}} L_\phi$. 

To understand the solution space, it is more natural to invert these relations and write the bulk parameters $r_\star$, $q$, $\mu$ in terms of the asymptotic boundary parameters $L_\phi$, $\Phi$. We have
\begin{equation}
q=r_{\star}^{2}\frac{\Phi}{L_\phi}, \qquad r_{\star}=\frac{\pi l^{2}}{2L_\phi}\left(  1\pm\sqrt{1-\frac{\Phi^{2}}%
{\Phi_{max}^{2}}}\right)  ,
\end{equation}
where $\Phi_{max}=\frac{\pi l}{\sqrt{2}}$. The parameter $\mu$ can then be written in terms of boundary data using \eqref{mu_eq}. For $\Phi<\Phi
_{max}$, there are two branches of solutions. As $\Phi\rightarrow0$, the $+$ branch of the solution approaches the
pure AdS soliton,  and the $-$ branch approaches
Poincar\'{e}-AdS. We are interested in the solution of the $+$ branch. The two branches coalesce at $\Phi=\Phi_{max}$.

As explained in appendix \ref{A2-susy}, preserving supersymmetry requires $\mu=0$. This occurs for $r_{\star}^{6}=q^{2}l^{2}$, that is, when
\begin{equation}
r_{\star}^{2}=\frac{\Phi^{2}l^{2}}{L_\phi^{2}}\quad\Rightarrow
\quad2-3\frac{\Phi^{2}}{\Phi_{max}^{2}}\pm2\sqrt{1-\frac{\Phi^{2}}{\Phi
_{max}^{2}}}=0.
\end{equation}
On the $+$ branch one finds that the supersymmetric point corresponds to  $\Phi_{S}=\frac{2\pi}{3}l$. We will assume this condition is satisfied on any occasion when a supersymmetric background is considered. 

From the dual field theory perspective, it is convenient to introduce a new parameter $Q$ such that the boundary gauge field is simply
\begin{equation}\label{eq:5dvector}
\mathcal{A}(r\to \infty)=Q\, d\phi
\end{equation}
at the supersymmetric point in the parameter space. This can be accomplished by re-parameterizing as $q=-Q^3 l^2$. The condition $\mu=0$ for the solution to preserve supersymmetry is then $r_\star^2= (Q l)^2$. All together, the supersymmetric soliton is given by
\begin{align}\label{susy_soliton}
    ds_5^{2}=\frac{r^{2}}{l^{2}}&(-dt^{2}+dx_1^{2}+dx_2^{2})+\frac{l^2}%
{r^{2}} \, \frac{dr^{2}}{f(r)}+\frac{r^{2}}{l^{2}} f(r)d\phi^{2}, \qquad f(r)=1-\left(\frac{l Q}{r}\right)^6,
\\
&\mathcal{A}= \left(Q- \frac{l^2 Q^3}{r^2}\right) d\phi, \qquad \mathcal{F}=d\mathcal{A}=  \frac{2l^2 Q^3}{r^3} dr \wedge d\phi. \nonumber
\end{align}
Note that the holonomy of the boundary gauge field is fixed as
\begin{equation}
    \Phi=\oint \mathcal{A}=\frac{2\pi}{3} l.
\end{equation}

\section{Deformed backgrounds in Type IIB}
\label{3-IIB}
As noted by Anabalon and Ross in \cite{Anabalon:2021tua}, the supersymmetric AdS$_5$ soliton summarized in the previous section can be obtained from a dimensional reduction of Type IIB supergravity. Here we review the reduction ansatz first given in  \cite{Anabalon:2021tua}, in which the soliton is embedded into the background AdS$_5\times S^5$, before presenting a new infinite family of solutions obtained from AdS$_5\times Y^{p,q}$ backgrounds. As explained in section \ref{1b}, the ten-dimensional solutions obtained in this fashion can be understood as implementing a particular deformation in the corresponding SCFT duals.

\subsection{AdS$_5 \times S^5$ embedding}
\label{3a-S5}
The embedding of the solution \eqref{soliton} into the Type IIB background  AdS$_5\times S^5$ yields the metric \cite{Cvetic:1999xp,Anabalon:2021tua}
\begin{align}\label{metric-AdS5xS5}
       ds^2 _{10}  = ds^2_{5}&+ l^2 \biggl\{ d\theta^2 + \sin^2\theta d\varphi^2+ \sin^2\theta\sin^2\varphi \left( d\varphi_1+ \frac{\mathcal{A}}{l} \right) ^2
       \\
           & + \sin^2\theta \cos^2\varphi \left( d\varphi_2 + \frac{\mathcal{A}}{ l} \right) ^2 + \cos^2\theta \left( d\varphi _3 + \frac{\mathcal{A}}{l} \right) ^2\biggr\}. \nonumber
\end{align}
Note that the three $U(1)$ factors in the Cartan of the $S^5$ isometry group are uniformly fibered with $\phi$-circle inside $ds_5^2$. In each example, we will see a particular $U(1)$ inside the isometry group for the internal manifold mixing with the $U(1)$ of the $S_\phi^1$. The smoothness and supersymmetry-preserving features (for $\mu=0$) of the five-dimensional solution are inherited by its embedding in ten dimensions. The range of the various angular coordinates is
\begin{equation}
    0 \leq \theta \leq \pi/2, \qquad 0 \leq \varphi \leq \pi/2, 
    \qquad 0 \leq \varphi_i \leq 2\pi \quad i=1,2,3.
\end{equation}
It's convenient to introduce,
\begin{eqnarray}
 \mu_1= \sin\theta \sin\varphi, \quad \mu_2= \sin\theta \cos\varphi, \quad \mu_3= \cos\theta,\label{mui}
\end{eqnarray}
satisfying $\mu_1^2+\mu_2^2+\mu_3^2=1$. We can then write the five-form field strength solving the Type IIB supergravity equations of motion for the metric \eqref{metric-AdS5xS5} as
\begin{align}\label{F5S5}
       F_5= G_5+\star G_5, \qquad G_5= \frac{4}{l} \text{vol}_5 &-l^2 \sum_{i=1}^{3} \mu_i d\mu_i \wedge \left(d\varphi_i+\frac{\mathcal{A}}{l}\right)\wedge \star_5 \mathcal{F}.
\end{align}
Here and throughout this work, we use $\star_5$ to denote the Hodge star with respect to the 5d metric in \eqref{soliton}. Explicitly we have
\begin{equation}\label{vol5_and_starF}
  \text{vol}_5=  \frac{r^3}{l^3} dt \wedge dx_1 \wedge dx_2 \wedge dr \wedge d\phi, \qquad \star_5 \mathcal{F}=-\frac{2q}{l^3} dt \wedge dx_1 \wedge dx_2.
\end{equation}
We can of course understand $F_5$ as sourced by a stack of D3-branes, whose number is related to the quantization of flux by
\begin{equation}
Q_{D3}= \frac{1}{(2\pi)^4 
} \int F_5=\frac{l^4}{4\pi 
},
\end{equation}
where we have suppressed factors of $g_s$ and $\alpha'=\ell_s^2$.
The field theory dual of this construction is a twisted compactification of $\mathcal{N}=4$ SYM, which still preserves four supercharges when we set $\mu=0$ in \eqref{soliton}.

\subsection{AdS$_5 \times Y^{p,q}$ embeddings}
\label{3b-Ypq}
Having reviewed the holographic realization of the deformation procedure for the dual of $\mathcal{N}=4$ SYM, we now consider its application to duals of SCFTs with less supersymmetry. In particular, we deform Type IIB backgrounds with Sasaki-Einstein internal manifolds $Y^{p,q}$ \cite{Gauntlett_2004}, 
 \begin{align}\label{undeformed_Ypq}
     ds_{10}^2=ds_{\text{AdS}_5}^2+l^2 ds_Y^2, &\qquad ds_Y^2=ds_B^2+\frac{1}{9}(d\psi +A)^2,
    \nonumber \\
     F_5=(1+\star)G_5,& \qquad  G_5= \frac{4}{l} \text{vol}_{\text{AdS}_5}
 \end{align}
Here $B$ is a four-dimensional K\"{a}hler-Einstein base. Its K\"{a}hler form $J$ is related to the one-form $A$ as $dA=6J$.  This $S_\psi^1$ fibration over $B$ corresponds to the $\mathrm{U}(1)$ R-symmetry of the dual $\mathcal{N}=1$ superconformal field theories, which are the IR fixed points of quiver gauge theories labelled by the integers $p$ and $q$. In each example we will provide explicit expressions for $ds_Y^2$.

The deformation is accomplished by embedding the soliton solution \eqref{soliton} into \eqref{undeformed_Ypq}. The appropriate reduction ansatz of Type IIB to minimal 5d gauged supergravity is given in \cite{Buchel_2007}, or in a more general form in \cite{Gauntlett:2007ma}, for example. In this case it reads
 \begin{align}\label{deformed_Ypq}
    & ds_{10}^2=ds_5^2+l^2\left[ds_B^2+\frac{1}{9}(d\psi +A+\frac{3}{l} \mathcal{A})^2 \right],
    \nonumber \\
   &  F_5=(1+\star)G_5, \qquad  G_5= \frac{4}{l} \text{vol}_{5} - l^2 \, J_2 \wedge \star_5 \mathcal{F},
 \end{align}
 with $ds_5^2$, $\mathcal{A}$, and $\mathcal{F}$ given by \eqref{soliton}. We provide explicit examples for $T^{1,1}$ (equivalently $Y^{1,0}$) and generic $Y^{p,q}$ below.

 \subsubsection*{$T^{1,1}$ example}
For the case of $T^{1,1}$, the base manifold $B$ is $S^2 \times S^2$, so that 
\begin{equation}
ds^2_Y=\frac{1}{6}\sum_{i=1}^2\left(d\theta_i^2+\sin^2\theta_i d\phi_i^2\right) + \frac{1}{9}\left( d\psi +\sum_{i=1}^2 \cos\theta_i d\phi_i\right)^2.
\end{equation}
The coordinates $(\theta_i,\phi_i)$ on the two-spheres have canonical $\pi$ and $2\pi$ periodicities, respectively, while $\psi\in[0,4\pi]$. The deformed ten-dimensional metric can then be expressed 
\begin{align}\label{deformed_T11}
     ds^2_{10}= ds^2_{5}+l^2\left[\frac{1}{6}\sum_{i=1}^2\left( d\theta_i^2+\sin^2\theta_i d\phi_i^2\right) + \frac{1}{9}\left( d\psi + \sum_{i=1}^2\cos\theta_i d\phi_i+\frac{3}{l}\mathcal{A}\right)^2\right],
\end{align}
again with  $ds_5^2$ and $\mathcal{A}$ as in \eqref{soliton}. The RR five-form field strength can be computed from
\begin{align}\label{F5T11}
      F_5= G_5+\star G_5, \qquad  G_5= \frac{4}{l} \text{vol}_5 &+ \frac{l^2}{6}( \sin \theta_1 ~ d\theta_1 \wedge d\phi_1+\sin \theta_2 ~ d\theta_2 \wedge d\phi_2) \wedge \star_5 \mathcal{F},
\end{align}
where $\text{vol}_5$ and $ \star_5 \mathcal{F}$ are as in \eqref{vol5_and_starF}.
The quantized charge associated to $F_5$ is found by integrating over $T^{1,1}$,
\begin{equation}
Q_{D3}= \frac{1}{(2\pi)^4 
} \int F_5=\frac{4l^4}{27\pi 
}.
\end{equation}
This can be interpreted as the number of D3-branes sitting at the tip of a cone over $T^{1,1}$. The field theory we are deforming via this holographic construction is the Klebanov-Witten CFT \cite{Klebanov:1998hh}.

  \subsubsection*{Generic $Y^{p,q}$ example}
For the more general case in which $M_5$ is any $Y^{p,q}$ manifold, the deformed metric reads
  \begin{equation}\label{deformed_Ypq}
\begin{split}
 ds_{10}^2= ds^2_{5}+ & l^2\left[\frac{1-y}{6}\left( d\theta^2+\sin^2\theta d\varphi^2\right)+\frac{1}{w(y)v(y)} dy^2+\frac{w(y)v(y)}{36}\left( d\beta+\cos\theta d\varphi\right)^2\right.\\
   & \left.+\frac{1}{9}\left( d\psi- \cos\theta d\varphi+y\left( d\beta+\cos\theta d\varphi\right)+ \frac{3}{l}\mathcal{A}\right)^2\right],\\
    \end{split}
\end{equation}
where the functions $w(y)$, $v(y)$ are controlled by a single parameter $a$,
\begin{equation}
    w(y)=\frac{2(a-y^2)}{1-y}\,\,,\,\,v(y)=\frac{a+2y^3-3y^2}{a-y^2}.
\end{equation}
The base space $B$ parameterized by coordinates 
\begin{equation}
    0 \leq \theta \leq \pi, \qquad  0 \leq \varphi \leq 2\pi, \qquad  y_1 \leq y \leq y_2, \qquad  0 \leq \beta \leq 2\pi, 
\end{equation}
is ensured to be smooth and compact if one chooses 
\begin{equation}
    0 < a < 1,
\end{equation}
and the endpoints $y_1,~ y_2$ are constrained appropriately--- see e.g. \cite{Gauntlett_2004}.
Note the integers $p,q$ are related to integrals of the K\"{a}hler form over two-cycles in $B$. This two-form is given by
\begin{equation}
    J_2=\frac{1}{6}(1-y) \sin \theta ~ d\theta \wedge d \varphi+\frac{1}{6} dy \wedge (d\beta+ \cos \theta ~ d\varphi), 
\end{equation}
and so the RR field strength $F_5$ can be expressed
\begin{align} \label{F5Ypq}
 F_5= G_5+\star G_5, \quad    G_5= \frac{4}{l} \text{vol}_5 &- \frac{l^2}{6}\left[(1-y) \sin \theta ~ d\theta \wedge d \varphi+ dy \wedge (d\beta+ \cos \theta ~ d\varphi)\right] \wedge \star_5 \mathcal{F}.
\end{align}
The associated quantized charge is found to be
\begin{equation}
Q_{D3}= \frac{1}{(2\pi)^4 
} \int F_5=\frac{2l^4}{27\pi 
}(y_1-y_2)(y_2+y_1-2).
\end{equation}
We have verified explicitly that the backgrounds presented above in eqs.\eqref{deformed_Ypq}-\eqref{F5Ypq} solve the full Type IIB supergravity equations of motion.

Let us close this section with brief words about the ${\cal N}=1$ SCFTs associated with any generic member of the $Y^{p,q}$ family. After the construction of this family of Sasaki-Einstein manifolds \cite{Gauntlett_2004}, the dual SCFTs were constructed in \cite{Benvenuti:2004dy}, see also \cite{Herzog:2004tr} for a summary. 

The quiver associated is constructed in terms of two basic 'unit-cells' denoted by $\sigma,\tau$ that are put together like Legos. $Y^{p,q}$ has $p$ unit cells, of which $q$ are of the $\sigma$-type. There are bifundamental fields $(U^\alpha, V^\alpha, Y, Z)$ with $\alpha=1,2$ is an $SU(2)$ index. All gauge nodes are $SU(N)$. The superpotential is constructed by summing cubic and quartic terms of the form
\begin{equation}
W\sim W_3+W_4,~~~W_3\sim \epsilon_{\alpha\beta}(U_L^\alpha V^\beta Y +U_R^\alpha V^\beta Y),~~~ W_4\sim \epsilon_{\alpha\beta}Z U_R^\alpha Y U_L^\beta.
\end{equation}
The cubic terms are associated with $\sigma$-cells and the quartic ones with $\tau$-cells. Imposing the vanishing of the NSVZ beta functions for the gauge groups and the superpotentials (consequently the vanishing of R-symmetry anomalies) leaves us with a two-dimensional space of anomalous dimensions. 

There are some known deformations of the conformal quivers. For example, a quasi-marginal deformation that triggers a cascade \cite{Herzog:2004tr}. Another possibility is to switch on a VEV represented by the parameter $q$, the R-symmetry fibration and the replacement of AdS$_5$ by $ds_5^2$ as indicated in (\ref{deformed_Ypq}). Let us now discuss a different infinite family of backgrounds.

\section{Deformed backgrounds in Type IIA}
\label{4-IIA}
Let us begin this section by reviewing the general form of supersymmetric AdS$_5$ backgrounds in massive Type IIA supergravity, dual to $\mathcal{N}=1$ SCFTs. Given such a background, we can embed the five-dimensional supergravity solution \eqref{soliton} using the framework of \cite{plethora}, which we will then summarize in a slightly modified form. Finally, we will present examples of several deformed backgrounds obtained in this way. The full massive IIA equations of motion and Bianchi identities have been explicitly checked for each new background discussed.

Supersymmetric AdS$_5$ solutions in massive Type IIA supergravity were first classified in \cite{afpt2015}. Subsequent work by the authors of \cite{bpt2017} reduced the number of BPS equations controlling the classification, writing the metric in terms of a pair of functions $D_s,\,D_u$ as
\begin{align}\label{mIIA_metric}
ds_{10}^2&=e^{2W}\left[ ds_{\text{AdS}_5}^2+e^{2A}(dv_1^2+dv_2^2)+\frac{1}{3}e^{-6 \lambda}
ds_3^2 \right] \\
ds_3^2&=-\frac{4}{\partial_s D_s} D\psi^2-\partial_s \widetilde{D}_s ds^2-2 \partial_u D_s du ds-\partial_u D_u du^2, \nonumber
\end{align}
where $D_s,\,D_u$ depend only on the coordinates $(v_1,v_2,s,u)$, and we have defined
\begin{align}
\widetilde{D}_s=D_s-\frac{3}{2}\log s, \hspace{2mm} \text{and} \hspace{3mm} D \psi= d \psi-\frac{1}{2} \star_2 d_2 D_s.
\end{align}
Note that the subscripts on the Hodge star $\star_2$ and total derivative $d_2$ restrict these operators to the coordinates $(v_1,v_2)$. In the examples we will discuss, this subspace describes a constant-curvature Riemann surface.
The angular coordinate $\psi$ parameterizes a U(1) isometry of the spacetime, corresponding to an R-symmetry in the SCFT dual. We can express the warp factors appearing in the metric in terms of $D_s,\,D_u$ as
\begin{align}
e^{4W}=\frac{-\partial_s D_s}{3 \text{det}(h)}, \hspace{3mm} e^{2A}=\frac{\text{det}(h) e^{D_s}}{24}, \hspace{3mm} e^{-6\lambda}=\frac{\text{det}(h)}{8s \, \text{det} (g)},
\end{align}
where the determinants refer to the $(u,s)$ subspace,
\begin{align}
\text{det} (g)=\partial_u D_u \partial_s \widetilde{D}_s-(\partial_u D_s)^2, \hspace{3mm} \text{det}( h)=\partial_u D_u \partial_s D_s-(\partial_u D_s)^2.
\end{align}
The dilaton takes the form
\begin{equation}
    e^{2\Phi}=e^{6W}e^{-6\lambda},
\end{equation}
and the RR and NSNS field strengths are given in terms of $D_s, \, D_u$ as
\begin{align}\label{mIIA_fluxes}
    &F_0=36 \frac{\sqrt{2s}}{\partial_2 D_s}\partial_u(\partial_s D_u -\partial_u D_s),
    \\
    &F_2=\frac{1}{3}F_0 \, \tilde{\xi} \wedge D\psi -d \left[\star_2 d_2 D_u+2\frac{\partial_u D_s}{\partial_s D_s}D\psi \right]+\Delta_2 D_u \, dv_1 \wedge dv_2 \nonumber
    \\
    &-\partial_u(s e^{D_s} \text{det}(g))dv_1 \wedge dv_2+\star_2 d_2(\partial_u D_s-\partial_s D_u)\wedge ds, \nonumber
    \\
    &F_4=\frac{1}{3}F_2 \wedge \tilde{\xi} \wedge D\psi -\frac{1}{36}d\left( \sqrt{2s} \, e^{D_s} \text{det}(h) dv_1\wedge dv_2 \wedge D\psi -2 \sqrt{2s}\, d(\star_2 d_2 D_s) \wedge D\psi \right) \nonumber
    \\
    &+\frac{1}{36\sqrt{2s}} \frac{\partial_u D_s}{\partial_s D_s}\left[2 \, du\wedge d(\star_2 d_2 D_s)+e^{D_s} \text{det}(h) du \wedge dv_1 \wedge dv_2+3 \, d(\partial_u D_s e^{D_s})\wedge dv_1 \wedge dv_2\right]\wedge D\psi \nonumber
    \\
    &H_3 = \frac{1}{3}d \left[\tilde{\xi} \wedge D\psi +\frac{1}{8}\frac{e^{D_s} }{\sqrt{2s}}\partial_u D_s \,dv_1 \wedge dv_2\right] +\frac{1}{36\sqrt{2s}}du \wedge d \star_2 d_2 D_s-\frac{e^{D_s}}{12}\text{det}(g) \, \tilde{\xi} \wedge dv_1 \wedge dv_2 \nonumber.
\end{align}
Borrowing from the notation in \cite{plethora},
we have written the field strengths in terms of a one-form $\tilde{\xi}$, 
\begin{align}
    \tilde{\xi} =-\frac{1}{6 \, \text{det}(g) \sqrt{2s}}\left(\frac{3}{2s} \partial_u D_s \, ds+\text{det}(h) \, du \right).
\end{align}
The functions $D_s$, $D_u$ are constrained by the Bianchi identities for the field strengths. 
The Bianchi identity for $F_0$ requires it to be piece-wise constant, yielding a first differential equation for  $D_s$, $D_u$. The Bianchi identities for $H_3$ and $F_2$, respectively, then translate to two additional equations,
\begin{align}\label{bianchis}
    \Delta_2D_s&=\partial_s(s \text{det}(g) e^{D_s} ) +\frac{F_0}{24}\frac{\partial_s(e^{D_s})}{\sqrt{2s}}, 
    \\
    \Delta_2(\partial_u D_u)&=\partial_u^2(s \text{det}(g) e^{D_s} ) +\frac{sF_0}{36} \frac{\partial_s(\text{det}(h) e^{D_s})}{\sqrt{2s}}.
\end{align}
The Bianchi for $F_4$ is satisfied automatically.

\subsection{Embedding into IIA duals of $\mathcal{N}=1$ SCFTs}
\label{4a}
As shown in \cite{plethora}, five-dimensional minimal gauged supergravity can be embedded into the massive IIA system presented above by making the following substitutions
\begin{align}\label{mIIA_embedding} 
    ds_{\text{AdS}_5}^2 \,\rightarrow \,ds_5^2, &
    \\
    F_4 \,\rightarrow \,F_4^g-\left(\star_5 \mathcal{F}-\frac{1}{3}\mathcal{F} \wedge \mathcal{D} \psi\right)& \wedge \frac{ds}{\sqrt{2s}}+\frac{\sqrt{2s}}{3}\mathcal{F}\wedge \text{vol}_\Sigma 
    \nonumber \\
    F_2,  \, H_3  \,\rightarrow  \, F_2^g, \,  H_3^g, \qquad F_6, \, F_8, & \,\rightarrow  \, -\star F_4^g, \, \star F_2^g,  
    \nonumber
\end{align}
where $ds_5^2$ and $\mathcal{F}$ are the line element and two-form field strength of a given solution to the five-dimensional action \eqref{eq:Lag5d}, and the superscripts $g$ ($g$ for gauging) denote the substitutions
\begin{equation}\label{5d_gauging}
    D\psi \, \rightarrow \, \mathcal{D}\psi=D\psi- 3\mathcal{A},
\end{equation}
taken inside of all potentials, before acting with total derivatives to compute the corresponding field strengths. In other words, if in the initial background one has
\begin{equation}\label{gauging_vs_potentials}
     H_3=dB_2, \quad F_2=F_0 B_2+dC_1, \quad F_4=dC_3+B_2 \wedge F_2-\frac{1}{2}F_0 \, B_2 \wedge B_2,
\end{equation}
then $H_3^g, ~ F_2^g, ~ F_4^g$ are computed here by first implementing \eqref{5d_gauging} on $B_2, ~C_1, ~ C_3$ to obtain $B_2^g, ~C_1^g, ~ C_3^g$, and then constructing the gauged field strengths from these new potentials. As an aside, note that we have chosen to write  \eqref{mIIA_embedding} in the form $F_4 \rightarrow F_4^g+ (\cdots)$, at the cost of introducing the term proportional to $\mathcal{F}\wedge \text{vol}_\Sigma$, which is not present in \cite[Eq.~(3.17)]{plethora}.\footnote{In presenting the general system in \eqref{mIIA_metric}-\eqref{mIIA_fluxes}, we have also written $F_4$ differently from \cite[Eqn.~(2.31)]{plethora}, to put it in a form where gauging inside all total derivatives appearing reproduces $F_4^g$ obtained from gauged potentials. } The results are the same.
We are also using a different normalization of the five-dimensional gauge vector $\mathcal{A}$, resulting in additional factors of 3 compared to \cite{plethora}.

The 5d solution we are interested in embedding is of course the $ds_5^2$, $\mathcal{A}$, and $\mathcal{F}$ given in \eqref{soliton}, as a mechanism to cap off the geometry and deform the dual SCFT. It is worth emphasizing that the calculations performed in \cite{plethora} ensure that for any supersymmetric AdS$_5$ solution in massive Type IIA, \eqref{mIIA_metric}-\eqref{bianchis}, one can make the substitutions \eqref{mIIA_embedding} to implement this deformation. For $\mu=0$ (in which case \eqref{soliton} can be expressed as \eqref{susy_soliton}) the resulting deformed SQFT dual will no longer be conformal, but will preserve some supersymmetry, and become effectively three-dimensional as it flows to the IR. We now consider several examples.

\subsection{Application to reductions of 6d theories on $\Sigma_2$}
\label{4b}
For the solutions to the system \eqref{mIIA_metric}-\eqref{bianchis} which we are interested in, the coordinates $(\psi,s,u)$ describe a submanifold which is fibered over a Riemann surface $\Sigma$ with coordinates $(v_1,v_2)$. This requires the warp factor $e^{2A}$ to separate into a product of functions on these two submanifolds, which we we can accomplish by adopting the ansatz
\begin{equation}\label{sep_ansatz}
    D_s=F_s(s,u)+2A_0(v_1,v_2), \qquad D_u =F_u(s,u)+2 
    \Tilde{A}_0(v_1,v_2)
\end{equation}
The Bianchi identity \eqref{bianchis} then implies
\begin{align}
\Delta_2 A_0=-\kappa e^{2A_0},
\end{align}
where $\Delta_2 = \partial_{v_1}^2+\partial_{v_2}^2$, and $\kappa$ is the curvature of the surface, which we parameterize as $ \kappa \in \{-1,0,1\}$. An appropriate solution for $A_0$ can be written 
\begin{align}
e^{A_0}=\frac{2}{1+\kappa(v_1^2+v_2^2)}.
\end{align}
We will focus specifically on the hyperbolic case $\kappa=-1$. The function $\Tilde{A}_0(v_1,v_2)$ is actually unconstrained, since it does not contribute to $s,u$ derivatives of $D_u$, and the instances in which $D_u$ is acted on only by $v_1,v_2$ derivatives cancel against one another. The local metric and volume form on the Riemann surface can then be written
\begin{align}
    ds_\Sigma^2=e^{2A_0}(dv_1^2+dv_2^2), \qquad \text{vol}_\Sigma=dA_\Sigma=e^{2A_0} \, dv_1 \wedge dv_2, \qquad A_\Sigma \equiv \star_2 d_2 A_0.
\end{align}
The one-form $A_\Sigma$ is precisely the connection appearing in $D\psi$,
\begin{equation}
    D\psi=d\psi-\frac{1}{2}\star_2 d_2 D_s=d\psi-A_\Sigma.
\end{equation}
Finally, note that for higher-genus examples, the volume $V_\Sigma$ of the Riemann surface is given by
\begin{equation}
 V_\Sigma=\int \text{vol}_\Sigma=4\pi(g-1).
\end{equation}

The separable ansatz \eqref{sep_ansatz} encompasses many interesting solutions for specific choice of $F_s$ and $F_u$, including not only those first introduced in \cite{afpt2015} and \cite{bpt2017}, but also the reductions to IIA of the GMSW solution \cite{gmsw2004}, the solution of BBBW \cite{bbbw2012}, and that of Maldacena-Nu\~nez \cite{mn2012}.

\subsubsection{AFPRT example}
First we consider deformations to an infinite family of backgrounds first introduced by Apruzzi, Fazzi, Passias, Rota and Tomasiello in \cite{afpt2015,afprt2015} (abbreviated AFPRT). These are dual to non-Lagrangian SCFTs, and can be interpreted as twisted compactifications of a D6-D8-NS5 system on a negative curvature Riemann surface $\Sigma$. 
The 6d `parent' SCFTs should be UV descriptions of linear quiver gauge theories.
To obtain the AFPRT solutions from the general AdS$_5$ massive IIA system, in  \eqref{sep_ansatz} take 
\begin{equation}
    F_s=\log (2\alpha(z)^2 \sin^2 \theta), \qquad F_u=\frac{\ddot{\alpha}(z)}{18^2\pi^2} \, \log \left( 2\, \frac{1+\cos \theta}{1 - \cos \theta} \right),
\end{equation}
and transform the coordinates $(s,u)$ as
\begin{equation}
    u=\alpha(z) \, \cos \theta, \qquad s=\frac{1}{2} \left(\frac{\dot{\alpha}(z)}{18 \pi}\right)^2.
\end{equation}
These solutions are controlled by a single function $\alpha(z)$, constrained as
\begin{equation}\label{alpha_eq}
    \dddot{\alpha}(z)=-162 \pi^3 \, F_0.
\end{equation}
Recall $F_0$ is the Romans mass. We allow it to be piece-wise constant with discontinuities at the positions of flavor D8 branes. The function $\alpha$ is related to the rank function $\mathcal{R}$ in the associated 6d linear quiver theory by
\begin{equation}
    \mathcal{R}(z)=-\frac{\Ddot{\alpha}}{81\pi^2}.
\end{equation}
Note the coordinate $z$ can be thought of like a continuum version of the direction along the quiver, in the sense developed in \cite{ct2015}.

Following the embedding procedure to lift the five-dimensional solution \eqref{soliton} into this background, we obtain the new deformed solution \cite{Chatzis:2024top}
\begin{eqnarray}\label{AFPRT_metric}
   \mathrm{d}s_{10}^2=&18\pi\sqrt{-\frac{\alpha}{6 \Ddot{\alpha}}}\left[\mathrm{d}s_5^2+\frac{1}{3} \mathrm{d}s_\Sigma^2-\frac{\Ddot{\alpha}}{6\alpha}\mathrm{d}z^2-\frac{\alpha \Ddot{\alpha}}{6 \dot{\alpha}^2-9 \alpha \Ddot{\alpha}} \left(\mathrm{d}\theta^2+\sin^2 \theta \mathcal{D}\psi^2\right)
    \right], \qquad \mathrm{d}s_\Sigma^2= \frac{4(\mathrm{d}v_1^2+\mathrm{d}v_2^2)}{(1- v_1^2-v_2^2)^2},\nonumber
    \\
    &   {\cal D}\psi= \mathrm{d}\psi - 3\mathcal{A} - A_\Sigma, \qquad 
     \mathcal{A}=q\left(\frac{1}{r^2}-\frac{1}{r_*^2} \right)  \mathrm{d}\phi, \qquad A_\Sigma= \frac{2(v_1 ~\mathrm{d}v_2- v_2 ~ \mathrm{d}v_1)}{1-v_1^2-v_2^2}, 
\end{eqnarray}
The dilaton is given by
\begin{equation}
    e^{-4\Phi}= \frac{1}{2^5 3^{17}\pi^{10}}\left( -\frac{\ddot{\alpha}}{\alpha}\right)^3 \left( 2\dot{\alpha}^2-3 \alpha \ddot{\alpha}\right)^2,
\end{equation}
and the fluxes are
\begin{align}\label{AFPRT_fluxes}
   H_3=\mathrm{d}B_2^g,& 
    \qquad 
   F_2 =F_0 B_2^g +\mathrm{d}C_1^g,
\\
   F_4 =\left(\mathrm{d}C_3^g +B_2^g  \wedge F_2 -\frac{1}{2} F_0\, B_2^g  \wedge B_2^g\right)&-\frac{\Ddot{\alpha}}{18\pi}\mathrm{d}z  \wedge \left(\star_5 \mathcal{F}-\frac{1}{3}\mathcal{F} \wedge \mathcal{D} \psi\right) -\frac{\dot{\alpha}}{54\pi}\mathcal{F}\wedge \text{vol}_\Sigma  , \nonumber
   \\
       B_2^g =\frac{1}{3} \tilde{\xi} \wedge \mathcal{D}\psi, 
       \qquad 
    C_1^g = &\frac{\Ddot{\alpha}}{162\pi^2}\cos \theta\, \mathcal{D}\psi, \qquad 
    C_3^g = \frac{\dot{\alpha}}{162\pi} \mathcal{D}\psi \wedge \text{vol}_\Sigma. \nonumber
\end{align}
We have expressed these in terms of the one-form
\begin{equation}
      \tilde{\xi}=3\pi \left( \cos\theta \mathrm{d}z-\frac{2\alpha \dot{\alpha}}{2\dot{\alpha}^2-3\alpha \Ddot{\alpha}} \sin \theta \mathrm{d}\theta \right).
\end{equation}
The coordinate $z$ takes values in a finite interval, while the angles $\theta$, $\psi$ range from $0$ to $\pi$ and $0$ to $2\pi$, respectively, parameterizing a deformed two-sphere. Depending on the boundary conditions for $\alpha$ on the $z$-interval, these solutions describe various brane systems, including:
\begin{itemize}
    \item D6-branes at endpoints where $\alpha$ has a single zero while $\ddot{\alpha} \neq 0$.  
    \item O6-D6 sources at endpoints where $\ddot{\alpha}$ has a single zero, with $\alpha \neq 0$.
    \item O8-D8 sources at single zeros of both $\Ddot{\alpha}$ and $2\dot{\alpha}^2-3 \alpha \Ddot{\alpha}$.  
\end{itemize}
If both $\alpha$ and $\Ddot{\alpha}$ have single zeros, the (deformed) $S^2$ shrinks smoothly and the metric is regular. For the moment, let us restrict attention to setups with these regular boundary conditions. So
\begin{equation}
    \alpha= \ddot{\alpha}=0 \quad \text{ at } \quad z=z_i,z_f, \qquad z \in [z_i,z_f ].
\end{equation}
In this case, the quantized $H_3$ field strength evaluates to
\begin{equation}\label{QNS_AFPRT}
     Q_{\text{NS5}}= \frac{1}{(2\pi 
     )^2} \int H_3 =\left. \left(z-\frac{2\alpha \dot{\alpha}}{2\dot{\alpha}^2-3 \alpha \ddot{\alpha}}\right) \right|_{z_i}^{z_f}=z_f-z_i
\end{equation}
Thus we restrict to $z$-intervals of integer-valued width. Moving to the Page charge associated to $F_2$, there are two-cycles on both $\Sigma$ and the $S^2$ defined by the coordinates $(\theta,\psi)$ to consider. Each are defined at fixed values $z=z_*$, and at fixed points on the complementary two-manifold. First, we compute
\begin{equation}
    Q_{\text{D6}}=\frac{1}{2\pi 
        } \int_{\Sigma} (F_2-B_2 F_0)=-\frac{\ddot{\alpha}|_{z_*}}{162 \pi^2} \cos \theta_* \, V_\Sigma.
\end{equation}
We also have 
\begin{equation} \label{QD6_AFPRT}
        Q_{\text{D6}}=\frac{1}{2\pi 
        } \int_{S^2} (F_2-B_2 F_0)= -\frac{\ddot{\alpha}|_{z_*}}{81\pi^2 
        }.
\end{equation} 
The interpretation of this flux quanta in terms of the number of D6-branes in the associated Hanany-Witten setup is somewhat subtle. Consider a linear quiver consisting of some number of rank-$N_k$ gauge nodes, described by
\begin{equation}\label{explicit_rank_func}
   \mathcal{R}(z)=-\frac{\Ddot{\alpha}}{81\pi^2}=N_k+(N_{k+1}-N_k)(z-k) \quad \text{for} \quad z \in [k,k+1], \quad k \in \mathbb{Z}.
\end{equation}
Equation \eqref{QD6_AFPRT} only appears to correctly count the number of D6-branes at the integer interval boundaries, $z_*=k$.  To exclude additional D6-charges induced on D8 sources, we can implement the large gauge transformation
\begin{equation}
    B_2 \rightarrow B_2 - d[\pi(z-k)\cos \theta \mathcal{D}\psi],
\end{equation}
which, after using \eqref{alpha_eq}, has the effect of shifting \eqref{QD6_AFPRT} to
\begin{equation}
    Q_{\text{D6}}=-\frac{\ddot{\alpha}|_{z_*}}{81\pi^2}+\frac{\dddot{\alpha}|_{z_*}}{81\pi^2}(z_*-k) .
\end{equation}
For the rank function in \eqref{explicit_rank_func} this evaluates to $N_k$, for any $z_*$ the $k^\text{th}$ interval. The Page charge associated to $F_4$ evaluates to zero, signalling the absence of D4-branes.
These remarks on the regularity and Page charges apply both to the original AdS$_5 \times \Sigma$ backgrounds and the deformed versions we have given in \eqref{AFPRT_metric}-\eqref{AFPRT_fluxes}. 

There are multiple sources of evidence that the undeformed backgrounds of this class are dual to compactifications of six-dimensional $\mathcal{N}=(1,0)$ SCFTs--- specifically UV-completions of linear quiver gauge theories.\footnote{This applies to the case of regular boundary conditions for $\alpha$. These can, however, be relaxed if the 6d quiver is suitably modified--- see the next example.} First, the conjectured duality between corresponding AdS$_7$ solutions and the six-dimensional linear quivers has passed a variety of checks, see for example \cite{ct2015}, \cite{Nunez:2023loo}. Second, very simple analytic maps between each AdS$_5 \times \Sigma$ solution and a corresponding AdS$_7$ solution exist \cite{afprt2015}, as well as holographic solutions interpolating along such a flow \cite{Merrikin:2022yho}. Third, direct comparison of observables between the AdS$_5 \times \Sigma$ solutions and the reductions of the 6d linear quivers shows a match in the appropriate limits (even though a quiver-like description of the 4d theories is not known) \cite{Merrikin:2022yho}. 

As an example, in appendix \ref{A3-anomaly} we show that the anomaly polynomial for the reductions of 6d linear quiver theories reproduces the holographic central charge result at leading order. Now, recall that the latter is essentially set by the lower-dimensional Newton constant ($G_5$, here). This was hinted at already in the 1980s--- see \cite{Brown:1986nw}. If we simply had a direct product space one would have $a$ proportional to $l^3/G_5$, with $1/G_5$ in turn proportional to the volume of the internal manifold. For metrics of the form \eqref{mIIA_metric}, the warp factor $e^W$ instead acts as the effective AdS$_5$ radius. Following the reasoning of \cite{ct2015} and others, we can deal with the dependence on internal coordinates by averaging over the space $M_5$. The resulting formula for $a$ from gravity gives
\begin{equation}\label{AFPRT_a}
a= \frac{\pi}{8}\frac{1}{G_{10}} \int e^{8W-2\Phi} \text{vol}_{M_5}=\frac{V_\Sigma}{(6\pi)^5 
} \int -\alpha \ddot{\alpha} dz,
\end{equation}
where the factors of $e^\Phi$ from the change between string and Einstein frames. Note that one has $e^{8W}$ rather than $e^{3W}$ because we have defined $\text{vol}_{M_5}$ as the volume form for $ds^2(M_5)$ without the overall $e^{2W}$ factor. For explicit choices of $\alpha(z)$ and the associated quiver, we find that the reduction of the 6d anomaly polynomial $I_8$ to the 4d polynomial $I_6$ gives the same answer for $a$ as \eqref{AFPRT_a} (in the appropriate holographic, long-quiver limit).

This check may also tell us something about the 4d SCFTs we are deforming (or at least what they are not). A heuristic 4d quiver proposal for these theories was given in \cite{Merrikin:2022yho}, but we find it fails to reproduce the leading-order coefficients agreed on by both the anomaly polynomial and holography. It seems likely to us that these 4d theories do not admit a quiver description, as described in more detail in appendix \ref{A3-anomaly}.

\subsubsection{BPT example}
The solutions found  by Bah, Passias and Tomasiello in \cite{bpt2017} (which we abbreviate as BPT) can be obtained from \eqref{sep_ansatz} by choosing 
\begin{align}
    F_s= \log & \left[8F_0^2 z^3 (1-k^3) p(z)\right], \qquad F_u=\frac{F_0}{6} \, zk \log \left[\frac{2p(z)}{z^3(1-k^3)}\right],
    \\
    &p(z)= (z_0-z)[3 \ell z_1^2-\kappa (z^2+z_0 z +z_0^2)], \label{eq:pofz}
\end{align}
 where we have introduced new coordinates $(z,k)$ related to $(s,u)$ by
\begin{equation}
    u=F_0 z^3 (2-k^3) 
    \qquad s =\frac{1}{8} \,(F_0 z^2k^2)^2.
\end{equation}
Note $\ell = \pm 1$ and $z_0,z_1$ are positive constants parameterizing this family of backgrounds. The physics of the solutions is quite rich, as they describe a system of D6-branes, D4-branes smeared in a continuous distribution, and O8-planes with D8-branes on them. In the following, we will restrict attention to the case
\begin{equation}\label{kappa_and_ell}
    \kappa=-1, \qquad \ell=+1
\end{equation}
for which the field-theoretic interpretation is best understood. \footnote{Taking, for example, $\ell=\kappa=1$, the scaling of the holographic $a$-anomaly suggests instead a five-dimensional `parent' theory rather than a six-dimensional origin as seen here.}

Applying the embedding procedure outlined earlier, we deform these BPT backgrounds to obtain new solutions. The new metric is
\begin{align} \label{BPT_metric}
   & ds_{10}^2=e^{2W}\left[ ds_5^2-\frac{p'(z)}{9z^2} ds_{\Sigma}^2+ds_3^2 \right],  \qquad \mathrm{d}s_\Sigma^2= \frac{4(\mathrm{d}v_1^2+\mathrm{d}v_2^2)}{(1- v_1^2-v_2^2)^2},
    \\
    &ds_3^2= -\frac{p'(z)}{3zp(z)} \,dz^2-\frac{z p'(z)}{3p(z)-z p'(z)}\left[ \frac{k\, dk^2}{1-k^3}+\frac{4}{3}\,\frac{p(z)(1-k^3)}{3p(z)-z p'(z)(1-k^3)} \, \mathcal{D}\psi^2\right],
\\
     & {\cal D}\psi= \mathrm{d}\psi -3\mathcal{A} - A_\Sigma, \qquad  \mathcal{A}=q\left(\frac{1}{r^2}-\frac{1}{r_*^2} \right)  \mathrm{d}\phi, \qquad A_\Sigma= \frac{2(v_1 ~\mathrm{d}v_2- v_2 ~ \mathrm{d}v_1)}{1-v_1^2-v_2^2},
    \\
    &  \qquad \qquad \qquad \qquad  e^{4W}=\frac{-z}{k \, p'(z)}[3p(z)-z p'(z)(1-k^3)],
\end{align}
where $ds_5^2$, $\mathcal{A}$ are as in  \eqref{soliton}, and the dilaton is given by
\begin{align}
    e^{-4\Phi}=F_0^4 z^3 k^5 \, \frac{-p'(3p-zp')^2}{[3p-z p'(1-k^3)]^3}.
\end{align}
The various mIIA fluxes are 
\begin{align}\label{BPT_fluxes}
   H_3=dB_2^g,
      \qquad 
      F_2=&F_0 B_2^g +dC_1^g,
\\
   F_4=\left(dC_3^g+B_2^g \wedge F_2-\frac{1}{2} F_0\, B_2^g \wedge B_2^g\right)&+\, F_0 z kd(zk) \wedge \left(\star_5 \mathcal{F}-\frac{1}{3}\mathcal{F} \wedge \mathcal{D} \psi\right) +\frac{F_0}{6}\, z^2 k^2\, \mathcal{F}\wedge \text{vol}_\Sigma , \nonumber
   \\
    B_2^g=\frac{1}{3}\tilde{\xi}\wedge \mathcal{D}\psi+&\frac{k}{9z}(3z^2 \kappa-p')\text{vol}_\Sigma, \nonumber
\\
     C_1^g=F_0 \frac{z k}{3}\frac{3p+(1-k^3)zp'}{3p-(1-k^3)zp'}\mathcal{D}\psi, 
     \qquad C_3^g&= \frac{F_0}{18} z^2k^2\left(\frac{4z^{-2}p'p+9p+(1-k^3)zp'}{3p-(1-k^3)zp'}\right) \mathcal{D}\psi \wedge \text{vol}_\Sigma, \nonumber
\end{align}
where the one-form $ \tilde{\xi}$ in this case is
\begin{equation}
    \tilde{\xi}=-k dz-\frac{3p+zp'}{3p-zp'}\, z dk.
\end{equation}
Positivity of the metric can be ensured by restricting the range of the coordinates as
\begin{equation}
    0 \leq k \leq 1, \qquad 0 \leq z \leq z_0.
\end{equation}
Note that for the choice of parameters \eqref{kappa_and_ell}, $z_0$ is the only real zero of $p$, and $p'=-3(z^2+z_1^2)$ has no real zeros.
In \cite{bpt2017} a careful study is made of the original backgrounds near the endpoints of the $z$ and $k$ intervals. The system of branes identified can be summarized as follows:
\begin{itemize}
    \item $2 Q_{\text{D8}}$-many D8-branes inside an O8 plane positioned at $k=0$, with 
    \begin{equation}
        Q_{\text{D8}}=8-2\pi F_0
    \end{equation}
    \item $Q_{\text{D6}}$-many D6-branes, positioned at the point ($k=1$, $z=z_0$) and extended on AdS$_5$ and $\Sigma$. The integer $Q_{\text{D6}}$ is related to the solution parameters by
    \begin{equation}
         Q_{\text{D6}}=\frac{1}{2\pi}\int_{\mathcal{C}_2}(F_2-B_2 F_0)= \frac{2 z_0}{3 
         } F_0,
    \end{equation}
    where $\mathcal{C}_2$ is the appropriate two-cycle, which is locally a sphere near ($k=1$, $z=z_0$).
    \item $ Q_{\text{D4}}$-many D4-branes at $z=0$. They are extended on AdS$_5$ and smeared in a continuous distribution over the Riemann surface. In the dual field theory, they play the role of simple punctures in the reduction on $\Sigma$. In terms of the volume $V_\Sigma=4\pi(g-1)$ of the Riemann surface
    \begin{equation}
       Q_{\text{D4}} =\frac{1}{(2\pi)^2} \int_{\mathcal{C}_4}(F_4-B_2 \wedge F_2 +\frac{1}{2}F_0 B_2 \wedge B_2)=\frac{V_\Sigma z_1^2}{18 \pi^2 
       }F_0,
    \end{equation}
    where $\mathcal{C}_4$ is a four-cycle at $z=0$ built from $\Sigma$ and a local two-sphere.
\end{itemize}
Note that at the other boundaries, so $z=z_0$ away from $k=0$ or $1$, and $k=1$ away from the $z=0$ or $z_0$, the $\psi$-circle shrinks smoothly. In addition to the quantized charges given above there is the number $Q_{\text{NS5}}$ of NS5-branes, whose near-horizon yields the AdS$_5$ background. This is found by integrating the NS-NS flux $H_3$ and is related to the other flux quanta by $2\pi F_0 Q_{\text{NS5}} = Q_{\text{D6}}$.

The regularity analysis and charge quantization is completely analogous for the deformed backgrounds, \eqref{BPT_metric}-\eqref{BPT_fluxes}.

Now, in the `punctureless' case $ Q_{\text{D4}}=0$ (obtained by setting the parameter $z_1$ to zero) these BPT backgrounds are an instance of \eqref{AFPRT_metric}-\eqref{AFPRT_fluxes}. Renaming the coordinate $z$ appearing there to $\zeta$, the function $\alpha$ required to get the BPT solution is
\begin{equation}\label{alpha_bpt}
    \alpha=(3\pi)^3F_0(\zeta_0^3-\zeta^3),
\end{equation}
with constants identified as $z_0= 3\pi \zeta_0$. To see this one should use the change of coordinates
\begin{equation}
    \zeta=\frac{zk}{3\pi}, \qquad \cos \theta =\frac{z_0^3-(2-k^3)z^3}{z_0^3-z^3 k^3}.
\end{equation}
Recall the six-dimensional origin of the AFPRT backgrounds.
With $n$ non-zero, the backgrounds of BPT describe an analogous reduction of a six-dimensional $(1,0)$ theory on a Riemann surface, but with simple punctures. These are the 4d SCFTs which we are deforming via the solutions in \eqref{BPT_metric}-\eqref{BPT_fluxes}. The parent six-dimensional theories are E-string theories coupled to quiver gauge theories, as described in \cite{bpt2017}. The appearance of an additional, exceptional flavour group in the 6d theories is related to the relaxation of regular boundary conditions for $\alpha$, interpreted as O8-D8 sources at that endpoint of the $z$-interval (at $\zeta=0$, in the example above).


\section{Deformed backgrounds in 11d}
\label{5-11d}
In this section, we study the embedding of the solution in \eqref{soliton} into the family of solutions introduced by Lin, Lunin and Maldacena (LLM) \cite{Lin:2004nb} and also those of  Gaiotto and Maldacena (GM) \cite{Gaiotto:2009gz} in the electrostatic form.
 These embeddings are accomplished using the fact that minimal, $D=5$ gauged supergravity is a consistent truncation of Romans' $D=5$ $SU(2)\times U(1)$ supergravity, which can in turn be uplifted to $D=11$ supergravity using the results of \cite{Gauntlett:2007sm}. Thus, we begin with a review of the bosonic content of the 5d Romans' supergravity and the embedding of \eqref{soliton} into it.
We will also describe the dual field theory interpretation of these constructions in terms of quiver theories.
\subsection{Embedding via 5d Romans supergravity}
\label{sec:Romans}
The bosonic field content of Romans' $D=5$ 
$SU(2) \times U(1)$ gauged
supergravity \cite{rom} is a scalar field $X$, a metric,  $U(1)$ and $SU(2)$ gauge
fields $B$ and $A^i$, $i=1,2,3$, and a complex two-form $C$ that
is charged under the $U(1)$ gauge field. The corresponding
field strengths for these potentials are
\begin{eqnarray}
    G&=&\dd B ,\nn F^i&=&\dd A^i-\tfrac{1}{\sqrt
2}m\epsilon_{ijk}A^j\wedge A^k , \nn F&=& \dd C+imB\wedge C ,
\end{eqnarray} 
where $m$ is related to the gauge coupling of the non-Abelian field. The equations of motion for the scalar and the
gauge fields will read
\bea 
\label{eomromans5d} \dd (X^{-1}\, {\star \dd X}) &=&\tfrac{1}{3}
X^4\, {\star G}\wedge G - \tfrac{1}{6} X^{-2} \, ({\star F^i}\wedge F^i +
{\star {C}}\wedge \bar C)\nn &&- \tfrac{4}{3}m^2\, (X^2 - X^{-1})\,
{\star \oneone} , \nonumber \\
\label{eomromans5d2} \dd (X^4\, {\star G}) &=& - \tfrac{1}{2} F^i\wedge F^i
-
\tfrac{1}{2} {\bar C}\wedge C , \nonumber \\
\label{eomromans3} D(X^{-2}\, {\star F^i})&=&- F^i\wedge G, \nonumber \\
 X^{2}\, {\star F} &=&  i\, m\, C \,, \label{eomromans4}
 \eea
in which $D(X^{-2}\star F^i)\equiv \dd (X^{-2}\star F^i)+{\sqrt
2}m\epsilon_{ijk}A^k\wedge (X^{-2}\star F^j)$, and
$\epsilon_{01234}=+1$ for the five-dimensional space. $\bar C$
is the complex conjugate of $C$. The Einstein equation
is
\bea R_{\m\n} &=& 3 X^{-2}\,  \partial_\m X\, \partial_\n X -
\tfrac{4}{3}m^2\,(X^2 + 2  X^{-1})\, g_{\m\n}\nn & & + \tfrac{1}{2}
X^4 \, (G_\m{}^\r G_{\n \r} -\tfrac{1}{6} g_{\m\n} \,
G_{\r\s}G^{\r\s}) + \tfrac{1}{2} X^{-2}\, (F^{i\ \r}_\m F^{i}_{\n\r}
- \tfrac{1}{6} g_{\m\n}\, F^i_{\r\s}F^{i\r\s})\nn & & + \tfrac{1}{2}
X^{-2}\,  ({C}_{(\m}{}^\r\,  \bar C_{\n)\r} - \tfrac{1}{6}
g_{\m\n}\, C_{\r\s}\bar C^{\r\s})\, . \label{eomromans5} \eea
These equations of motion can be derived from the following 5d
Lagrangian
\bea {\cal L} &=& R\, {\star \oneone} - 3 X^{-2} {\star  \dd X}\wedge \dd X
-\tfrac{1}{2} X^4\,  {\star G}\wedge G -\tfrac{1}{2} X^{-2}\,
({\star F^i}\wedge F^i + \star C_\2\wedge \bar C)\nn & & -\frac{i}{2m} C\wedge
\bar F- \tfrac{1}{2} F^i\wedge F^i\wedge B + 4m^2(X^2 + 2 X^{-1})\,
{\star \oneone}\, . \label{laromans} \eea
Note that if we restrict the various fields as
\begin{equation}
    X=1, \qquad F^1=F^2=C=0, \qquad F^3={\sqrt 2} G \label{RomansAR1}
\end{equation}
the equations of motion \eqref{eomromans5d} truncate to
the equations of motion for minimal $D=5$ gauged
supergravity, \eqref{eq:5dEin} with the U(1) gauge fields $\mathcal{A}$ and $B$ identified. We are interested in the soliton solution in \eqref{soliton}, which we repeat here for convenience:
\begin{align} 
ds^{2}=\frac{r^{2}}{l^{2}}&(-dt^{2}+dv_1^{2}+dv_2^{2})+\frac{l^2 dr^{2}}%
{r^{2} f(r)}+\frac{r^{2}}{l^{2}} f(r)d\phi^{2}, \qquad f(r)=1-\frac{\mu l^2}{r^{4}}-\frac{q^{2} l^2}{r^{6}}, 
\nonumber \\
&B=\mathcal{A}=q\left(  \frac{1}{r^{2}}-\frac{1}{r_{\star}^{2}}\right)  d\phi, \qquad G=\mathcal{F}=-\frac{2q}{r^3} dr \wedge d\phi.  \label{RomansAR2}
\end{align}
Now we present the uplift to the LLM setup.

\subsection{LLM embedding}
\label{5b}

The geometry of the most general AdS$_5$ solutions in $D=11$
supergravity dual to 4d $\mathcal{N}=2$ SCFTs was first presented in \cite{Lin:2004nb}. It was proven
that these supergravity solutions are determined by solutions to a
continuous three-dimensional Toda equation.  In
\cite{Gauntlett:2006ux}, the same
conditions were found from a different perspective. 

Following \cite{Gauntlett:2007sm}, we consider the embedding of the solution of the previous section (\ref{RomansAR1}, \ref{RomansAR2}) into the LLM background in 11d. The result is
\begin{align}
\frac{ds_{11}^2}{\kappa^{2/ 3}}&=e^{2 \lambda}\left[4 d s_5^{2}+y^2 e^{-6 \lambda} \, D \mu_i \, D \mu^i+\frac{4}{1-y\partial_y D_0}D \chi^2-\frac{\partial_y D_0}{y} d y^2-\frac{\partial_y e^{D_0}}{y}(dv_1^2+dv_2^2)\right]\label{GM-LLM}
\nonumber \\
&D\chi =d \chi+a_1+\mathcal{A}, \qquad \mathcal{A}=q \left(\frac{1}{r^2}-\frac{1}{r_{\star}^2}\right) d \phi, \qquad  a_1=\frac{1}{2}\left(\partial_{v_2} D_0 \, d v_1-\partial_{v_1} D_0 \, d v_2 \right).
\end{align}
Where $\k$ is the Newton constant in 11d, $d s_5^{2}$ is the line element in \eqref{RomansAR2}, the functions $\lambda(v_2,v_2,y)$ and $D_0(v_1,v_2,y)$ are related as
\begin{equation}
    e^{-6 \lambda}=\frac{-\partial_y D_0}{y(1-y\partial_y D_0)},
\end{equation}
and we have parameterized the deformed two-sphere by
\begin{align}
& D \mu_1=d \mu_1+\sqrt{2} \mu_2 A^{(3)} =d \mu_1+2 \mu_2 \mathcal{A} \\ 
& D \mu_2=d \mu_2-\sqrt{2} \mu_1 A^{(3)} =d \mu_2-2\mu_1 \mathcal{A} \\
& D \mu_3=d \mu_3,
\label{GM-LLM2}\end{align}
with the $\mu_i$ are subject to the constraint $\delta^{ij}\mu_i \mu_j=1$. For example, one can select
\begin{align}\label{mus}
    & \mu_1=\sin \theta \cos \varphi, \quad \mu_2=\sin \theta \sin \varphi, \quad \mu_3=\cos \theta.
\end{align}
The four-form field strength is given by
\begin{equation}
    G_4=G_4^g+\b_2 \wedge \mathcal{F}+\b_1 \wedge \star_5 \mathcal{F} ,
\end{equation}
where $G_4^g$ can be obtained from the four-form field strength of the LLM background by taking $d\chi +a_1 \rightarrow D\chi =d\chi+a_1+\mathcal{A}$ and $d\mu_i \rightarrow D\mu_i$. Namely
\begin{align} 
& G_4^g= 2\kappa \, \text{vol } \tilde{S}^2 \wedge \left[D\chi \wedge d(y^3 e^{-6\lambda})+y(1-y^2 e^{-6 \lambda}) da_1-\frac{1}{2} \partial_y e^{D_0}\, dv_1 \wedge dv_2\right]
\\
& \text { vol } \tilde{S}^2=\frac{1}{2} \varepsilon^{i j k} \mu_i D\mu_j \wedge D\mu_k ,
\end{align}
and we have defined 
\begin{align}\label{beta_eqs}
    &\b_1= 4 \kappa \,  d(\mu_3 y),
    \\
    &\b_2= -4\kappa \left[\frac{y^3}{2} e^{-6\lambda} \text {vol }\tilde{S}^2 + \left(\mu_3 dy+y(1-y^2 e^{-6 \lambda}) d\mu_3\right) \wedge D\chi +\frac{\mu_3}{2} \partial_y e^{D_0}\, dv_1 \wedge dv_2 \right].
\end{align}
When our deformation is turned off ($\m=q=\mathcal{A}=0$), the 4d $N=2$ SCFT duals enjoy an $SU(2)\times U(1)$ R-symmetry, which is manifest as the isometries of the internal metric. The isometries of the two-sphere parameterised by the $\mu^i$ corresponds to the $SU(2)$ part of the R-symmetry. The isometries $S^1_\chi$ corresponds to the $U(1)$ R-symmetry. The embedding we have presented twists a particular $U(1)$ inside of this original $SU(2)\times U(1)$ symmetry with the $U(1)_\phi$ of the 5d soliton solution. This is reflected in the coefficients by which the one-form $\mathcal{A}$ contributes to $D\chi$ and the $D\mu_i$.

\subsection{The electrostatic case}
\label{5c}

Let us move the new background in (\ref{GM-LLM})-(\ref{beta_eqs}) to the electrostatic notation of e.g. \cite{Gaiotto:2009gz}. This entails a transformation from the variables $[y, v_1,v_2, D_0(v_1,v_2,y)]$ to $[\sigma,\eta, V(\sigma,\eta)]$, assuming an additional isometry in the $[v_1,v_2]$ plane. In fact, defining
\begin{equation}
v_1= R \cos\beta,~~v_2=R \sin\beta, \qquad \tilde{\chi}=\chi+\beta,~~\tilde{\beta}=-\beta   
\end{equation}
and imposing that $\beta$ is the isometry direction, we have the following Backl\"und transformation,
\begin{equation}
R^2 e^{D_0(R,y)}=\sigma^2,~~y=\sigma\partial_\sigma V=\dot{V},~~\log R=\partial_\eta V=V'.\label{changeLL-GM} 
\end{equation}
Following the steps detailed in the appendix A of the paper \cite{Macpherson:2024frt} we find that the eleven dimensional metric reads
\begin{align}
\frac{ds^2_{11}}{\kappa^{2/3}}=& f_1\Big[4 ds_5^2 + f_2 D\mu_i D\mu^i + f_3 (d \tilde{\chi} + \mathcal{A})^2 + f_4(d\sigma^2+d\eta^2)+ f_5\left(d \tilde{\beta} + f_6 d\tilde{\chi} + f_6 \mathcal{A} \right)^2  \Big] ,\nonumber\\
&  f_1=\bigg(\frac{\dot{V}\tilde{\Delta}}{2V''}\bigg)^{\frac{1}{3}},~~~~f_2 = \frac{2V''\dot{V}}{\tilde{\Delta}},~~~~f_3=\frac{4\sigma^2}{\Lambda},~~~~f_4 = \frac{2V''}{\dot{V}},\nonumber\\
& f_5=\frac{2\Lambda V''}{\dot{V}\tilde{\Delta}},
    ~~~f_6=\frac{2\dot{V}\dot{V}'}{V''\Lambda},~~~
    f_7=-\frac{\dot{V}^2V''}{\tilde{\Delta}},~~f_8=\frac{1}{2}\bigg(\frac{\dot{V}\dot{V}'}{\tilde{\Delta}}-\eta\bigg),\nonumber
\\ & \tilde{\Delta} =\Lambda(V'')^2+(\dot{V}')^2,~~~~~~~~~\Lambda=\frac{2\dot{V}-\ddot{V}}{V''}.\label{def-GM}
\end{align}
Defining $D\tilde{\chi} = d\tilde{\chi}+\mathcal{A}$ the four-form field strength can be written
\begin{align} 
G_4= & 4\kappa \, d\left[ f_7 D \tilde{\chi}+f_8 d\tilde{\beta}\right] \wedge \text{vol}\tilde{S}^2+4\kappa  \,d(\mu_3  \dot{V}) \wedge \star_5 \mathcal{F}
\\
&+8\kappa(f_7 D \tilde{\chi}+f_8 d\tilde{\beta}) \wedge d\mu_3 \wedge \mathcal{F}-4\kappa \left[ d(\mu_3 \dot{V}) \wedge D\tilde{\chi}+d(\mu_3 \eta) \wedge d\tilde{\beta}\right]\wedge \mathcal{F} \nonumber.
\end{align}
Then, reducing to Type IIA along the $\beta$-direction to preserve SUSY, as explained in \cite{Macpherson:2024frt}, we find the string frame background,
\beq\label{eq:N=2} 
\begin{aligned}
ds^2&= f_1^{\frac{3}{2}} f_5^{\frac{1}{2}}\bigg[4ds^2_5+f_2 D\mu_iD\mu^i+f_4(d\sigma^2+d\eta^2)+f_3  D\tilde{\chi}^2\bigg], \\[2mm]
e^{\frac{4}{3}\Phi}&=  f_1 f_5
,~~~~  H_3 = 4\k ~d\left[f_8 \text{vol}\tilde{S}^2-\eta  \mu_3 \mathcal{F}\right],~~~~C_1=  f_6 D\tilde{\chi},~~~~,\\[2mm]
C_3&=4\k f_7 D\tilde{\chi}\wedge\text{vol}\tilde{S}^2+4\k ~\mu_3 \dot{V} \left(\star_5 \mathcal{F}-D\tilde{\chi} \wedge \mathcal{F}\right).
\end{aligned}
\eeq
We have used the relation $d[\text{vol}\tilde{S}^2]=-2d\mu_3 \wedge \mathcal{F}$. Hence 
\beq\label{eq:F4} 
\begin{aligned}
F_4&=d C_3 - H_3\wedge C_1 = 4\k ~d \left[ f_7 D\tilde{\chi} 
 \wedge\text{vol}\tilde{S}^2+\mu_3 \dot{V} \left(\star_5 \mathcal{F}-D\tilde{\chi} \wedge \mathcal{F}\right)\right]-H_3\wedge C_1.\nn 
\end{aligned}
\eeq
In the $r\to\infty$ limit, the background asymptotes to the original GM background. In this limit, the 5d subspace in \ref{eq:N=2} with metric $ds_5^2$ reduces to AdS$_5$, and the fibered sphere $\tilde{S}^2$ can be written as a round sphere $S^2$ by absorption of the constant factors in the fibered metric. This procedure can be done locally in the $r\to\infty$ limit. Page charges of the background solution can be calculated in this limit following \cite{Gaiotto:2009gz, Macpherson:2024frt}, which we will review in the next section. These will be the only relevant brane charges present in the background, as the deformation does not introduce any new cycles carrying relevant fluxes.

\subsubsection{Rank function and Page charges}\label{sec:GMPage}

Boundary conditions for the function $V(\sigma, \eta)$ encode the data of a particular dual linear-quiver field theory, via a Rank function. The formalism is described in \cite{Nunez:2019gbg} and \cite{Macpherson:2024frt}, and we provide a very brief summary here.
It is in this logic that we have an infinite number of background solutions in this family, each one of them associated with
a distinct 4d linear quiver ${\cal N}=2$ SCFT. In our configuration, these UV fixed points are deformed by compactification on the $\phi$-circle into a confining QFT.

We are interested in solutions of the Laplace equation 
\begin{equation}\label{eqn:laplace}
\frac{1}{\sigma}\partial_\sigma (\sigma \partial_\sigma V)+\partial_\eta^2 V\equiv\ddot{V}+\sigma^2V''=0.
\end{equation}
Suitable boundary conditions must be imposed on $V(\sigma,\eta)$. We choose the following conditions
\beq
\dot{V}\Big|_{\eta=0,P} =0,~~~~\dot{V}|_{\sigma=0} =\mathcal{R}(\eta), ~~~~V\Big|_{\sigma\to\infty}=0 \label{eq:boundarycondtions}
\eeq
We can reproduce a linear quiver consisting of $P-1$ nodes associated to $SU(N_i)$ vector multiplets by choosing the rank function $\mathcal{R}$ \cite{Nunez:2019gbg,Macpherson:2024frt} as
\begin{equation}\label{eqn:Rkoriginal}
\mathcal{R}(\eta) =
    \begin{cases}
     ~~~~~~~~~~~~~~~ N_1\eta & \eta\in[0,1]\\
      N_i+(N_{i+1}-N_i)(\eta-i) & \eta\in[i,i+1] ~~~ i=1,\cdots,P-2\\
   ~~~~~~~~~   N_{P-1}(P-\eta)  & \eta\in[P-1,P],
    \end{cases}     
\end{equation}
so that the range of $\eta$ is between $0$ and the integer $P$.  For this choice of rank function, one has
\begin{align}
V(\sigma,\eta) &=-\sum_{n=1}^\infty {\cal R}_n \sin\bigg(\frac{n\pi}{P}\eta\bigg) K_0\bigg(\frac{n\pi}{P}\sigma\bigg),\label{eqn:potential}
\\
{\cal R}_n&=\frac{2}{P}\int_{0}^P \mathcal{R}(\eta)\sin\bigg(\frac{n\pi}{P}\eta\bigg)d\eta =\frac{2P}{(n\pi)^2}\sum_{k=1}^Pb_k\sin\left(\frac{n\pi k}{P}\right),~~~b_k=2N_k-N_{k+1}-N_{k-1},\nonumber
\end{align}
with $K_0(\sigma)$ the modified Bessel function of the second kind and $N_i=0$  for $i=0$ and $i= P$. 
Now we quote the Page charges associated to different D-branes in the bulk geometry. The details of the calculation can be found in Appendix \ref{app:GM}. 
\begin{itemize}
    \item There are quantized charges of the NS5 branes in the background,
\beq
Q_{\text{NS5}}=-\frac{1}{(2\pi)^2}\int_{S^3} H_3=P.
\eeq
with $S^3$ being parameterised by the $(\eta,\tilde{S}^2)$ coordinates in the metric of \ref{eq:N=2}
 \item There are D6 branes in each interval of $\eta=[i,i+1]$ with the corresponding Page charges 
\beq
Q^i_{D6}=-\frac{1}{2\pi}\int_{\tilde{S}^2}F_2=2N_i-N_{i-1}-N_{i+1}.
\eeq
The integration contour in given in the Appendix \ref{app:GM}. These branes support the flavour degrees of freedom in the dual quiver field theory
.
\item We can also define a quantized Page charge for D4 branes
\beq
Q^i_{D4}=-\frac{1}{(2\pi)^3}\int_{\text{S}^2\times \tilde{\text{S}}^2}\hat F_4= -\frac{1}{(2\pi)^3}\int_{\text{S}^2\times \tilde{\text{S}}^2}(F_4-B_2 \wedge F_2)=N_i-N_{i-1},
\eeq
The choice of integration contour is discussed in Appendix \ref{app:GM}. This is the total amount of charge of D4 branes, including the `true' colour D4 present in the background and the D4 charge induced on the D6 and NS branes. The `true' D4 charge in the interval $[i,i+1]$, related to the colour degrees of freedom in the QFT, is $N_i$. 
 These are interpreted as colour branes and support the vector multiplets in the dual QFT. 
\end{itemize}
Let us provide a brief interpretation of the associated field theory.
The electrostatic backgrounds in (\ref{eq:N=2}) whose functions are written in terms of a rank function, as in (\ref{eqn:potential}) admit a very simple field theory dual. In fact, they can be thought as the field theory associated with the low-energy limit of a Hanany-Witten set-up consisting of $P$ NS five branes, that bound $N_k$ D4 branes in the k-th interval. There are $F_k=(2 N_k-N_{k+1}- N_{k-1})$ D6 branes in the same interval. The dual UV-SCFT is an ${\cal N}=2$ quiver with $(P-1)$ gauge nodes of rank $N_k$ and $F_k$ flavours. These UV SCFTs are deformed by the presence of a VEV (proportional to the parameter $q$). The deformation introduces the fibration of a diagonal U(1) group inside the R-symmetry with the compact $\phi$-direction. This is the interpretation of the infinite family of backgrounds in (\ref{eq:N=2}).

The SCFTs associated with the more generic backgrounds in (\ref{GM-LLM}) is slightly more elaborated, in the sense that the UV SCFT is non-Lagrangian. These QFTs were uncovered by Gaiotto in \cite{Gaiotto:2009we}. The deformation on those SCFTs is again via a VEV for an $U(1)$ R-symmetry current. This is holographically realised, as above, by the fibration between a diagonal $U(1)$ group in the R-symmetry and the compact $\phi$-direction. This triggers a flow ending in a confining and gapped system.

Let us now study non-perturbative aspects of all of the RG-flows in QFTs associated with the backgrounds in the previous sections.

\section{Observables}
\label{6-observables}
In this section we will calculate a plethora of observables in the field theories dual to the new families of supergravity backgrounds introduced in sections \ref{3-IIB} and \ref{4-IIA}. We will review how the calculations are realized holographically and then present the results for each case. An important feature highlighted here is the \textit{universality} of many of the observables. Although from the geometric point of view this is to be expected by construction, we appreciate the fact that even though the dual $\mathrm{QFT}s$ are very different, they all bear certain common characteristics. A significant example being the key signatures of confinement exhibited by all of the QFTs obtained from our deformation procedure. This feature can be attributed as a consequence of the Gauntlett and Varela conjecture \cite{Gauntlett:2007ma} about consistent Kaluza-Klein truncations of wrapped products of $\mathrm{AdS}$ spaces with Riemannian manifolds. We will comment further on this in the discussion section.
\subsection{Wilson loop}
\label{6a}
To compute the Wilson loop holographically, we embed a probe string with fixed endpoints at $x_1=\pm L/2$ and $r=\infty$, which enters the bulk in a $\mathrm{U}$-shaped fashion (see \cite{Sonnenschein:1999if}) while keeping the rest of the coordinates constant. Choosing to parameterise the worldsheet as $\tau=t$, $x_1 = \sigma$ while letting $r=r(x_1)$, in each example the induced metric on the string is:

\begin{equation}\label{indmetric-Wilson}
	\mathrm{d}s^2 _{\mathrm{ind}}= - \frac{r^2}{l^2} \mathrm{d}t^2 + \left( \frac{r^2}{l^2}+ \frac{l^2}{r^2f(r)}r ^{\prime 2} \right) \mathrm{d}x_1^2.
\end{equation}

We emphasize that because the string embedding extends only in the $\widehat{\mathrm{AdS}}_5$ subspace, common to all of the backgrounds, \eqref{indmetric-Wilson} is going to be the same for every case covered in this paper, modulo an overall constant factor depending on the internal coordinates. This is responsible for the \textit{universal} behaviour of the Wilson loop. The Nambu-Goto action for the probe string takes the form:

\begin{equation}\label{SNG}
	S _{\mathrm{NG}}= \frac{\mathcal{T}}{2\pi\alpha ^{\prime}}\int _{-L/2} ^{L/2}\mathrm{d}x_1 \sqrt{F^2(r) + G^2(r) r ^{\prime 2}} ,
\end{equation}

where we introduce the functions:

\begin{equation}
	F(r) = \frac{r^2}{l^2}\,\, , \,\, G(r) = \frac{1}{ \sqrt{f(r)}}.
\end{equation}
The fact that $F(r_{\star})=r_{\star}^2/l^2\neq0$ reveals a nonzero effective tension of the chromoelectric string and already hints at confining behaviour \cite{Kol:2014nqa,Faedo:2013ota,Faedo:2014naa,Kinar:1998vq}. Following \cite{Kol:2014nqa,Faedo:2014naa} we can write an approximate expression for the length of separation $\hat{L}$ of the quark anti-quark pair, as a function of the turning point of the string (denoted by $r_0$):

\begin{equation}\label{Lapp-Wilson}
 \hat{L}(r_0) = \left.\frac{\pi G(r)}{F^{\prime}(r)}\right| _{r_0}= \frac{l^2\pi}{2r\sqrt{f(r_0)}}= \frac{\pi l^2 r_0^2}{2 \sqrt{r_0^6-\mu l^2 r_0^2 - q^2l^2}}.
\end{equation}

Studying this expression we get further evidence for confinement: As the string reaches the end of the spacetime, that is $r_0\to r_{\star}$, $\hat{L}(r_0)$ diverges, allowing for an infinite separation of the pair. We can then calculate the effective potential, see \cite{Nunez_2010}, which controls the equation of motion of the string\footnote{the equation of motion reads: $$\dv{r}{x}=\pm V_{\mathrm{eff}}(r),$$ where the $\pm$ sign refers to the left and right branches of the string around $r_0$.}:

\begin{equation}
	V _{\mathrm{eff}}(r) = \frac{F(r)}{F(r_0)G(r)}\sqrt{F^2(r)-F^2(r_0)}= \frac{\sqrt{r^4-r_0^4}}{r_0^2 l^2 r} \sqrt{r^6-l^2 (\mu r^2 + q^2)}.
\end{equation}
This diverges $\sim r^4$ as $r\to\infty$ and vanishes as $r\to r_0$, satisfying the conditions for confinement \cite{Nunez_2010}. Then one can write the separation and the energy\footnote{where we normalize the expression by subtracting the contribution of an infinitely massive quark and antiquark.} of the pair as the following integrals: 

\begin{equation}\label{L-Wilson}
	L _{QQ}(r_0) =2 \int _{r_0}^{\infty}\frac{\mathrm{d}r}{V_{\mathrm{eff}}(r)}= 2 l^2 r_0^2 \int _{r_0} ^{\infty} \frac{\mathrm{d}r r}{\sqrt{\left( r^4-r_0^4 \right) \left[ r^6-l^2(q^2+r^2\mu) \right] }},
\end{equation}

\begin{equation}\label{energy-Wilson}
	E _{QQ}(r_0) = \frac{r_0^2}{l^2}L _{QQ}(r_0) + 2\int _{r_0}^{\infty} \mathrm{d}r\frac{\sqrt{r^4-r_0^4}}{r^2 \sqrt{f(r)}} - 2 \int _{r _{\star}} ^{\infty} \frac{\mathrm{d}r }{\sqrt{f(r)}},
\end{equation}

where in the case of $q=\mu=0$ we see that the length and energy reduce to the standard $\mathrm{AdS}_5$ expressions:

\begin{equation}
\begin{split}
&	L _{\mathrm{AdS}_5}(r_0) = \frac{2l^2}{r_0} \int _1 ^{\infty} \frac{\mathrm{d}y}{y^2 \sqrt{y^4-1}}=\frac{2l^2}{r_0}\frac{\sqrt{2}\pi^{3/2}}{\Gamma^2\left(\frac{1}{4}\right)},\\
&E_{\mathrm{AdS}_5}(r_0)=2r_0\left[\int_1^{\infty}\mathrm{d}y\left( \frac{y^2}{\sqrt{y^4-1}}-1\right)-1\right],\\
&E_{\mathrm{AdS}_5}(L) = -\frac{8\pi^2 \sqrt{2g_{\mathrm{YM}}^2N}}{\Gamma^{4}\left(\frac{1}{4}\right)}\frac{1}{L_{\mathrm{AdS}_5}}.
\end{split}
\end{equation}

\begin{figure}
    \centering
    \includegraphics[width=0.5\textwidth]{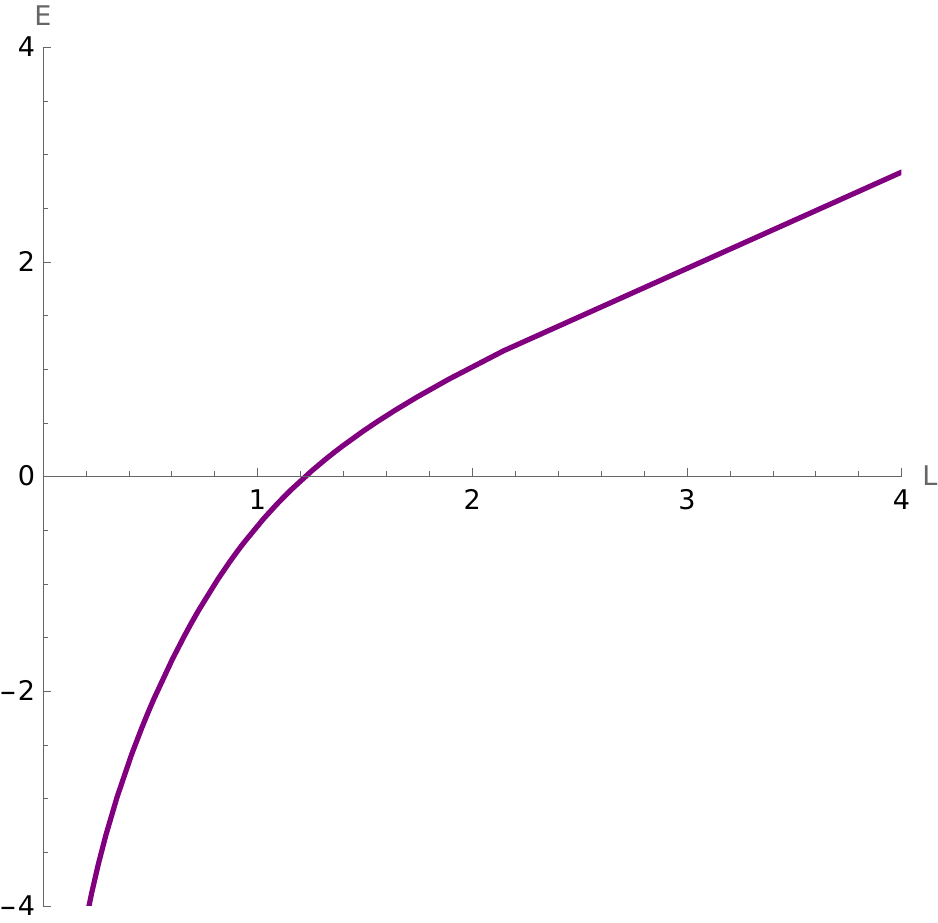}       
    \caption{Parametric plot of the energy \eqref{energy-Wilson} with respect to the length of the separation of the pair for the deformed $\mathrm{AdS}\times\mathrm{S}^5$ with $l=\mu=q=1$. This interpolates between a Coulomb-like behaviour dictated by conformality and a linear behaviour for large values of $L$, signaling confinement.}
    \label{plotE(L)-Wilson}
\end{figure}

We make use of numerical methods in order to study \eqref{L-Wilson} and \eqref{energy-Wilson} and parametrically plot the energy as a function of the length in Figure \ref{plotE(L)-Wilson}. Indeed, we see a linear behaviour for large separations that is characteristic of confinement.\\

\begin{figure}[t]
  \centering
  \includegraphics[width=0.4\textwidth]{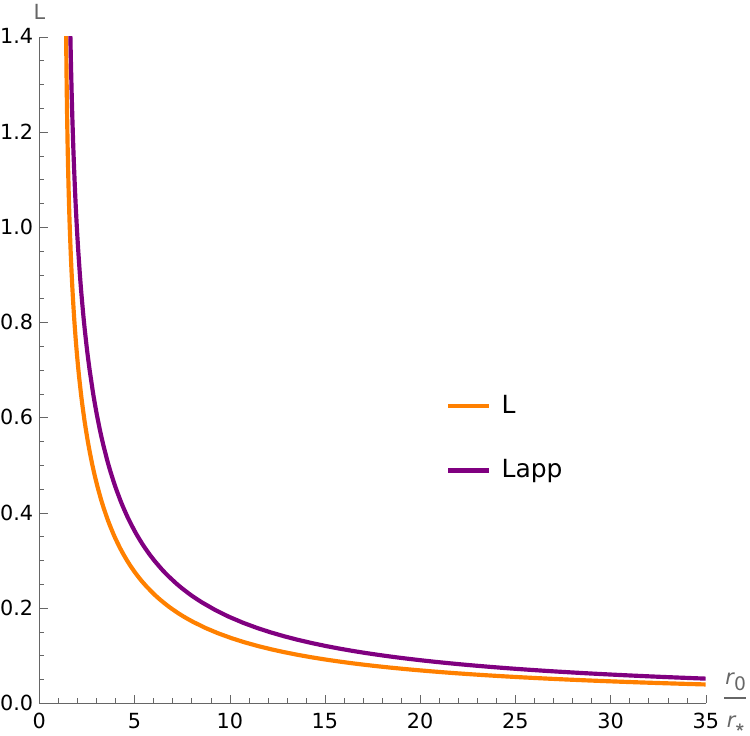}
 \hspace{0.3cm} 
 \centering
  \includegraphics[width=0.4\textwidth]{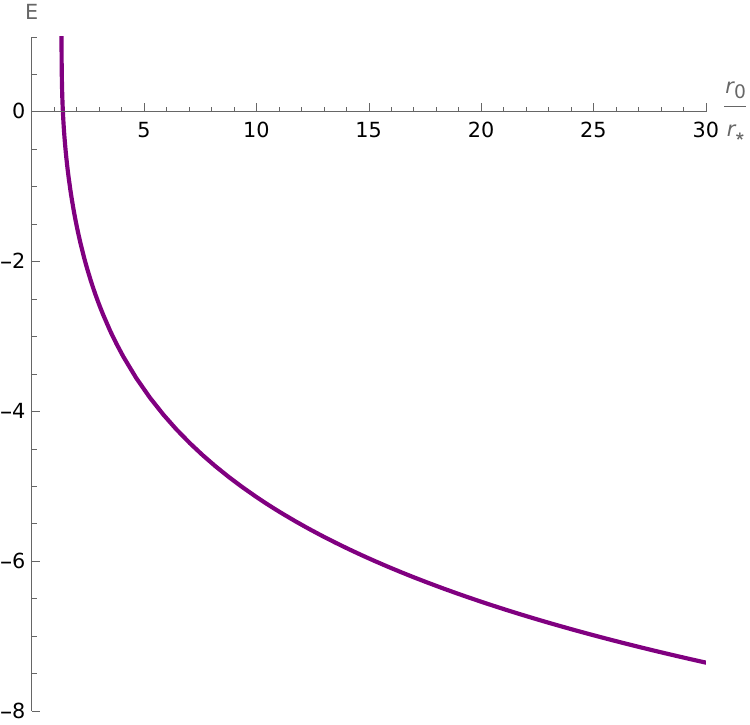}\\ \vspace{0.2cm}
  \centering
\caption{Plots of the length and energy of separation of the quark-anti-quark system \eqref{L-Wilson} as a function of the turning point of the string $r_0$, for the deformed $\mathrm{AdS}\times\mathrm{S}^5$ case. The parameters are fixed to $l=\mu=q=1$. We emphasize that the results depicted in these plots, as well as Figure \ref{plotE(L)-Wilson}, are the same for all the families of solutions we presented.}
\label{PlotsforL(r0)E(r0)E(L)-Wilson-AdS5xS5}
\end{figure}

One can address the question regarding the stability of the $\mathrm{U}$-shaped embedding that gives \eqref{indmetric-Wilson}. Although this question should be addressed by studying the equations of motion of fluctuations around the embedding, this being the most trusted way that can also detect instabilities originating from internal directions of the background, see \cite{Chatzis:2024dlt, Avramis:2006nv, Avramis:2007mv}, a usual and convenient way of extracting information about this is by calculating the derivative of the length function with respect to the turning point $r_0$ and looking at its sign \cite{Kinar:1998vq,Bachas:1986,Faedo:2013ota,Faedo:2014naa,Nunez:2023nnl}. Since here our functions $F$ and $G$ do not depend on internal space coordinates, we can use the approximate expression \eqref{Lapp-Wilson} to get:

\begin{equation}\label{Z}
Z(r_0):=\frac{\mathrm{d}\hat{L}(r_0)}{\mathrm{d}r_0}=- \frac{l^2\pi\left[2f(r_0)+r_0f^{\prime}(r_0)\right]}{4r^2f^{3/2}(r_0)}= - \frac{l^4\pi }{4r_0^3 f^{3/2}(r_0)}\left( \frac{4q^2}{r_0^5}+ \frac{2r_0}{l^2} + \frac{2\mu}{r_0^3}\right) <0,
\end{equation}
which is negative and therefore the embedding is deemed stable. The reasoning behind this criterion is that there are two conditions that need to be satisfied: That the derivative of $E(L)$ with respect to the length is positive, producing an attractive force (this is satisfied, as it is proportional to $F(r)>0$) and also that the force is a non increasing function of the separation (concavity condition) \cite{Bachas:1986}. The sign of $E^{\prime \prime}(L)$ is governed by $L^{\prime}(r_0)=Z(r_0)$. Thus the regions of $r_0$ where $Z(r_0)>0$ are unphysical and destabilize the embedding. 

We make note that the universal character of the results presented here is surprising from a purely QFT perspective. By following our deformation procedure and tuning $\mu=0$, one obtains many families of supersymmetric QFTs that are in general very different from each other (some of them not even admitting Lagrangian descriptions). Yet we have shown they all have the same confining behaviour. This showcases once more the power of holography, as from the gravity perspective the common geometric origin of this behaviour is very clear. One more thing to emphasize is that it would be interesting to probe other directions in the Wilson loop embeddings, in particular those that span $M_5$. In that we might extract new information about the QFTs and new physical phenomena.

\subsection{'t Hooft loop}
\label{6b}
To calculate the 't Hooft loop, that is the magnetic analogue of the Wilson loop, we probe the background with either a $\mathrm{D}3$ or a $\mathrm{D}4$ brane, depending on whether the background is in Type $\mathrm{IIB}$ or $\mathrm{IIA}$, that extends at least on a $\mathbb{R}^{1,2}[t,x,\phi]$ 
submanifold of the background with $r=r(x)$, where $x$ is a field theory direction and $\phi$ the shrinking circle. Then, we calculate the Dirac-Born-Infeld action for the $\mathrm{D}$-brane using the induced metric. After integrating the compact coordinates, which give a different overall factor for each background, we get a similar expression to the integral in \eqref{SNG} for the effective $2\mathrm{d}$ string:

\begin{equation}\label{S_t}
S_{\mathrm{D}p}=T_{\mathrm{D}p}\int\mathrm{d}^{p+1}\sigma e^{-\Phi}\sqrt{-\mathrm{det}(g_{\mathrm{ind}})}\propto \int _{-\frac{L}{2}} ^{\frac{L}{2}}\mathrm{d}x\sqrt{F_t^2(r)+G_t^2(r)r^{\prime 2}},\quad
 p=3,4,
\end{equation}
 but with different functions $F_t$ and $G_t$. By probing the circular direction $\phi$ with the brane, we get a factor of the  function $f(r)$ in the determinant which leads to screening, expressed as the vanishing of the function $F_t$ in the $\mathrm{IR}$: $\mathrm{T}_{\mathrm{eff}}=F_t(r_{\star})=0$.\\
 
 This fact is also true for the Entanglement Entropy studied in the next section, which has a phase transition as well. This can be attributed as a generic feature of cigar-like geometries \cite{Kol:2014nqa}. \\

Below we list the calculation of \eqref{S_t} carried out for each background, emphasising the \textit{universal} character given by the integral above.\\

\paragraph{Deformed $\mathrm{\mathbf{AdS}_{\mathbf{5}}\mathbf{\times}\mathrm{\mathbf{S}}^{\mathbf{5}}}$} We will study the dynamics of the 't Hooft loop for the background \eqref{metric-AdS5xS5} by introducing a $\mathrm{D}3$ brane on the subspace spanned by the coordinates $[t,x_1,\phi,\varphi_3]$ keeping $\theta=0$ and $r=r(x_1)$. The induced metric on this subspace then reads:

\begin{equation}
	\mathrm{d}s^2 _{\mathrm{ind}} = \frac{r^2}{l^2}\left[-\mathrm{d}t^2+\left(1+\frac{l^4}{r^4f(r)}r^{\prime 2}\right)\mathrm{d}x_1^2+f(r)\mathrm{d}\phi^2\right]+ l^2 \left( \mathrm{d}\varphi_3 + \frac{\mathcal{A}}{l}  \right)^2.  
\end{equation}
The Dirac-Born-Infeld action for this metric and a trivial dilaton, reduces to the following:\footnote{here, $L _{\phi}$ and $L_{\varphi_3}$ are the periods of each cycle.}

\begin{equation}
\begin{aligned}
	S _{\mathrm{D}3} &= \int \mathrm{d}^4\sigma e ^{-\Phi} \sqrt{-\mathrm{det}(g _{\mathrm{ind}})}=T _{\mathrm{D}3}L _{\phi}L _{\varphi_3}\mathcal{T}\int _{-L/2} ^{L/2} \mathrm{d}x_1 \sqrt{F_t^2(r)+ G_t^2(r) r^{\prime 2}}\,,
 \end{aligned}
\end{equation}
where the functions $F_t$ and $G_t$ read:
\begin{equation}\label{FG-tHooft}
	F_t(r) = \frac{r^3 \sqrt{f(r)}}{l^2}\,\, , \,\, G_t(r) = r.
\end{equation}
As mentioned, $F_t$ vanishes as $r_0\to r _{\star}$, giving off a zero effective tension. We then expect the system to screen. To prove this, we use \eqref{FG-tHooft} to calculate the approximate length $\hat{L}_{\mathrm{MM}}$, effective potential, length $L_{\mathrm{MM}}$ and energy $E_{\mathrm{MM}}$ functions of $r_0$, in the same fashion as in the Wilson loop case. These are given below: 
\begin{align}
    \hat{L}_{\mathrm{MM}} (r_0) &= \frac{\pi l^2 r_0^3}{3r_0^4-l^2\mu} \sqrt{f(r_0)}, \label{Lapp-tHooft-AdS5xS5}
    \\
    V _{\mathrm{eff}} (r) &= \frac{r^2 \sqrt{f(r)}}{l^4 F_t(r_0)}\sqrt{r^6-l^2(q^2+r^2\mu) -F_t^2(r_0)l^4},
    \\
    L _{\mathrm{MM}}(r_0) &= 2l^4 F_t(r_0) \int _{r_0} ^{\infty} \frac{\mathrm{d}r}{r^2\sqrt{f(r)\left[r^6-l^2(q^2+\mu r^2)-l^4F_t^2(r_0)  \right] }},
    \\
    E _{\mathrm{MM}}(r_0) &= \frac{\sqrt{f(r_0)}r_0^3}{l^2}L _{\mathrm{MM}}(r_0) + 2 \int _{r_0}^{\infty}\mathrm{d}r \frac{\sqrt{f(r)r^6-f(r_0)r_0^6}}{r^2 \sqrt{f(r)}} - 2\int _{r _{\star}} ^{\infty} \mathrm{d}r r. \label{E(r0)-tHooft}
\end{align}

\begin{figure}[t]
  \centering
\includegraphics[width=0.4\textwidth]{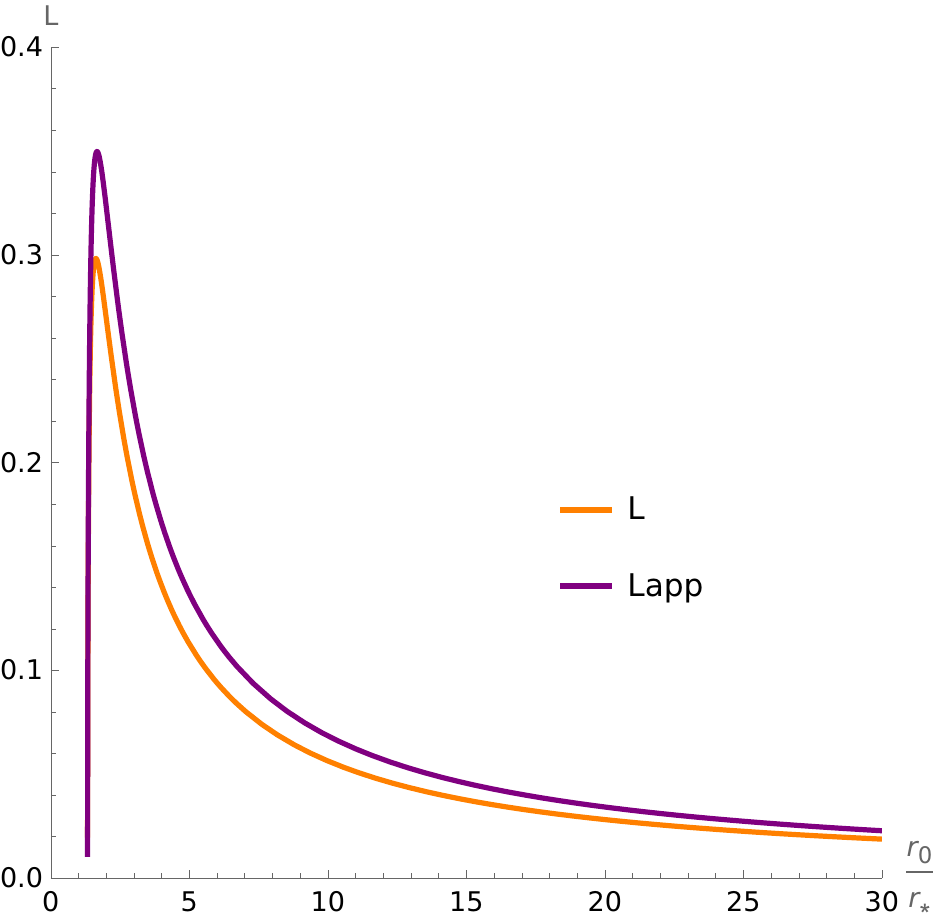}
 \hspace{0.4cm} \centering
\includegraphics[width=0.4\textwidth]{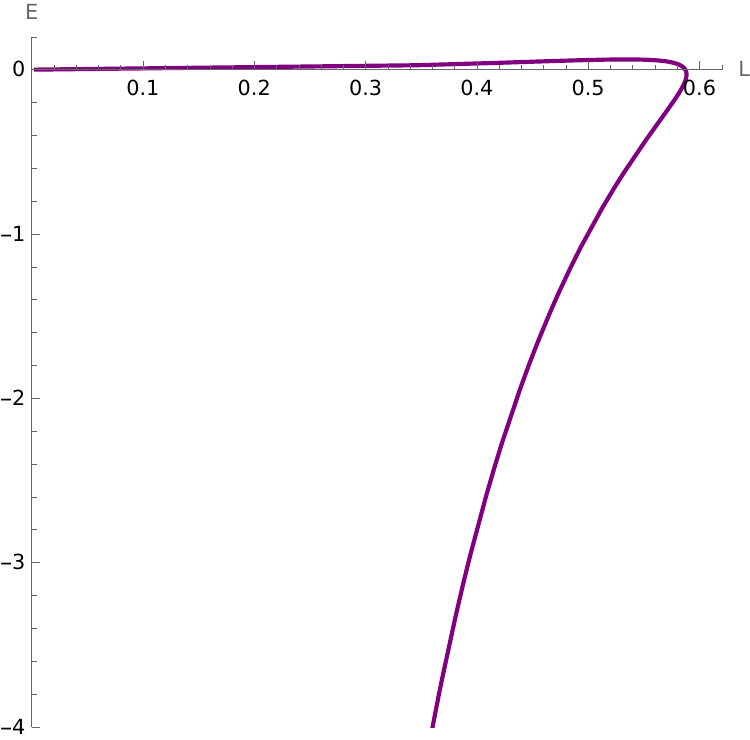}
\caption{Plots of the length function, its approximate expression and the energy as a function of the length, for the monopole-anti-monopole pair for the 't Hooft loop, with $l=\mu=q=1$ fixed. We see the double-valuedness of the energy, which expresses screening as a phase transition.}
\label{fig:L(r0)E(r0)-thooft-AdS5xS5}
\end{figure}

Notice again the factor $\sqrt{f(r_0)}$ multiplying the first term in \eqref{E(r0)-tHooft} which is responsible for the vanishing of the effective tension \cite{Kinar:1998vq,Sonnenschein:1999if}: The monopole-antimonopole pair can be arbitrarily separated with no energy cost. This can be also seen as a phase transition: The system favours the disconnected embedding of two straight strings extending from $r=\infty$ to $r=r_*$, over the $\mathrm{U}$-shaped one described in the Wilson loop section. This embedding expresses two infinitely massive probe monopoles moving freely.\\

To figure out the stability of this embedding, following the logic explained around \eqref{Lapp-tHooft-AdS5xS5}:

\begin{equation}
\begin{split}
    Z(r_0)=\dv{\hat{L}_{\mathrm{MM}}}{r_0}&=-\frac{l^2\pi\left[12f^2(r_0)-r_0^2f^{\prime 2}(r_0)+2r_0f(r_0)\left(5f^{\prime}(r_0)+r_0f^{\prime\prime}(r_0)\right) \right]}{r_0^2\sqrt{f(r_0)}\left(6f(r_0)+r_0f^{\prime}(r_0)\right)^2}\\
    &=\frac{\pi l^2 \left[l^4\mu^2-3r_0^8+6l^2(2q^2r_0^2+r_0^4\mu)\right]}{(l^2r\mu-3r_0^5)^2\sqrt{f(r_0)}},
\end{split}
\end{equation}
which changes sign around its single root. The root can be seen in Figure \ref{fig:L(r0)E(r0)-thooft-AdS5xS5} as the maximum of $L$: The left branch describes an unstable embedding ($Z(r_0)>0$) while the right branch, which corresponds to the zero energy configuration, is stable ($Z(r_0)<0$). As was discussed in \cite{Chatzis:2024dlt}, the study of small linear fluctuations agrees with these regions being unphysical, as these are the regions of $r_0$ for which their spectrum has negative eigenfrequencies.\\

\paragraph{Deformed $\mathrm{\mathbf{AdS}_{\mathbf{5}}\mathbf{\times}\mathrm{\mathbf{T}}^{\mathbf{1,1}}}$} We take a $\mathrm{D}3$ probe brane in \eqref{deformed_T11} extended on the submanifold spanned by $[t,x_1,\phi,\psi]$ and keep $\theta_1=0=\theta_2$, $\phi_1=\mathrm{const}$, $\phi_2=\mathrm{const}$, $r=r(x_1)$. The induced metric on the $\mathrm{D}3$ reads:
\begin{equation}\label{induced-metric-tHooft-T11}
    \mathrm{d}s^2 _{\mathrm{ind}} =-\frac{r^2}{l^2}\mathrm{d}t^2 +\left[ \frac{r^2}{l^2}+ \frac{l^2r^{\prime 2}}{r^2f(r)}\right]\mathrm{d}x_1^2 + \frac{r^2f(r)}{l^2} \mathrm{d}\phi^2+ \frac{l^2}{9}\left(\mathrm{d}\psi+\frac{\mathcal{A}}{l}\right)^2,
\end{equation}
and the action takes the form:
\begin{equation}
    S _{\mathrm{D}3}=\frac{T_{\mathrm{D}3}\mathcal{T}L_{\psi}L_{\phi}}{3}\int _{-L/2}^{L/2}\mathrm{d}x_1\sqrt{F_t^2(r)+r^{\prime 2}G_t^2(r)},
\end{equation}
having exactly the same functions as in the $\mathrm{AdS}_5\times \mathrm{S}^5$ uplift \eqref{FG-tHooft}, so the results for the approximate length, length and energy functions will be the same as in \eqref{Lapp-tHooft-AdS5xS5}-\eqref{E(r0)-tHooft}. We again have a vanishing $F_t(r_{\star})$ which signals screening.\\

\paragraph{Deformed $\mathrm{\mathbf{AdS}_{\mathbf{5}}\mathbf{\times}\mathrm{\mathbf{Y}}^{\mathbf{p,q}}}$} We will use in \eqref{deformed_Ypq} the same embedding for the $\mathrm{D}3$ brane as in the $\mathrm{T}^{1,1}$ case that is extended in the coordinates $[t,x_1,\phi,\psi]$ while setting $r=r(x_1)$, $\theta=0$ and $\varphi,y,\beta$ constant. The induced metric is then:
\begin{equation}
    \mathrm{d}s^2 _{\text{ind}}= \frac{r^2}{l^2}\left[ -\mathrm{d}t^2+\left(1+\frac{l^4\, r^{\prime 2}}{r^4}\right)\mathrm{d}x_1^2+f(r)\mathrm{d}\phi^2\right]+ \frac{l^2}{9}\left(\mathrm{d}\psi+\frac{A_1}{l}\right)^2,
\end{equation}
which is the same as \eqref{induced-metric-tHooft-T11} and we therefore get the exact same expressions for the 't Hooft loop in the $\mathrm{Y}^{p,q}$ uplift.\\

\paragraph{Deformed $\mathrm{\mathbf{AdS}_{\mathbf{5}}\mathbf{\times}\mathrm{\mathbf{AFPRT}}}$} Now turning to the type $\mathrm{IIA}$ solutions, we consider a probe $\mathrm{D}4$ brane in the background \eqref{AFPRT_metric} extended in $[t,x_1,\phi,\theta,\psi]$ with $r=r(x_1)$ and $x_2,z,v_1,v_2=\mathrm{const}$. We can then write the induced metric:

\begin{equation}
 \begin{split}
        \mathrm{d}s^2_{\mathrm{ind}}&=18\pi\sqrt{-\frac{\alpha}{6\ddot{\alpha}}}\left\{ \frac{r^2}{l^2}\left[ -\mathrm{d}t^2 +\left( 1+ \frac{l^4r^{\prime 2}}{r^4f(r)}\right)\mathrm{d}x_1^2 + f(r) \mathrm{d}\phi^2\right]-\frac{\alpha\ddot{\alpha}}{6\dot{\alpha}^2-9\alpha\ddot{\alpha}}\left(\mathrm{d}\theta^2+\sin^2\theta D\psi^2\right)\right\}.
    \end{split}
    \end{equation}

In this case, the action takes the form:

\begin{equation}
    \begin{split}
        \mathrm{S}_{\mathrm{D}4}=\frac{4\pi}{3}\mathrm{T}_{\mathrm{D}4}\mathcal{T}L_{\phi} \sqrt{- \frac{\alpha^3(z_{0})\ddot{\alpha}(z_{0})}{6\dot{\alpha}^2(z_{0})-9\alpha(z_{0})\ddot{\alpha}(z_{0})}} \int _{-L/2}^{L/2}\mathrm{d}x_1\sqrt{F_t^2(r)+r^{\prime 2}G_t^2(r)},
  \end{split}
\end{equation}

where $z_{0}$ is a constant value and now $F_t$ and $G_t$ differ from \eqref{FG-tHooft} by a factor of $l^{-1}$:

\begin{equation}\label{FGtl-1}
    F_t(r) = \frac{r^3\sqrt{f(r)}}{l^3}\,\, , \,\, G_t(r) = \frac{r}{l},
\end{equation}

\paragraph{Deformed $\mathrm{\mathbf{AdS}_{\mathbf{5}}\mathbf{\times}\mathrm{\mathbf{GM}}}$} For the background \eqref{eq:N=2}, we will again consider a $\mathrm{D}4$ brane that extends in the submanifold $[t,x_1,\phi,\varphi,\chi]$ setting $\theta=\frac{\pi}{2}$ and all the other coordinates constant, as well as $r=r(x_1)$. The induced metric on the $\mathrm{D}4$ is then\footnote{where we used that $D\mu_i D\mu_i \Big|_{\theta=\pi/2}=\sin^2\varphi(\mathrm{d}\varphi-2\mathcal{A})^2+\cos^2\varphi(\mathrm{d}\varphi -2\mathcal{A})^2=(\mathrm{d}\varphi -2\mathcal{A})^2$ from \eqref{GM-LLM2} and \eqref{mus}.}:

\begin{equation}
\mathrm{d}s^2_{\mathrm{ind}}= f_1^{3/2}f_5^{1/2}\Bigg\{ 4 \frac{r^2}{l^2}\left[-\mathrm{d}t^2 \left(1+ \frac{l^4 r^{\prime 2}}{r^4 f(r)}\right) \mathrm{d}x_1^2 + f(r)\mathrm{d}\phi^2\right]+ f_2 (\mathrm{d}\varphi-2\mathcal{A})^2+f_3D\chi^2\Bigg\}.
\end{equation}

From this, we get the following action, that is again indicating screening:

\begin{equation}
\begin{split}
S_{\mathrm{D}4} = 32\mathrm{T}_{\mathrm{D}4} \mathcal{T}L_{\phi}\pi^2f_1(\sigma_{0},\eta_{0})^3\sqrt{f_2(\sigma_{0},\eta_{0})f_3(\sigma_{0},\eta_{0})f_5(\sigma_{0},\eta_{0})}\int_{-L/2}^{L/2}\mathrm{d}x_1\sqrt{F_t^2(r)+G_t^2(r)r^{\prime 2}},
\end{split}
\end{equation}
where $(\sigma_{0},\eta_{0})$ are constant values of the coordinates and $F_t$ and $G_t$ are the same as \eqref{FGtl-1}.\\

\subsection{Entanglement Entropy}\label{6c}
For the calculation of the Entanglement Entropy, we will consider a codimension 2 manifold $\Sigma_8$ connecting two entangled regions and take one of them to be of width $L_{\mathrm{EE}}$ while the remaining space comprises the other region \cite{Ryu:2006bv,Kol:2014nqa,Klebanov:2007ws}. Then we write down the action, that is the square root of the determinant of the induced metric on $\Sigma_8$ multiplied by a power of the dilaton (in the cases where $\Phi\neq 0$). One integrates out the 7 coordinates and ends up with an integral expression:

\begin{equation}\label{S_EE}
    S _{\mathrm{EE}} = \frac{1}{4G_N} \int _{\Sigma _8}\mathrm{d}^8\sigma \sqrt{e^{-4\Phi}\mathrm{det}\left( g _{\Sigma _8} \right) }\propto \int _{-\frac{L}{2}} ^{\frac{L}{2}}\mathrm{d}x_1\sqrt{F_{\mathrm{EE}}^2(r)+G_{\mathrm{EE}}^2(r)r^{\prime 2}},
\end{equation}
where $G_N$ denotes Newton's constant in $10$ dimensions.\\

We will now present the results for each background, calculating the action to observe that $F_{\mathrm{EE}}$ and $G_{\mathrm{EE}}$ are the same functions as in the 't Hooft loop \eqref{FGtl-1}. This will entail a phase transition in the Entanglement Entropy. This has been argued to be yet another signal for confinement \cite{Klebanov:2007ws}.\\

\paragraph{Deformed $\mathrm{\mathbf{AdS}_{\mathbf{5}}\mathbf{\times}\mathrm{\mathbf{S}}^{\mathbf{5}}}$} To calculate the Entanglement Entropy for this background, we consider the 8 dimensional manifold $\Sigma _8[x_1,x_2,\phi,\theta,\varphi,\varphi_1,\varphi_2,\varphi_3]$ and $r=r(x_1)$. The induced metric is:

\begin{equation}
	\mathrm{d}s^2 _{\mathrm{ind}} = \frac{r^2}{l^2}\left[\left(1+\frac{l^4  r^{\prime 2}}{r^4f(r)}\right)\mathrm{d}x_1^2+\mathrm{d}x_2^2+f(r)\mathrm{d}\phi^2\right]+ l^2 \mathrm{d}\tilde{\Omega}_5,
\end{equation}

where $\mathrm{d}\tilde{\Omega}_5$ denotes the element on the fibered five-sphere spanned by $[\varphi,\theta,\varphi_1,\varphi_2,\varphi_3]$. After computing the determinant, the action \eqref{S_EE} reads:

\begin{equation}\label{dimi10}
\begin{split}
		S _{\mathrm{EE}} &=\mathrm{V}_{\tilde{\mathrm{S}}^5} \frac{L _{x_2}L _{\phi}l^5}{4G_N}\int_{-L/2}^{L/2} \mathrm{d}x_1~\sqrt{\frac{r^6}{l^6} f(r) \left[1+ \frac{l^4 r'^2}{r^4 f(r)} \right]}\\
  &=\frac{l^5 L_{x_2} \pi^3 L_\phi}{4G_N} \int_{-L/2}^{L/2} \mathrm{d}x_1~\sqrt{\frac{r^6}{l^6} f(r) \left[1+ \frac{l^4 r'^2}{r^4 f(r)} \right]}.
\end{split}
\end{equation}

Here, $\mathrm{V}_{\tilde{\mathrm{S}}^5}$ denotes the volume of the five-sphere mentioned above\footnote{We make note that the volume of the internal spaces is preserved by the fibration, in this case, $\mathrm{V}_{\mathrm{S}^5}=\mathrm{V}_{\tilde{\mathrm{S}}^5}$. We also use throughout this paper the convention that $\mathrm{vol}_{\mathcal{M}}$ denotes the volume form of a manifold $\mathcal{M}$, while $\mathrm{V}_{\mathcal{M}}$ its integrated volume.} and $L _{x_2}$ the extent of the variable $x_2$. We see that one gets the same functions as in \eqref{FGtl-1} (we chose to keep the extra $l$ factor here by having $l^5$ in the coefficient), therefore the length function and the behaviour will be the same. That is, $\mathrm{T}_{\mathrm{eff}}=0$ in the $\mathrm{IR}$ and the system exhibits a phase transition, which is a sign of confinement.\\


\paragraph{Deformed $\mathrm{\mathbf{AdS}_{\mathbf{5}}\mathbf{\times}\mathrm{\mathbf{T}}^{\mathbf{1,1}}}$} We define the codimension two submanifold spanned by all the compact and two spatial directions of the five-dimensional solution and $\mathrm{T}^{1,1}$: $\Sigma_8 [x_1,x_2,\phi,\theta_1,\theta_2,\phi_1,\phi_2,\psi]$ with $r=r(x_1)$. The metric induced on $\Sigma_8$ is:
\begin{equation}
\begin{split}
\mathrm{d}s^2_{\mathrm{ind}}&=\left[ \frac{r^2}{l^2}+ \frac{l^2r^{\prime 2}}{r^2f(r)}\right]\mathrm{d}x_1^2+ \frac{r^2}{l^2}\mathrm{d}x_2^2+\frac{l^2f(r)}{r^2}\mathrm{d}\phi^2\\
&+\frac{l^2}{6}\sum_{i=1}^2\left(\mathrm{d}\theta_i^2+\sin^2\theta_i\mathrm{d}\phi_i^2\right) + \frac{l^2}{9}\left(\mathrm{d}\psi +\sum_{i=1}^2 \cos\theta_i\mathrm{d}\phi_i+ \frac{\mathcal{A}}{l}\right)^2.
\end{split}
\end{equation}
We now calculate the action, for which we find:
\begin{equation}
S_{\mathrm{EE}}= \frac{l^5L_{\phi}L_{x_2}\mathrm{V}_{\mathrm{T^{1,1}}}}{4G_N}\int _{-L/2}^{L/2}\mathrm{d}x_1 \sqrt{F_{\mathrm{EE}}^2(r)+G_{\mathrm{EE}}^2(r)r^{\prime 2}},
\end{equation}
where the volume is:
\begin{equation}
   \mathrm{V}_{\mathrm{T}^{1,1}}=\frac{1}{108}\int_{\mathrm{T}^{1,1}}\mathrm{d}\theta_1\mathrm{d}\theta_2\mathrm{d}\phi_1\mathrm{d}\phi_2\mathrm{d}\psi \sin\theta_1\sin\theta_2=\frac{16\pi^3}{27},
\end{equation}

and once again we get the same functions as in \eqref{FGtl-1}. Using the same arguments, we conclude that a phase transition takes place, signaling confinement.\\

\paragraph{Deformed $\mathrm{\mathbf{AdS}_{\mathbf{5}}\mathbf{\times}\mathrm{\mathbf{Y}}^{\mathbf{p,q}}}$} In the same manner as previously, we define the eight dimensional manifold to be $\Sigma_8[x_1,x_2,\phi,\theta,\varphi,y,\beta,\psi]$, $r=r(x_1)$ and the induced metric reads:
\begin{equation}
    \begin{split}
\mathrm{d}s^2 _{\mathrm{ind}}&= \frac{r^2}{l^2}\left[\left(1+\frac{l^4\, r^{\prime 2}}{r^4}\right)\mathrm{d}x_1^2+\mathrm{d}x_2^2+f(r)\mathrm{d}\phi^2\right]+\frac{(1-y)l^2}{6}\left(\mathrm{d}\theta^2+\sin^2\theta\mathrm{d}\varphi^2\right)\\
&+\frac{l^2}{w(y)v(y)}\mathrm{d}y^2+\frac{w(y)v(y)l^2}{36}\left(\mathrm{d}\beta+\cos\theta\mathrm{d}\varphi\right)^2\\
   & +\frac{l^2}{9}\left(\mathrm{d}\psi- \cos\theta\mathrm{d}\varphi+y\left(\mathrm{d}\beta+\cos\theta\mathrm{d}\varphi\right)+\frac{A_1}{l}\right)^2,\\
    \end{split}
\end{equation}
which yields the following action:
\begin{equation}
\begin{split}
    S_{\mathrm{EE}}&= \frac{L_{x_2}L_\phi L_{\beta}L_{\psi}\pi l^5}{108G_N}\int _{y_1}^{y_2} \mathrm{d}y (y-1) \int _{-L/2}^{L/2} \mathrm{d}x_1 \sqrt{\frac{r^6}{l^6}f(r)+\frac{r^2}{l^2}r^{\prime 2}}\\
    &= \frac{L_{x_2}L_\phi l^5}{ 4G_N}\mathrm{V}_{\mathrm{Y}^{p,q}}\int _{-L/2}^{L/2} \mathrm{d}x_1 \sqrt{F_{\mathrm{EE}}^2(r)+G_{\mathrm{EE}}^2(r)r^{\prime 2}},
    \end{split}
\end{equation}
which once again gives off \eqref{FGtl-1}. We can also substitute the formula of the volume of the Sasaki-Einstein manifold for completeness, that is given in terms of $p$ and $q$ \cite{Martelli:2004wu}:
\begin{equation}\label{volYpq}
\begin{split}
\mathrm{V}_{\mathrm{Y}^{p,q}}&= \frac{1}{108}\int _{\mathrm{Y}^{p,q}}\mathrm{d}\theta\mathrm{d}\varphi\mathrm{d}y\mathrm{d}\beta\mathrm{d}\psi (y-1)\sin\theta=\frac{8\pi^3}{27}\int_{y_1}^{y_2}\mathrm{d} y (y-1)\\
&=\frac{q^2\left(2p+\sqrt{4p^2-3q^2}\right)\pi^3}{3p^2\left(3q^2-2p^2+p\sqrt{4p^2-3q^2}\right)},
\end{split}
\end{equation}
where the reader can check that this reproduces the volume of $\mathrm{T}^{1,1}$ as $(p,q)\to(1,0)$. We can therefore write the general expression:

\begin{equation}\label{EE_Ypq}
S_{\mathrm{EE}}=\frac{L_{x_2}L_\phi q^2\left(2p+\sqrt{4p^2-3q^2}\right)\pi^3l^5}{12 G_Np^2\left(3q^2-2p^2+p\sqrt{4p^2-3q^2}\right)}\int _{-L/2}^{L/2} \mathrm{d}x_1 \sqrt{F_{\mathrm{EE}}^2(r)+G_{\mathrm{EE}}^2(r)r^{\prime 2}}.
\end{equation}

\paragraph{Deformed $\mathrm{\mathbf{AdS}_{\mathbf{5}}\mathbf{\times}\mathrm{\mathbf{AFPRT}}}$} For this calculation, we choose the eight-manifold as $\Sigma_8[x_1,x_2,\phi,\theta,\psi,v_1,v_2,z]$ with $r=r(x_1)$. The induced metric on $\Sigma_8$ is written as:

\begin{equation}
\begin{split}
    \mathrm{d}s_{\mathrm{ind}}^2= 18\pi\sqrt{-\frac{\alpha}{6\ddot{\alpha}}}&\biggl\{  \frac{r^2}{l^2}\left[ \left(1+ \frac{l^4 r'^2}{r^4 f(r)} \right)\mathrm{d}x_1^2+ \mathrm{d}x_2^2+ f(r)\mathrm{d}\phi^2 \right] 
+\frac{1}{3} \mathrm{d}s_\Sigma^2-\frac{\Ddot{\alpha}}{6\alpha}\mathrm{d}z^2\\
 & -\frac{\alpha \Ddot{\alpha}}{6 \dot{\alpha}^2-9 \alpha \Ddot{\alpha}} \left(\mathrm{d}\theta^2+\sin^2 \theta \mathcal{D}\psi^2\right)
\biggr\},
\end{split}
\end{equation}
and the action reads:

\begin{equation}
S_{\mathrm{EE}}=\frac{L_{x_2}L_{\phi}}{486 G_N}\mathrm{V}_{\Sigma} \int_0^P \mathrm{d}z \left(-\alpha\ddot{\alpha}\right)\int _{-L/2}^{L/2}\mathrm{d}x_1\sqrt{F_{\mathrm{EE}}^2(r)+r^{\prime 2}G_{\mathrm{EE}}^2(r)},
\end{equation}
where the volume of the Riemann surface is given by
\begin{equation}
    \mathrm{V}_{\Sigma}=4\int    \frac{\mathrm{d}v_1\mathrm{d}v_2}{(v_1^2+v_2^2-1)^2}.
\end{equation}
Here $P$ denotes the number of nodes in the dual quiver description and the functions $F_{\mathrm{EE}}$ and $G_{\mathrm{EE}}$ are again the same as in \eqref{FGtl-1}. Note that the factor $\int\mathrm{d}z\alpha\ddot{\alpha}$ is seen also in \eqref{AFPRT_a} and will make its appearance in the calculation of the flow central charge and holographic complexity studied in the upcoming sections.

\paragraph{Deformed $\mathrm{\mathbf{AdS}_{\mathbf{5}}\mathbf{\times}\mathrm{\mathbf{GM}}}$} Here we take the 8-manifold to extend in $\Sigma_8[x_1,x_2,\phi,\theta,\varphi,\sigma,\eta,\tilde{\chi}]$ and let $r=r(x_1)$ as usual. The metric on $\Sigma_8$ then reads:

\begin{equation}
    \begin{split}
        \mathrm{d}s^2_{\mathrm{ind}}= f_1^{3/2}f_5^{1/2}&\Bigg\{ 4 \frac{r^2}{l^2}\left[ \left(1+ \frac{l^4 r^{\prime 2}}{r^4 f(r)}\right) \mathrm{d}x_1^2 + \mathrm{d}x_2^2 + f(r)\mathrm{d}\phi^2\right] \\
        &+ f_2 D\mu_i D\mu_i + f_4 (\mathrm{d}\sigma^2+\mathrm{d}\eta^2) + f_3 D\tilde{\chi}^2\Bigg\}.
    \end{split}
\end{equation}
Then, the action can be written using the definitions of the various $f_i$'s from \eqref{def-GM}, in the following form:

\begin{equation}\label{EE_GM}
    S_{\mathrm{EE}} = \frac{64\pi^2L_{\phi}L_{x_2}}{G_N}\int_0^P\mathrm{d}\eta\int_0^{\infty}\mathrm{d}\sigma \sigma \dot{V}V^{\prime\prime}\int _{-L/2}^{L/2}\mathrm{d}x_1 \sqrt{F_{\mathrm{EE}}^2(r)+G_{\mathrm{EE}}^2(r)r^{\prime 2}},
\end{equation}
where again, we end up with the same functions \eqref{FGtl-1} in the $x_1$ integral.\\

Concluding this section, we emphasize that our results for the Entanglement entropy are all of the form:

\begin{equation}\label{EE_formula_Dimitris}
S_{\mathrm{EE}}=\frac{\mathcal{N}_{\mathrm{i}}}{4}\int _{-L/2} ^{L/2} \mathrm{d}x_1 \sqrt{F^2_{\mathrm{EE}}(r) + G^2_{\mathrm{EE}}(r) r^{\prime 2}},
\end{equation}
where the functions $F_{\mathrm{EE}}$ and $G_{\mathrm{EE}}$ are found to be the same as in the case of the 't Hooft loop. This implies a phase transition which following \cite{Klebanov:2007ws,Kol:2014nqa} is indicative of confinement. We will see that the constants $\mathcal{N}_{i}$, that depend on the details of each background, are related to the counting of the degrees of freedom and complexity of the system as analysed more closely in sections \ref{6d},\ref{6e}. These are given along with formulas like \eqref{EE_formula_Dimitris} for the observables describing the aforementioned properties in table \ref{table_Dimitris} towards the end of subsection \ref{6e}.


\subsection{Flow central charge}
\label{6d}
In a generic $\mathrm{CFT}$ the central charge of the theory captures the number of degrees of freedom and is equivalent to the free energy of the system. The observable studied here called the holographic flow central charge, is a monotonic function that extends this notion in holography for a flow across dimensions. The quantity takes a constant value at the $\mathrm{CFT}$ fixed points and describes the number of degrees of freedom as the system flows between them \cite{Macpherson:2014eza,Bea:2015fja}. We make note that the dimension of the QFT changes along the $\mathrm{RG}$ flow in the backgrounds studied here. The flow central charge is capable of  detecting if the theory is gapped as well as the existence of any conformal fixed points.  \\

 At the $\mathrm{CFT}$ points, the central charge is proportional to the volume of the internal space $M_5$. The flow central charge is defined following this fact. We start by writing a background dual to a $(d+1)$-dimensional $\mathrm{QFT}$ in the form \cite{Bea:2015fja}:

\begin{eqnarray}
& & \mathrm{d}s^2=-\alpha_0 \mathrm{d}t^2 +\alpha_1 \mathrm{d}x_1^2+ \alpha_2 \mathrm{d}x_2^2+\dots+ \alpha_d \mathrm{d}x_d^2 +\left(\alpha_1\alpha_2\cdots\alpha_d\right)^{\frac{1}{d}}\beta(r) \mathrm{d}r^2+ g_{ij}(\mathrm{d}\theta^i-A^i)(\mathrm{d}\theta^j-A^j),\nonumber\\
& &\Phi=\Phi(r,\theta^i),\label{flow-back}
\end{eqnarray}

for some functions $\alpha_i(r),\beta(r)$. The last part expresses the internal manifold. We then define the submanifold that includes the field theory directions and the internal space: 

\begin{equation}
		G_{ij}\mathrm{d}\xi^i \mathrm{d}\xi^j= \alpha_1 \mathrm{d}x_1^2+ \alpha_2 \mathrm{d}x_2^2+....+ \alpha_d \mathrm{d}x_d^2 + g_{ij}(\mathrm{d}\theta^i-A^i)(\mathrm{d}\theta^j-A^j).
  \end{equation}
For our purpose we have $d=3$. Then, the flow central charge can be calculated by first working out the following quantity:

\begin{equation}\label{Hfunction}
	H(r) = \left[\int \mathrm{d}^5\theta \sqrt{e ^{-4\Phi}\mathrm{det}\left( G _{ij} \right) }\right]^2.
\end{equation}
Following \cite{Bea:2015fja,Merrikin:2022yho}, we have that:

\begin{equation}
    c _{\mathrm{flow}} = \frac{d^{d}\beta_0 ^{d/2}H^{\frac{2d+1}{2}}}{ G ^{(10)} _{N} \left( H^{\prime} \right) ^d}.
\end{equation}

We list the calculation of $c_{\mathrm{flow}}$ for all the different backgrounds below, where once more we find the different dual $\mathrm{QFT}$s to have a \textit{universal} character for this observable as well.\\
 
\paragraph{Deformed $\mathrm{\mathbf{AdS}_{\mathbf{5}}\mathbf{\times}\mathrm{\mathbf{S}}^{\mathbf{5}}}$} We can rewrite the background in the form \eqref{flow-back}, where: 

\begin{equation}
	\alpha _0 = \frac{r^2}{l^2}= \alpha_2\,\,  , \,\, \alpha_1 = \frac{r^2}{l^2}f(r)\,\, , \,\, \beta_0 = \frac{l^4}{r^4}f^{-4/3}(r),
\end{equation}
and the last part expresses the fibered five-sphere:
\begin{equation}
\begin{split}
	g _{ij}\left(\mathrm{d}\theta ^{i}- A^{(i)}\right)\left(\mathrm{d}\theta ^{j}-A^{(j)}\right)=l^2\mathrm{d}\tilde{\Omega}^2_5&= l^2 \left\{ \mathrm{d}\theta^2+ \sin^2\theta \mathrm{d}\varphi ^2 + \sin^2\theta\sin^2\varphi \left( \mathrm{d}\varphi_1 + \frac{A_1}{l} \right) ^2\right.\\
	+&\left.\sin^2\theta \cos^2\varphi \left( \mathrm{d}\varphi_2 + \frac{A_1}{ l} \right) ^2 + \cos^2\theta \left( \mathrm{d}\varphi_3 + \frac{A_1}{l} \right) ^2\right\}.
\end{split}
\end{equation}

We now define the subspace:    
\begin{equation}
	\begin{split}
		G _{ij}\mathrm{d}\xi^i \mathrm{d}\xi ^j &= \alpha_1 \mathrm{d}\phi^2 + \alpha_2 (\mathrm{d}x_1^2+\mathrm{d}x_2^2)+ g _{ij}(\mathrm{d}\theta ^i- A^{(i)} )( \mathrm{d}\theta ^{j} -A^{(j)} )\\
		&= \frac{r^2}{l^2}f(r) \mathrm{d}\phi ^2 + \frac{r^2}{l^2}\left( \mathrm{d}x_1^2+ \mathrm{d}x_2^2 \right) + l^2 \mathrm{d}\widehat{\Omega}_5,
	\end{split}
\end{equation}
after taking the determinant and calculating \eqref{Hfunction} to be:

\begin{equation}
	H(r)=\left(\mathrm{V}_{\mathrm{S}^5}l^2r^3\sqrt{f(r)}  \right) ^2=l^4\pi^6r^6f(r),
\end{equation}
 we find:
\begin{equation}\label{cflowS5}
	c _{\mathrm{flow}} =\frac{l^8\mathrm{V}_{\tilde{\mathrm{S}}^5}}{8G_N}\left[\frac{\sqrt{f(r)}}{f(r)+\frac{r}{6}f^{\prime}(r)}\right]^3 = \frac{27 l^8\pi^3r^3}{8G_N} \frac{\left( r^6-l^2r^2\mu - l^2q^2 \right) ^{3/2}}{\left( 3r^4-l^2\mu \right) ^{3}}.
\end{equation}
We immediately notice by power counting that this function takes a fixed value at infinity:
\begin{equation}
    c_{\mathrm{UV}}=\lim _{r\to\infty}c(r)=\frac{l^8\pi^3}{8G_N},
\end{equation}
which is indicative of the conformal symmetry of the 4-dimensional QFT in the $\mathrm{UV}$, and depends on the details of the internal space for each background ($\tilde{\mathrm{S}}^5$, in the present example). We also notice as the system flows towards the $\mathrm{IR}$, $c_{\mathrm{flow}}$ vanishes due to the $f$ factor in the numerator. This expresses the absence of dynamical degrees of freedom in the deep $\mathrm{IR}$, where the system is gapped and governed by a $\mathrm{TQFT}$. These facts are true for all the solutions studied here, all having a similar functional form as \eqref{cflowS5}.\\

\paragraph{Deformed $\mathrm{\mathbf{AdS}_{\mathbf{5}}\mathbf{\times}\mathrm{\mathbf{T}}^{\mathbf{1,1}}}$} Here, we have again
\begin{equation}
	\alpha _0 = \frac{r^2}{l^2}= \alpha_2\,\,  , \,\, \alpha_1 = \frac{r^2}{l^2}f(r)\,\, , \,\, \beta_0 = \frac{l^4}{r^4}f^{-4/3}(r),
\end{equation}
and the internal space is written as:

\begin{equation}
   g_{ij}(\mathrm{d}\theta^i-A^i)(\mathrm{d}\theta^j-A^j)=l^2\left\{\frac{1}{6}\sum_{i=1}^2\left(\mathrm{d}\theta_i^2+\sin^2\theta_i\mathrm{d}\phi_i^2\right) + \frac{1}{9}\left(\mathrm{d}\psi + \sum_{i=1}^2\cos\theta_i\mathrm{d}\phi_i+\frac{3}{l}\mathcal{A}\right)^2\right\}.
\end{equation}

  \begin{figure}[t]
  \centering
\includegraphics[width=0.5\textwidth]{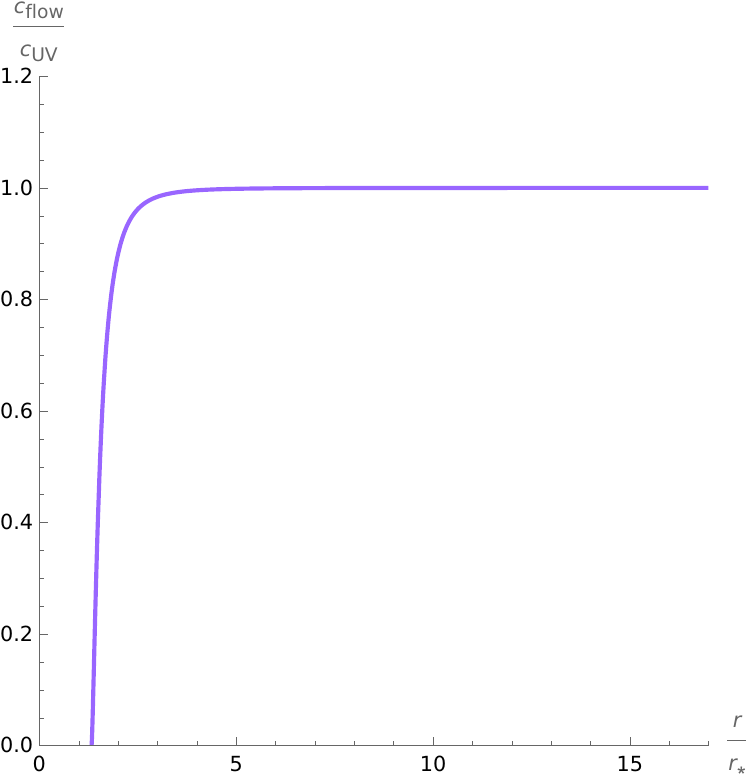}
 \hspace{0.3cm} 
 \centering
\caption{Plot of the flow central charge normalized to its $\mathrm{UV}$ value for the deformed $\mathrm{AdS}_5\times \mathrm{T}^{1,1}$ (keeping $l=\mu=q=1$): It interpolates between an IR gapped 3d system (expressing a $\mathrm{TQFT}$ in the deep IR) and the 4d $\mathrm{SCFTs}$ in the UV, where the value it takes depends on the details of the internal space of the theory. This behaviour is similar for all the supergravity solutions in this work.}
\label{cflowT11plot}
\end{figure}

Then, we consider the following 8-dimensional subspace
\begin{equation}
\begin{split}
G _{ij}\mathrm{d}\xi^i \mathrm{d}\xi ^j &= \frac{r^2}{l^2}f(r) \mathrm{d}\phi ^2 + \frac{r^2}{l^2}\left( \mathrm{d}x_1^2+ \mathrm{d}x_2^2 \right)+\frac{l^2}{6}\sum_{i=1}^2\left(\mathrm{d}\theta_i^2+\sin^2\theta_i\mathrm{d}\phi_i^2\right) \\
&+ \frac{l^2}{9}\left(\mathrm{d}\psi +\sum_{i=1}^2 \cos\theta_i\mathrm{d}\phi_i\right)^2,
\end{split}
\end{equation}

and in a very similar calculation, we find $H$ to be:

\begin{equation}
    H(r)= \left(\mathrm{V}_{\mathrm{T}^{1,1}}l^2r^3\sqrt{f(r)}\right)^2=\mathrm{V}_{\mathrm{T}^{1,1}}^2l^4r^6f(r),
\end{equation}

This leads to the following flow central charge:
\begin{equation}\label{cflowT11}
c_{\mathrm{flow}}=\frac{l^8\mathrm{V}_{\mathrm{T}^{1,1}}}{8G_N}\left[\frac{\sqrt{f(r)}}{f(r)+\frac{r}{6}f^{\prime}(r)}\right]^3=\frac{2l^8\pi^3r^3\left(r^6-l^2r^2\mu-l^2q^2\right)^{3/2}}{G_N(3r^4-l^2\mu)^3}.
    \end{equation}
We emphasize that the form of this expression is nearly the same as in \eqref{cflowS5} with the difference being the volume of the internal manifold. Here the $\mathrm{UV}$ value reads:
\begin{equation}
    c_{\mathrm{UV}}= \frac{2l^8\pi^3}{27},
\end{equation}
 and we present the plot of $c_{\mathrm{flow}}/c_{\mathrm{UV}}$ for  this case in Figure \ref{cflowT11plot} where the previously described \textit{universal} behaviour is apparent.\\

\paragraph{Deformed $\mathrm{\mathbf{AdS}_{\mathbf{5}}\mathbf{\times}\mathrm{\mathbf{Y}}^{\mathbf{p,q}}}$} For the $\mathrm{Y}^{p,q}$ calculation, the internal space is:

\begin{equation}
    \begin{split}
 g_{ij}(\mathrm{d}\theta^i-A^i)(\mathrm{d}\theta^j-A^j)=& l^2\left[\frac{1-y}{6}\left(\mathrm{d}\theta^2+\sin^2\theta\mathrm{d}\varphi^2\right)+\frac{1}{w(y)v(y)}\mathrm{d}y^2+\frac{w(y)v(y)}{36}\left(\mathrm{d}\beta+\cos\theta\mathrm{d}\varphi\right)^2\right.\\
   & \left.+\frac{1}{9}\left(\mathrm{d}\psi- \cos\theta\mathrm{d}\varphi+y\left(\mathrm{d}\beta+\cos\theta\mathrm{d}\varphi\right)+ \frac{3}{l}\mathcal{A}\right)^2\right],\\
    \end{split}
\end{equation}
and the functions $\alpha_0,\alpha_1,\alpha_2,\beta$ are the same. Then, by defining:

\begin{equation}
\begin{split}
G _{ij}\mathrm{d}\xi^i \mathrm{d}\xi ^j &= \frac{r^2}{l^2}f(r) \mathrm{d}\phi ^2 + \frac{r^2}{l^2}\left( \mathrm{d}x_1^2+ \mathrm{d}x_2^2 \right)+l^2\left[\frac{1-y}{6}\left(\mathrm{d}\theta^2+\sin^2\theta\mathrm{d}\varphi^2\right)+\frac{1}{w(y)v(y)}\mathrm{d}y^2\right.\\
&\left.+\frac{w(y)v(y)}{36}\left(\mathrm{d}\beta+\cos\theta\mathrm{d}\varphi\right)^2+\frac{1}{9}\left(\mathrm{d}\psi- \cos\theta\mathrm{d}\varphi+y\left(\mathrm{d}\beta+\cos\theta\mathrm{d}\varphi\right)+ \frac{3}{l}\mathcal{A}\right)^2\right],
\end{split}
\end{equation}
we find
\begin{equation}
    H(r) = \left(\mathrm{V}_{\mathrm{Y}^{p,q}}l^2r^3\sqrt{f(r)}\right)^2=\mathrm{V}_{\mathrm{Y}^{p,q}}^2l^4r^6 f(r),
\end{equation}
with the volume of $\mathrm{Y}^{p,q}$ given in \eqref{volYpq}. Then the expression for the flow central charge is found to be the generalisation of \eqref{cflowT11} for a generic $\mathrm{Y}^{p,q}$:

\begin{equation}\label{cflow_Ypq}
\begin{split}
c_{\mathrm{flow}}&=\frac{l^8\mathrm{V}_{\mathrm{Y}^{p,q}}}{8G_N}\left[\frac{\sqrt{f(r)}}{f(r)+\frac{r}{6}f^{\prime}(r)}\right]^3= \frac{q^2\left(2p+\sqrt{4p^2-3q^2}\right)\pi^3}{3p^2\left(3q^2-2p^2+p\sqrt{4p^2-3q^2}\right)}\frac{l^8r^3\left( r^6-l^2r^2\mu-l^2q^2\right)^{3/2}}{8G_N (3r^4-l^2\mu)^3},
\end{split}
\end{equation}
where we make note that this expression is also of the same form as \eqref{cflowS5} and \eqref{cflowT11}.
\paragraph{Deformed $\mathrm{\mathbf{AdS}_{\mathbf{5}}\mathbf{\times}\mathrm{\mathbf{AFPRT}}}$} In this case we have the following functions: 

\begin{equation}
    \alpha _1 = \alpha_2 = 18\pi \sqrt{-\frac{\alpha}{\ddot{\alpha}}}\frac{r^2}{l^2}\,\, ,\,\, \alpha_3=18\pi \sqrt{-\frac{\alpha}{\ddot{\alpha}}}\frac{r^2}{l^2}f(r)\,\, ,\,\, \beta= \frac{l^4}{r^4f^{4/3}(r)},
\end{equation}
and the internal space metric is:

\begin{equation}
       g_{ij}(\mathrm{d}\theta^i-A^i)(\mathrm{d}\theta^j-A^j)=  18\pi\sqrt{-\frac{\alpha}{6 \Ddot{\alpha}}}\left[\frac{1}{3} \mathrm{d}s_\Sigma^2-\frac{\Ddot{\alpha}}{6\alpha}\mathrm{d}z^2-\frac{\alpha \Ddot{\alpha}}{6 \dot{\alpha}^2-9 \alpha \Ddot{\alpha}} \left(\mathrm{d}\theta^2+\sin^2 \theta \mathcal{D}\psi^2\right)
    \right]. 
\end{equation}
We then take the subspace
\begin{equation}
	\begin{split}
&G _{ij}\mathrm{d}\xi^i \mathrm{d}\xi ^j = \alpha_1 \mathrm{d}\phi^2 + \alpha_2 (\mathrm{d}x_1^2+\mathrm{d}x_2^2)+ g _{ij}(\mathrm{d}\theta ^i- A^{(i)} )( \mathrm{d}\theta ^{j} -A^{(j)} )\\
		&=\alpha_1 \mathrm{d}\phi^2 + \alpha_2 (\mathrm{d}x_1^2+\mathrm{d}x_2^2)+18\pi\sqrt{-\frac{\alpha}{6 \Ddot{\alpha}}}\left[\frac{1}{3} \mathrm{d}s_\Sigma^2-\frac{\Ddot{\alpha}}{6\alpha}\mathrm{d}z^2-\frac{\alpha \Ddot{\alpha}}{6 \dot{\alpha}^2-9 \alpha \Ddot{\alpha}} \left(\mathrm{d}\theta^2+\sin^2 \theta \mathcal{D}\psi^2\right)
    \right],
	\end{split}
\end{equation}
and compute its determinant weighted by the dilaton to get:

\begin{equation}
    \begin{split}
        H(r)&=\left( \int_{\mathrm{S}^2}\mathrm{d}\theta\mathrm{d}\psi\sin\theta\int _{\Sigma}\mathrm{d}v_1\mathrm{d}v_2 \int_0^P \mathrm{d}z \frac{2r^3\sqrt{f(r)}\sin\theta\alpha\ddot{\alpha}  }{243l^3\pi(v_1^2+v_2^2-1)^2} \right)^2\\
        &=\left(\frac{2r^3\mathrm{V}_{\Sigma}\sqrt{f(r)}  }{243l^3} \int_{0}^P  \mathrm{d}z\alpha\ddot{\alpha}  \right)^2.
    \end{split}
\end{equation}

This $H$ function leads to the flow central charge:

\begin{equation}
    c_{\mathrm{flow}}=\frac{l^3\mathrm{V}_{\Sigma}}{972G_N}\left[\frac{\sqrt{f(r)}}{f(r)+\frac{r}{6}f^{\prime}(r)}\right]^3\int_0^P\mathrm{d}z\left(-\alpha\ddot{\alpha}\right).
\end{equation}
We notice the expression $\int\mathrm{d}z(-\alpha\ddot{\alpha})$ that was foreshadowed in \eqref{AFPRT_a}.

\paragraph{Deformed $\mathrm{\mathbf{AdS}_{\mathbf{5}}\mathbf{\times}\mathrm{\mathbf{GM}}}$} In the Gaiotto-Maldacena case, \eqref{eq:N=2} gives off:

\begin{equation}
    \alpha_1=\alpha_2=4(f_1^3f_5)^{1/2}\frac{r^2}{l^2},\,\, \alpha_3=4(f_1^3f_5)^{1/2}\frac{r^2}{l^2}f(r),\,\, \beta = \frac{l^4}{r^4f^{4/3}(r)},
\end{equation}
while the 5-dimensional internal metric has the form:
\begin{equation}
       g_{ij}(\mathrm{d}\theta^i-A^i)(\mathrm{d}\theta^j-A^j)=  f_1^{\frac{3}{2}} f_5^{\frac{1}{2}}\left[f_2 D\mu_iD\mu_i+f_4(\mathrm{d}\sigma^2+\mathrm{d}\eta^2)+f_3 D\tilde{\chi}^2\right].
\end{equation}
By defining
\begin{equation}
	\begin{split}
G _{ij}\mathrm{d}\xi^i \mathrm{d}\xi ^j =  f_1^{3/2}f_5^{1/2}&\Bigg\{ 4 \frac{r^2}{l^2}\left[ \mathrm{d}x_1^2 + \mathrm{d}x_2^2 + f(r)\mathrm{d}\phi^2\right] \\
        &+ f_2 D\mu_i D\mu_i + f_4 (\mathrm{d}\sigma^2+\mathrm{d}\eta^2) + f_3 (\mathrm{d}\tilde{\chi}+\mathcal{A})^2\Bigg\},
	\end{split}
\end{equation}
and computing the $H$ function to be:

\begin{equation}
    \begin{split}
        H(r)&=\left( \int _{\mathrm{S}^2}\mathrm{d}\theta\mathrm{d}\phi \sin\theta \int \mathrm{d}\tilde{\chi}\int _0^{\infty}\mathrm{d}\sigma\int _0^P \mathrm{d}\eta \frac{8f_1^{9/2}f_2f_4r^3\sqrt{f_3f_5f(r)}}{l^3}\right)^2\\
        &=\left(\frac{256\pi^2r^3\sqrt{f(r)}}{l^3}\int _0^{\infty}\mathrm{d}\sigma\int _0^P \mathrm{d}\eta \sigma\dot{V}V^{\prime\prime} \right)^2,
    \end{split}
\end{equation}
we find the central charge to be:

\begin{equation}\label{cflow_GM}
    c_{\mathrm{flow}}=\frac{32 l^3\pi^2}{G_N }\left[\frac{\sqrt{f(r)}}{f(r)+\frac{r}{6}f^{\prime}(r)} \right]^3 \int _0^{\infty}\mathrm{d}\sigma\int _0^P \mathrm{d}\eta \sigma\dot{V}V^{\prime\prime}.
\end{equation}
As in the previous section, we will make note that all of the results for $c_{\mathrm{flow}}$ are captured by the following formula using the factors defined in the table \ref{table_Dimitris}:

\begin{equation}
    c_{\mathrm{flow}}=\frac{s_i l^3}{8}\left( \frac{\mathcal{N}_i}{L_{x_2}L_{\phi}}\right)\left[ \frac{\sqrt{f(r)}}{f(r)+\frac{r}{6}f^{\prime}(r)}\right]^3,
\end{equation}
where $s_i=1$ if $i\in\{\mathrm{S}^5,\mathrm{Y}^{p,q},\mathrm{AFPRT}\}$ and $s_i=8$ for $i=\mathrm{GM}$.
 
\subsection{Holographic complexity}\label{6e}
Here we will calculate  another quantity called holographic complexity, which is very related to the holographic central charge. In general, given a quantum circuit, computational complexity measures how many elementary gates one needs in order to construct a specific state in the Hilbert space from a reference one. Let us review the prescription to calculate this observable in holography, given in \cite{Fatemiabhari:2024aua}, as for the type of backgrounds we work in here a special prescription is needed. We will use the CV conjecture (a review of different conjectures is given in \cite{Reynolds:2017lwq}). The bottom line is that writing the backgrounds in the form
\begin{equation}
 \mathrm{d}s_{10,\mathrm{st}}^2=-g_{tt} \mathrm{d}t^2 + \mathrm{d}s_9^2,
\end{equation}
we should compute, according to \cite{Fatemiabhari:2024aua},
\begin{eqnarray}
  & & \mathcal{C}_V=\frac{1}{G_N}\int \mathrm{d}^9 x \sqrt{ \frac{e^{-4\Phi}\det(g_9)}{{\mathrm{A}} }},
\end{eqnarray}
where $A$ denotes an overall conformal factor in the metric of each background. Let us perform the explicit calculation for the three families of solutions we have.\\
\paragraph{Deformed $\mathrm{\mathbf{AdS}_{\mathbf{5}}\mathbf{\times}\mathrm{\mathbf{S}}^{\mathbf{5}}}$} We start by writing a constant time slice of the 10-dimensional metric \eqref{metric-AdS5xS5}, that is:

\begin{equation}
\begin{split}
    \mathrm{d}s_9^2& = \frac{r^2}{l^2} \left[\mathrm{d}x_1^2+\mathrm{d}x_2^2+f(r) \mathrm{d}\phi^2 \right] +\frac{l^2 \mathrm{d}r^2}{r^2 f(r)} +l^2 \biggl\{ \mathrm{d}\theta^2 + \sin^2\theta \mathrm{d}\varphi^2+ \sin^2\theta\sin^2\varphi \left( \mathrm{d}\varphi_1+ \frac{\mathcal{A}}{l} \right) ^2
       \\
           & + \sin^2\theta \cos^2\varphi \left( \mathrm{d}\varphi_2 + \frac{\mathcal{A}}{ l} \right) ^2 + \cos^2\theta \left( \mathrm{d}\varphi _3 + \frac{\mathcal{A}}{l} \right) ^2\biggr\}.
\end{split}
\end{equation}
while the conformal prefactor $\mathrm{A}$ and the dilaton factor are trivial: $e^{-4\Phi}={\mathrm{A}}=1
$. Then, the square root of the determinant yields:
\begin{equation}
   \sqrt{\mathrm{det}(g_9)}=l^3r^2\cos\theta\sin^3\theta\sin\varphi
\end{equation}

\begin{equation}
    \mathcal{C}_V= \frac{l^3\pi^3L_{x_1}L_{x_2}L_{\phi}}{G_N}\int _{r_{\star}} ^{\Lambda _{\mathrm{UV}}} \mathrm{d}r r^2 = \frac{l^3\pi^3L_{x_1}L_{x_2}L_{\phi}}{3G_N}(\Lambda_{\mathrm{UV}}^3-r^3_{\star}).
\end{equation}
Compare this result with \eqref{dimi10} and \eqref{cflowS5} to find the common features mentioned previously.

\paragraph{Deformed $\mathrm{\mathbf{AdS}_{\mathbf{5}}\mathbf{\times}\mathrm{\mathbf{Y}}^{\mathbf{p,q}}}$} By looking at the metric \eqref{deformed_Ypq} of the 10d background, we write:
\begin{equation}
\begin{split}
    \mathrm{d}s_9^2&=\frac{r^2}{l^2} \left[\mathrm{d}x_1^2+\mathrm{d}x_2^2+f(r) \mathrm{d}\phi^2 \right] +\frac{l^2\mathrm{d}r^2}{r^2 f(r)}  + l^2\left[\frac{1-y}{6}\left(\mathrm{d}\theta^2+\sin^2\theta\mathrm{d}\varphi^2\right)+\frac{1}{w(y)v(y)}\mathrm{d}y^2\right.\\
    &\left.+\frac{w(y)v(y)}{36}\left(\mathrm{d}\beta+\cos\theta\mathrm{d}\varphi\right)^2+\frac{1}{9}\left(\mathrm{d}\psi- \cos\theta\mathrm{d}\varphi+y\left(\mathrm{d}\beta+\cos\theta\mathrm{d}\varphi\right)+ \frac{3}{l}\mathcal{A}\right)^2\right],\\
\end{split}
\end{equation}
and again $A$ and the dilaton are trivial. We then have:
\begin{equation}
   \sqrt{\mathrm{det}(g_9)}=\frac{l^3r^2(y-1)\sin\theta}{108},
\end{equation}
which gives off the following complexity:
\begin{equation}
    \mathcal{C}_V= \frac{L_{x_1}L_{x_2}L_{\phi}l^3}{G_N}\mathrm{V}_{\mathrm{Y}^{p,q}}\int _{r_{\star}}^{\Lambda_{\mathrm{UV}}} \mathrm{d}r r^2=\frac{L_{x_1}L_{x_2}L_{\phi}l^3}{3G_N}\mathrm{V}_{\mathrm{Y}^{p,q}}(\Lambda_{\mathrm{UV}}^3-r_{\star}^3),
\end{equation}
where we used a cutoff $\Lambda_{\mathrm{UV}}$ to regularize the expression. This can also be compared with equations \eqref{EE_Ypq} and \eqref{cflow_Ypq}. The result for $\mathrm{T}^{1,1}$ can be obtained by taking $(p,q)\to(1,0)$. The fact that this is proportional to the $\mathrm{UV}$ cutoff reflects the infinite dimensional space of states of the system and the difficulty of connecting different states.\\

\paragraph{Deformed $\mathrm{\mathbf{AdS}_{\mathbf{5}}\mathbf{\times}\mathrm{\mathbf{AFPRT}}}$} In this case, comparing with \eqref{AFPRT_metric} we have:

\begin{equation}
    \begin{split}
        \mathrm{d}s^2_9 &=18\pi\sqrt{-\frac{\alpha}{6\ddot{\alpha}}}\left\{ \frac{r^2}{l^2}\left[ \mathrm{d}x_1^2+\mathrm{d}x_2^2 + f(r)\mathrm{d}\phi^2\right] + \frac{l^2\mathrm{d}r^2}{r^2f(r)}+ \frac{1}{3}\mathrm{d}s^2_{\Sigma} - \frac{\ddot{\alpha}}{6\alpha}\mathrm{d}z^2-\frac{\alpha\ddot{\alpha}}{6\dot{\alpha}^2-9\alpha\ddot{\alpha}}\left(\mathrm{d}\theta^2+\sin^2\theta\mathcal{D}\psi^2\right)\right\},\\
        &\mathrm{A}=18\pi\sqrt{-\frac{\alpha}{6\ddot{\alpha}}}, \,\,\, e^{-4\Phi} = \frac{1}{2^5 3^{17}\pi^{10}}\left( -\frac{\ddot{\alpha}}{\alpha}\right)^3 \left( 2\dot{\alpha}^2-3 \alpha \ddot{\alpha}\right)^2.
    \end{split}
\end{equation}

Given this, we find the holographic complexity to be:

\begin{equation}
\begin{split}
    \mathcal{C}_V&= \frac{L_{x_1}L_{x_2}L_{\phi}\mathrm{V}_{\mathrm{S}^2}\mathrm{V}_{\Sigma}}{1458 \pi l^2 G_N }\int_0^P\mathrm{d}z\left(-\alpha\ddot{\alpha}\right)\int _{r_{\star}}^{\Lambda_{\mathrm{UV}}}\mathrm{d}r r^2\\
    &= \frac{L_{x_1}L_{x_2}L_{\phi}\mathrm{V}_{\mathrm{S}^2}\mathrm{V}_{\Sigma}}{1458\pi l^2G_N}\int_0^P\mathrm{d}z\left(-\alpha\ddot{\alpha}\right)\left(\Lambda_{\mathrm{UV}}^3-r_{\star}^4\right).
    \end{split}
    \end{equation}
\paragraph{Deformed $\mathrm{\mathbf{AdS}_{\mathbf{5}}\mathbf{\times}\mathrm{\mathbf{GM}}}$} For this solution, we have from \eqref{eq:N=2}:

\begin{equation}
    \begin{split}
        &\mathrm{d}s^2_9 = (f_1^3f_5)^{1/2}\left\{\frac{4r^2}{l^2}\left[ \mathrm{d}x_1^2+\mathrm{d}x_2^2+f(r)\mathrm{d}\phi^2\right] +\frac{4l^2\mathrm{d}r^2}{r^2f(r)}+ f_2D\mu_iD\mu_i + f_4(\mathrm{d}\sigma^2+\mathrm{d}\eta^2)+f_3D\tilde{\chi}^2\right\}, \\
        & \mathrm{A}=(f_1^3f_5)^{1/2}, \,\,\, e^{-4\Phi}=(f_1f_5)^{-3},
    \end{split}
\end{equation}
and the complexity reads:
\begin{equation}
    \mathcal{C}_V= \frac{64L_{x_1}L_{x_2}L_{\phi}L_{\tilde{\chi}}\mathrm{V}_{\mathrm{S}^2}}{3l^2G_N}\left(\Lambda_{\mathrm{UV}}^3-r_{\star}^3\right)\int_{0}^{\infty} \mathrm{d}\sigma\int_0^P\mathrm{d}\eta \dot{V}V^{\prime\prime}\sigma,
\end{equation}
and can be compared with \eqref{EE_GM} and \eqref{cflow_GM}. Finally, we provide the following formula that summarizes the results of this subsection:
\begin{equation}
    \mathcal{C}_{\mathrm{V}}= \frac{m_i}{3l^2}\left( L_{x_1}\mathcal{N}_i\right) \left(\Lambda^3-r_{\star}^3\right),
\end{equation}
where $m_i=1$ for $i\in \{\mathrm{S}^5,\mathrm{Y}^{p,q}\}$, $m_i=1/2$ for $i=\mathrm{AFPRT}$ and $m_i=2$ for $i=\mathrm{GM}$ and $\mathcal{N}_i$ can be found in table \ref{table_Dimitris}.

\begin{landscape}
\begin{table}
\centering
\caption{Summary of the results for the EE, flow central charge and complexity.}\label{table_Dimitris}\vspace{0.2cm}
\begin{tabularx}{1.45\textwidth} { ||
  >{\raggedright\arraybackslash}m{1cm}| X X >{\centering\arraybackslash}X >{\centering\arraybackslash}X ||}
\hline
 & \(\displaystyle\rule{0pt}{5ex}\widehat{\mathrm{AdS}}_5\times\widehat{\mathrm{S}}^5\) & \(\displaystyle\rule{-8ex}{0pt}\widehat{\mathrm{AdS}}_5\times\widehat{\mathrm{Y}}^{p,q}\) & \(\displaystyle\rule{-35ex}{0pt}\widehat{\mathrm{AdS}}_5\times\widehat{\mathrm{AFPRT}}\) & \(\displaystyle\rule{-15ex}{0pt}\widehat{\mathrm{AdS}}_5\times\widehat{\mathrm{GM}}\) \\ [5ex] 
 \hline\hline
\(\displaystyle\rule{6pt}{0pt}\mathcal{N}_i\) & \(\displaystyle\rule{0pt}{7ex}\frac{l^5L_{x_2}L_{\phi}\mathrm{V}_{\mathrm{S}^5}}{G_N}\) & \(\displaystyle\rule{-8ex}{0pt}\frac{l^5L_{x_2}L_{\phi}\mathrm{V}_{\mathrm{Y}^{p,q}}}{G_N}\) & \(\displaystyle\rule{-35ex}{0pt}\frac{L_{x_2}L_{\phi}\mathrm{V}_{\mathrm{S}^2}\mathrm{V}_{\Sigma}}{486\pi G_N}\int_0^{P}\mathrm{d}z(-\alpha\ddot{\alpha})\) & \(\displaystyle\rule{-15ex}{0pt}\frac{32L_{x_2}L_{\phi}L_{\Tilde{\chi}}\mathrm{V}_{\mathrm{S}^2}}{G_N}\int_0 ^P\mathrm{d}\eta\int_0^{\infty}\mathrm{d}\sigma \sigma \dot{V}V^{\prime\prime}\) \\[8ex]
 \hline
\(\displaystyle\rule{6pt}{0pt}\mathrm{EE}\)&  & \(\displaystyle\rule{0pt}{8ex}\frac{\mathcal{N}_i}{4}\int _{-L/2}^{L/2} \mathrm{d}x_1 \sqrt{\frac{r^6}{l^6}\left[ f(r)+ \frac{l^4r^{\prime 2}}{r^4}\right]}\) & & \\[8ex]
 \hline
\(\displaystyle\rule{6pt}{0pt}c_{\mathrm{flow}}\) & & \(\displaystyle\rule{0pt}{8ex}\frac{s_i l^3}{8}\left( \frac{\mathcal{N}_i}{L_{x_2}L_{\phi}}\right)\left[ \frac{\sqrt{f(r)}}{f(r)+\frac{r}{6}f^{\prime}(r)}\right]^3\) & \(\displaystyle\rule{10ex}{0pt} \begin{aligned}[t]
    s_i= \begin{cases}
        1, & i\in\{\mathrm{S}^5,\mathrm{Y}^{p,q},\mathrm{AFPRT}\}, \\
        8, & i=\mathrm{GM}
    \end{cases}
\end{aligned}\) &  \\[8ex]
 \hline
 \(\displaystyle\rule{6pt}{0pt}\mathcal{C}_V\) & &  \(\displaystyle\rule{0pt}{8ex}\frac{m_i}{3l^2}\left( L_{x_1}\mathcal{N}_i\right) \left(\Lambda _{\mathrm{UV}}^3-r_{\star}^3\right)\) & \(\displaystyle\rule{10ex}{0pt} \begin{aligned}[t]
    m_i= \begin{cases}
        1, & i\in\{\mathrm{S}^5,\mathrm{Y}^{p,q},\mathrm{AFPRT}\}, \\
        2, & i=\mathrm{GM}
    \end{cases}
\end{aligned}\) &  \\[8ex]
 \hline
\end{tabularx}
\end{table}
\end{landscape}

\subsection{Gauge coupling}
\label{6f}
In this subsection, we attempt to define a gauge coupling and see how it flows to the IR of the dual $\mathrm{QFT}$. 
All of our backgrounds correspond to a dual $\mathrm{QFT}$ that at high energies extends in $[t,x_1,x_2,\phi]$, being $\phi$ the periodic direction that we mix with the R-symmetry in a  twisting-like fashion. We will use the Dirac-Born-Infeld action for a probe Dp-brane that extends along the field theory directions and contains a gauge field on its worldvolume $\Sigma_{p+1}$. We will read the gauge coupling as the coefficient of the gauge kinetic term $-\frac{1}{4}F_{\mu\nu}^2$, while the $r$ dependence in our results will be interpreted as the running of the coupling constant. We will briefly review the procedure which was nicely summarized in \cite{Nunez:2023nnl}, the main idea being that the general action for the Dp-brane consists of two terms: The DBI\footnote{we use the convention where there is no $2\pi$ in front of $F$} and the Wess-Zumino term,

\begin{equation}
    S_{\mathrm{D}_p}=  S_{\mathrm{D}_p,\mathrm{DBI}} + S_{\mathrm{D}_p,\mathrm{WZ}},
\end{equation}
which for a zero NS $B$ field read
\begin{equation}\label{DBI_action}
       S_{\mathrm{D}_p,\mathrm{DBI}}= T_{\mathrm{D}_p}\int _{\Sigma_{p+1}}\mathrm{d}^{p+1}\sigma\sqrt{-e^{-2\Phi}\mathrm{det}(h+F)},
       \end{equation}
       \begin{equation}
        S_{\mathrm{D}_p,\mathrm{WZ}}=- T_{\mathrm{D}_p}\int _{\Sigma_{p+1}}\mathcal{C}\wedge e^{-F},
\end{equation}
where $h$ is the induced metric on $\Sigma_{p+1}$, $F= F_{\mu\nu}\mathrm{d}x^{\mu}\wedge\mathrm{d}x^{\nu}$ ($\mu,\nu=0,\dots,p+1)$ the field strength of the gauge field on $\Sigma_{p+1}$ and $\mathcal{C}$ denotes the RR polyform. Although for our purpose we will focus solely on the $\mathrm{DBI}$ part of the action, we note that the $\mathrm{WZ}$ term contains important information regarding the BPS status of the brane and the $\theta$-term in the field theory. We will consider a small field strength expansion which brings \eqref{DBI_action} to the form:

\begin{equation}\label{SDBI_expansion}
S_{\mathrm{D}_p,\mathrm{DBI}} = T_{\mathrm{D}_p}\int _{\Sigma_{p+1}}\mathrm{d}^{p+1}\sigma\sqrt{-e^{-2\Phi}\mathrm{det}(h)}\left(1+\frac{1}{4}F_{\mu\nu}F^{\mu\nu}+\mathcal{O}(F^4)\right),
\end{equation}
with the indices being raised with $h_{\mu\nu}$. For the cases studied in this paper, we can without loss of generality turn on only the $t-x_1$ component of $F_{\mu\nu}$ and define the part of the induced metric extending over the field theory directions, which is of the form:

\begin{equation}
    h^{(3)}_{\mu\nu}\equiv h_{ij}= K(r,\theta^a)\eta_{ij},
\end{equation}
where we introduced the flat indices $i,j\in\{t,x_1,x_2\}$ and a function $K$ of the holographic coordinate that for the type $\mathrm{IIA}$ solutions also depends on the internal space coordinates, denoted as $\{\theta^a\}$. Then the quadratic in $F_{\mu\nu}$ term on $\Sigma_{p+1}$ gives off: 

\begin{equation}
    F_{\mu\nu}F^{\mu\nu}\equiv F_{ij}F^{ij}=h^{ik}h^{jl}F_{kl}F_{ij}=\frac{2}{K^2}F_{tx_1}F_{tx_1}
\end{equation}

By performing the $\phi$ integral and separating the remaining integral on the worldvolume to the field theory part and the rest of the directions in \eqref{SDBI_expansion}, we get the following Maxwell term included in the $\mathrm{DBI}$ action\footnote{here $\Sigma_{\mathrm{int}}$ stands for the part of the worldvolume consisting of internal coordinates, depending on the specific probe.}:

\begin{equation}
    \begin{split}
S_{\mathrm{D}_p,\mathrm{DBI}}&\supset L_{\phi}\mathrm{T}_{\mathrm{D}_p}\int \mathrm{d}^{2+1}x\int_{\Sigma _{\mathrm{int}}} \mathrm{d}^{p-3}\sigma \sqrt{-e^{-2\Phi}\mathrm{det}(h)}\frac{1}{2}K^{-2}F_{tx_1}^2\\
    &= \frac{1}{4}\left[L_{\phi}\mathrm{T}_{\mathrm{D}_p}\int_{\Sigma _{\mathrm{int}}} \mathrm{d}^{p-3}\sigma \sqrt{-e^{-2\Phi}\mathrm{det}(h)}K^{-2}\right]2\int \mathrm{d}^{2+1}x F_{tx_1}^2.
    \end{split}
\end{equation}
From this term one can read off the object inside the square brackets tho be the Yang-Mills gauge coupling of the 3-dimensional effective theory:

\begin{equation}\label{gYM_formula}
    \frac{1}{g_{\mathrm{YM}}^2}= L_{\phi}\mathrm{T}_{\mathrm{D}_p}\int _{\Sigma _{\mathrm{int}}}\mathrm{d}^{p-3}\sigma \sqrt{-e^{-2\Phi}\mathrm{det}(h)}K^{-2}(r,\theta^a)
\end{equation}

Let us see the details of applying this formula in each of the backgrounds.\\

\paragraph{Deformed $\mathrm{\mathbf{AdS}_{\mathbf{5}}\mathbf{\times}\mathrm{\mathbf{S}}^{\mathbf{5}}},\mathrm{\mathbf{Y}^{p,q}}$} For the deformed uplifts of the five-sphere and Sasaki-Einstein manifolds, we consider a $\mathrm{D}3$ probe brane extended along $[t,x_1,x_2,\phi]$ and switch on a field strength in the $t-x_1$ directions as described above. The induced metric on the brane for the $\mathrm{S}^5$ background can be read from \eqref{metric-AdS5xS5} to be\footnote{by $\mathcal{A}_{\phi}$ we denote the component of the one form $\mathcal{A}$.}:

\begin{equation}
   \mathrm{d}s^2 _{\mathrm{ind},\mathrm{D}3}=h_{\mu\nu}\mathrm{d}x^{\mu}\mathrm{d}x^{\nu}=\frac{r^2}{l^2}\left[-\mathrm{d}t+\mathrm{d}x_1^2+\mathrm{d}x_2^2+(f(r)+\mathcal{A}_{\phi}^2)\mathrm{d}\phi^2\right],
\end{equation}
where the quadratic term of the component of the gauge field originates from the $\propto \mathrm{d}\phi^2$ part of the fibration. The 3d metric then reads:
\begin{equation}
    h_{ij}=K(r)\eta_{ij}, \,\, K(r)=\frac{r^2}{l^2},
\end{equation}
We see that in this case there are no remaining internal coordinates in the worldvolume to integrate over and the dialton is trivial, therefore \eqref{gYM_formula} yields the following 3d coupling:
\begin{equation}\label{gYM_S5}
    \frac{1}{g_{\mathrm{YM}}^2}=L_{\phi}T_{\mathrm{D}_3}\sqrt{f(r)+\frac{l^2}{r^2}\mathcal{A}_{\phi}}.
\end{equation}
 This expression reveals the strongly coupled nature governing the $\mathrm{IR}$ dynamics of the $\mathrm{QFT}_3$ since for $r\to r_{\star}$ the coupling constant diverges. As for the $\mathrm{UV}$ limit $r\to \infty$ we note that this takes the value of the classical gauge coupling of the circle reduced $\mathrm{QFT}_4$. Finally, we comment that the case for the $\mathrm{Y}^{p,q}$ uplift is extremely similar, an{d one can derive it by setting $\mathcal{A}_{\phi}\to3\mathcal{A}_{\phi}$ in \eqref{gYM_S5}.

\paragraph{Deformed $\mathrm{\mathbf{AdS}_{\mathbf{5}}\mathbf{\times}\mathrm{\mathbf{AFPRT}}}$} Here we will use a $\mathrm{D}4$ brane to probe the $[t,x_1,x_2,\phi,\psi]$ directions with $\theta,r,v_1,v_2,z=\mathrm{const.}$ We switch on the same field strength as before in the worldvolume ($F_{tx_1}\neq0$) and the induced metric on the $\mathrm{D4}$ is:

\begin{equation}
    \mathrm{d}s^2 _{\mathrm{ind},\mathrm{D}4}=18\pi\sqrt{- \frac{\alpha}{\ddot{\alpha}}}\left\{ \frac{r^2}{l^2} \left(-\mathrm{d}t^2+\mathrm{d}x_1^2+\mathrm{d}x_2^2+f(r)\mathrm{d}\phi^2\right) - \frac{\alpha\ddot{\alpha}\sin^2\theta_{0}}{6\dot{\alpha}^2-9\alpha\ddot{\alpha}}(\mathrm{d}\psi-3\mathcal{A})\right\}.
\end{equation}

 In this case the conformal factor of the 3d field theory metric reads:

 \begin{equation}
     K(r)=18\pi \sqrt{-\frac{\alpha(z_{0})}{6\ddot{\alpha}(z_{0})}}\frac{r^2}{l^2}.
 \end{equation}

Then, applying \eqref{gYM_formula} we find the gauge coupling of the 3d field theory to be:

\begin{equation}\label{gYM_AFPRT}
    \frac{1}{g^2 _{\mathrm{YM}}}= -\frac{\sin\theta_{0}L_{\phi}L_{\psi}T_{\mathrm{D}4}\ddot{\alpha}(z_{0})}{162\pi^2}\sqrt{f(r)}.
\end{equation}
We notice that this also diverges in the $r\to r_{\star}$ limit and takes a constant value in the $\mathrm{UV}$. This expression also depends on the node of the quiver in the dual description, which amounts to focusing on one of the colour groups.

\paragraph{Deformed $\mathrm{\mathbf{AdS}_{\mathbf{5}}\mathbf{\times}\mathrm{\mathbf{GM}}}$} For the last case of the deformed Gaiotto-Maldacena background, we will also use a $\mathrm{D}4$ brane with $F_{tx_1}\neq 0$ that extends along $[t,x_1,x_2,\phi,\tilde{\chi}]$ keeping the rest of the coordinates constant. The induced metric reads:
\begin{equation}
\mathrm{d}s^2_{\mathrm{ind},\mathrm{D}4}= f_1^{3/2}f_5^{1/2}\left\{ 4\frac{r^2}{l^2}\left(-\mathrm{d}t^2+\mathrm{d}x_1^2+\mathrm{d}x_2^2 +f(r)\mathrm{d}\phi^2\right)+4f_2\mathcal{A}^2 + f_3 (\mathrm{d}\tilde{\chi}+\mathcal{A})^2 \right\},
\end{equation}
and the field theory prefactor function is:
\begin{equation}
    K(r) = 4f^{3/2}_1(\sigma_{0},\eta_{0})f^{1/2}_5(\sigma_{0},\eta_{0}) \frac{r^2}{l^2}.
\end{equation}
Using this, we get the expression:
\begin{equation}
    \frac{1}{g_{\mathrm{YM}}^2} =  L_{\phi}T_{\mathrm{D}4}L_{\tilde{\chi}}\sqrt{\frac{f_3(\sigma_{0},\eta_{0})}{f_5(\sigma_{0},\eta_{0})}}\sqrt{f(r) + f_2(\sigma_{0},\eta_{0}) \frac{l^2}{r^2}\mathcal{A}_{\phi}^2},
\end{equation}
enjoying the same $\mathrm{IR}$ and $\mathrm{UV}$ behaviour as \eqref{gYM_S5} and \eqref{gYM_AFPRT} and also depending on the choice of a certain point on the quiver.
\subsection{Global $U(1)$ symmetry breaking}\label{6g}

Some SUSY field theories possess a classical $U(1)$ R-symmetry that is broken into a discrete subgroup by quantum mechanical effects. The holographic dual to this field theory should contain this information. The fact that the Ramond potentials are gauge invariant under the R-symmetry is essential in this context. Since the global symmetries are realized as gauged symmetries in the bulk, the breaking of the global R-symmetry in the QFT appears as a spontaneous symmetry breaking in the bulk. The vector field in the bulk corresponding to the R-symmetry current should acquire a mass. We check this phenomenon in our backgrounds. 
For simplicity and to uncover the universal behaviour of symmetry breaking, we focus on the 5d background introduced in Section \ref{2-5d}. 

We perturb the $U(1)_\phi$ symmetry of the metric and $\mathcal{F}_{\mu\nu}$. Then, the Lagrangian up to the second order in fluctuations and the equations of motion for these fluctuations up to the first order are derived. 
The appearance of a massive gauge field from the fluctuation is interpreted as the symmetry breaking in supergravity. 

Following \cite{Klebanov:2002gr}, we gauge the isometry in the $\phi$ direction by replacing in the metric of (\ref{ARmetric})
    \begin{equation}
        d\phi \rightarrow d\phi + \mathfrak{a},\quad
        \phi \rightarrow \phi + \theta.
    \end{equation}
A gauge transformation as $ \phi \rightarrow \phi + \theta$ is absorbed by  $\mathfrak{a}\rightarrow \mathfrak{a}+ d \theta$. $\mathfrak{a}$ is a new one-form with legs only in $(t,x,y)$ direction. Here, one can choose a gauge in which $\theta=0$, which simplifies the expressions.
Now, we derive the Lagrangian for the fluctuating gauge fields $\mathfrak{a}$ by replacing them in the action (\ref{eq:Lag5d}). The action reads
\begin{equation} 
S\left(  g, \mathcal{A}, \mathfrak{a}\right) =S\left(  g,\mathcal{A}, \mathfrak{a}=0\right) +\frac{1}{16\pi G}\int d^{5}x\left(  \sqrt{-g}f(r)\left[-\frac{1}{2} \mathfrak{f}_{\mu\nu} \mathfrak{f}^{\mu\nu}-6q^2\mathfrak{a}_{\mu}\mathfrak{a}^{\mu}\right]\right) \label{eq:Lag5dper}%
\end{equation}
The equation of motion for the fluctuation $\mathfrak{f}$ is
\begin{equation} 
d\left[\star \mathfrak{f}\right]=-\frac{12q^2}{r^6}\mathfrak{a}, \label{eq:eom5dper}%
\end{equation}
which features a mass term for the new gauge field. This equation is derivable from the $\mu \phi$ component of the Einstein equations (\ref{eq:5dEin}).

Note that the introduction of the background gauge field $\mathcal{A}$ enables us to preserve SUSY in our solution. The Killing spinors given in (\ref{Kspinor}), depend explicitly on the $\phi$ coordinate. This is reminiscent of the fact that the isometry of the $\phi$ circle is acting as an R-symmetry generator, mixing with the other SUSY algebra elements. In this line of thought, the spontaneous breaking of the gauged symmetry in bulk corresponds to the $U(1)$ global R-symmetry in the dual field theory breaking.

\section{Discussion}
\label{7-discussion}
In the preceding sections, we use holography to implement a SUSY-preserving deformation to a variety of 4d SCFTs. In essence this deformation is a twisted compactification, triggering a flow to a gapped three-dimensional field theory. Our method should apply to any SCFT whose holographic dual admits a consistent truncation to $d=5$ minimal gauged supergravity. The field theories encountered in this work can be divided into two basic categories.
\begin{itemize}
\item{Field theories that admit a weakly coupled description in terms of elementary fields and a Lagrangian. Examples in this category are ${\cal N}=4$ SYM and the $\mathcal{N}=2$ theories of Gaiotto-Maldacena (in the electrostatic case). After the deformation, the dual description of those QFTs at strong coupling are given by the backgrounds in (\ref{metric-AdS5xS5}) and (\ref{def-GM}), respectively.}
\item{Field theories that do not admit a weakly coupled description and/or do not have a Lagrangian description. The field theories associated with the BPT, AFPRT and LLM backgrounds are examples in this category. The QFTs and associated UV-SCFTs are non-Lagrangian and strongly coupled. In contrast, the field theories dual to the $T^{1,1}$ and $Y^{p,q}$ backgrounds are intrinsically strongly coupled (as the anomalous dimensions of the elementary fields are large), but one can still write down a superpotential. }
\end{itemize}
Here we study both sets of examples using holographic observables.
A summary of the key results is shown in table \ref{table_Dimitris}.
Since the string theoretic duals are always weakly curved, the QFTs described in our work are all at strong coupling.
For the QFTs in the first category one can also gain insight from the perturbative description, as done in \cite{Kumar:2024pcz} for the $\mathcal{N}=4$ SYM case. It would be interesting to apply a similar analysis to the Lagrangian Gaiotto-Maldacena theories.
For the QFTs in the second category, additional holographic observables or perhaps algebraic methods could be used to learn more.

Let us summarize some of the key conclusions of the paper.
\begin{itemize}
    \item {
    When the deformation parameter $\mu$ in (\ref{soliton}) is set to zero, the dual QFT preserves four supercharges. This SUSY-preservation in the ten- or eleven-dimensional background is `inherited' from the preservation of four supercharges in the 5d minimal gauged supergravity solution.  We also know that in each case, the R-symmetry is broken by the VEV of a current in the QFT. This breaking is controlled by the parameter $q$ in eq.(\ref{soliton}).}
    \item{
    The density of degrees of freedom in the deformed QFTs is captured by a monotonic quantity, which we called c$_{flow}$.  It decreases along the flow to lower energies, interpolating between the undeformed SCFT value and zero. }
    \item{The QFTs are strongly coupled all along the flow. There are no degrees of freedom in the deep IR and so a TQFT must describe the system there.}
    \item{We found that the deformed QFTs confine external non-dynamical quarks, at least in the case where the Wilson loops do not explore the internal manifold.  As expected from this, 't Hooft loops display a screening behaviour.}
    \item{We found that some of our observables (Entanglement Entropy, c$_{flow}$, Complexity) contain two multiplicative contributions: one coming from the UV SCFT and another related to the flow. One might say the UV contribution is of a kinematical character while the flow contribution is of dynamical character. This `factorization' of observables could be a consequence of the conjecture by Gauntlett and Varela \cite{Gauntlett:2007ma}. In particular, fields in the current multiplet of the SCFT may be responsible for the `dynamical' contribution depending only on the truncated 5d supergravity solution (\ref{soliton}).}
\end{itemize}
The seed used to construct all of the new backgrounds presented here was the  Anabal\'on-Ross soliton in $d=5$. Analogous solutions exist in gauged supergravities in other dimensions. It would be straightforward to apply the same methodology, embedding those solitons into string theoretic backgrounds with appropriate AdS$_d$ factor. This is one natural direction for future research.
It would be informative, wherever possible, to study this holographic implementation of the twisted $S^1_\phi$ compactification alongside a Lagrangian account of the deformation. In particular,  understanding the effect of the compactification on fields transforming in the fundamental representation would be desirable. 

The deep IR of the deformed QFTs is also worthy of further study. Based on \cite{Cassani:2021fyv}, one expects the 3d IR description to reveal a Chern-Simons theory of level given by the number of colour branes. It would be nice to understand in detail how this story applies to the IR endpoints of the flows described here. Finally, it would be interesting to identify observables which do not `factorize' into a contribution from the flow and another from the lift. Probes which extend on both the (deformed) AdS directions and the internal directions in more non-trivial ways could help break this pattern.

\section*{Acknowledgments} For discussions, comments on the manuscript, we wish to thank: Federico Castellani, S. Prem Kumar, Ricardo Stuardo. We are supported by the grants  ST/X000648/1 and ST/T000813/1. D.C. has been supported by the STFC Consolidated  Grant ST/Y509644-1. A.F. has been supported by the STFC Consolidated  Grant ST/V507143/1 and the EPSRC Standard Research Studentship (DTP)  EP/T517987/1.

\appendix

\section{SUSY variations in 5d}
\label{A2-susy}
In this appendix we study the SUSY properties of the 5d solution provided in eq.(\ref{ARmetric}).
The supersymmetry transformation of the gravitino for the 5d gauged supergravity, which sets the equation for the Killing spinors,  is \cite{Klemm:2000gh}

\begin{align} \label{eq:susyvar}
\delta \psi _{\mu }dx^{\mu }&=\left( d+W\right) \Psi  =0 \equiv {\mathcal D}\, \Psi\ ,
\end{align}
with%
\begin{align}
A &= A_{\mu }dx^{\mu } \, \\
W&=\frac{1}{4}\omega _{ab}\gamma ^{ab}-%
\frac{i\sqrt{3}}{2l}A+\frac{i\sqrt{3}}{4!}\left( \gamma _{c}\gamma ^{ab}-6\delta
_{c}^{a}\gamma ^{b}\right) e^{c} F_{ab}+%
\frac{1}{2 l}\gamma _{c}e^{c}\ .  \notag
\end{align}%

The relation between the complex spinor $\Psi $ and the symplectic
Majorana spinor $\epsilon ^{a}$ is $\Psi =\epsilon ^{1}+i\epsilon ^{2}$ (see  \cite{Gutowski:2004yv}). The basis for the Clifford
algebra used for 5d calculations is:
\begin{eqnarray}
\gamma ^{0} &=&-i\left( 
\begin{array}{cc}
0 & \sigma _{2} \\ 
\sigma _{2} & 0%
\end{array}%
\right) \ ,\quad \gamma ^{1}=-\left( 
\begin{array}{cc}
\sigma _{3} & 0 \\ 
0 & \sigma _{3}%
\end{array}%
\right) \ ,\quad \gamma ^{2}=i\left( 
\begin{array}{cc}
0 & -\sigma _{2} \\ 
\sigma _{2} & 0%
\end{array}%
\right) \ ,  \notag \\
\gamma ^{3} &=&\left( 
\begin{array}{cc}
\sigma _{1} & 0 \\ 
0 & \sigma _{1}%
\end{array}%
\right) \ ,\quad \gamma ^{4}=i\gamma ^{0}\gamma ^{1}\gamma ^{2}\gamma ^{3}\ .
\end{eqnarray}

By acting with ${\mathcal D}$ for the second time on the eq.(\ref{eq:susyvar}), one can reach to a consistency condition.
The so-called 2-form integrability conditions are defined as
\begin{equation}
   {\mathcal D}\wedge{\mathcal D}\Psi= (dW+W\wedge W)\Psi =0.\label{5D integrability conditions}
\end{equation}

This equation has a non-trivial solution only if the determinant of the components of $dW+W\wedge W$ is zero.

This condition makes us choose $\m=0$ to have a supersymmetric solution.
At this supersymmetric point, we introduce a change of coordinate
\begin{equation}
r=\left\vert q\ell\right\vert ^{2/3}\cosh(\rho)^{1/3},
\end{equation}
and the metric and gauge field will read%

\begin{align}
ds^{2}  &  =\frac{\ell^{2}}{9}d\rho^{2}+\frac{r_\star^{2}}{\ell^{2}}\left[
\cosh\left(  \rho\right)  ^{2/3}\left(  -dt^{2}+dy^{2}+dz^{2}\right)
+\frac{\sinh\left(  \rho\right)  ^{2}}{\cosh(\rho)^{4/3}}d\phi^{2}\right]  ,\\
A  &  =\frac{\sqrt{3}q}{r_\star^{2}}\left(  \frac{1}{\cosh(\rho)^{2/3}%
}-1\right)  d\phi.
\end{align}
Here we have $r_\star=\left\vert q\ell\right\vert ^{1/3}$ and
$L_{\phi}=\frac{2\pi\ell^{5/3}}{3q^{1/3}}$. 

The solution to the Killing spinor equations
on our background is 
\begin{align} \label{Kspinor}
\Psi &  =\cosh(\rho)^{-1/3}e^{-i \pi\frac{\phi}{L_{\phi}}}\left(
\begin{array}
[c]{c}%
e^{\r/2} c_1\\
e^{-\r/2} c_2\\
-ie^{\r/2} c_2\\
-ie^{-\r/2} c_1
\end{array}
\right)  \text{ },
\end{align}
which has two complex integration constants $(c_1,c_2)$.

\section{$a$-anomaly match for reductions of 6d linear quivers}
\label{A3-anomaly}

The 't Hooft anomalies of a $d$-dimensional field theory can be represented with a $d+2$-form called the anomaly polynomial,\footnote{In this appendix we suppress wedge products for convenience.}
\begin{equation}
    I_{d+2}= \hat{A}(T) \text{ch}(E) |_{d+2}.
\end{equation}
The object $\hat{A}(T)$, or A-hat genus, is an expansion in the Pontryagin classes of the tangent bundle $T$ to the space.  $\text{ch}(E)$ is the Chern character of the bundle $E$ composed of all the gauge and global symmetries of the theory. Crucially this object is invariant along RG flows. 

For the six-dimensional linear quiver theories mentioned in section \ref{4b}, the anomaly polynomial is an eight-form $I_8$ which can be written as an expansion in the second Chern class of the R-symmetry bundle $C_2(R)$ (a four-form) and the first Pontryagin classes of the tangent bundle, $p_1$ and $p_2$ (four- and eight-forms, respectively). The coefficients encode the anomalies. As explained in \cite{ct2015}, one can follow the methods developed in \cite{Intriligator:2014eaa,Ohmori:2014kda} for a linear quiver consisting of $(P-1)$-many $SU(N_i)$ gauge nodes connected to $SU(f_i)$ flavor nodes to obtain
\begin{align}\label{I8}
 I_8=&\frac{1}{2}\sum_{ij} C_{ij}^{-1} N_i N_j C_2(R)^2+\frac{1}{24}\left[2(P-1)-\sum N_i^2\right] C_2(R)^2   \\
&+\frac{1}{48}\left[2(P-1)-\sum N_i^2\right]C_2(R) p_1 \nonumber \\& +(P-1)\frac{23p_1^2-116p_2}{5760}+\frac{1}{2}\left[2(P-1)-\sum N_if_i\right]\frac{7p_1^2-4p_2}{5760}, \nonumber
\end{align}
where $C_{ij}^{-1}$ is the inverse of the Cartan matrix for $A_{P-1}$,
\begin{equation}
    C_{ij}=2 \delta_{i,j}-\delta_{i,j-1}-\delta_{i,j+1}.
\end{equation}
To derive this result, in addition to the anomaly contributions from each species in the quiver theory, one needs to account for a Green-Schwarz term arising from couplings between the self-dual tensor and various gauge fields. Its form must be inferred from the cancellation of gauge anomalies.

Here we are interested in reductions of these 6d theories on a negative-curvature Riemann surface, of volume
\begin{equation}\label{Vsigma}
    V_\Sigma= \int \text{vol}_\Sigma =4\pi(g-1),
\end{equation}
with a topological twist appropriate to preserve $\mathcal{N}=1$ SUSY in four dimensions. To implement this twisted reduction in the anomaly polynomial, we can follow the same approach used in \cite{Bah:2017gph} in the context of theories from wrapped M5-branes probing $\mathbb{C}^2/ \mathbb{Z}_k$ singularities, shifting the Chern root of the R-symmetry bundle as
\begin{equation}
C_2(R) \rightarrow -\left[C_1(R')+\frac{\text{vol}_\Sigma}{4\pi}\right]^2.
\end{equation}
The prime in $C_1(R')$ indicates that this symmetry need not always coincide with the four-dimensional R-symmetry, a subtlety which we can overlook in the present example. Following this procedure in \eqref{I8} and integrating over the Riemann surface using \eqref{Vsigma}, we find
\begin{align}\label{I6}
 \int_{\Sigma_g}I_8=&\frac{1}{6}(g-1)\left[12\sum_{ij} C_{ij}^{-1} N_i N_j +2(P-1)-\sum N_i^2\right]C_1(R')^3   \\
&-\frac{1}{24}(g-1)\left[2(P-1)-\sum N_i^2\right]C_1(R') p_1 .\nonumber 
\end{align}
For a 4d SCFT with anomaly polynomial of the form
\begin{equation}
I_6^{\text{SCFT}}=\frac{1}{6} \alpha \, C_1(R)^3-\frac{1}{24} \beta \, C_1(R) \, p_1,
\end{equation}
the relationship between the $a$ and $c$ anomalies and the 't Hooft anomalies for the R-symmetry allows us to compute\cite{Anselmi:1997am}
\begin{equation}\label{a_from_I6}
a=\frac{3}{32}\left(3 \alpha-\beta \right), \qquad c=\frac{1}{32}\left(9 \alpha-5 \beta\right).
\end{equation}
Interpreting \eqref{I6} as the anomaly polynomial for the 4d theories in question, this allows us to make a check of the conjectured duality with the AFPRT AdS$_5$ backgrounds.
Now, for the purposes of comparison with the holographic result, we should focus on the leading-order contribution at large $P$.  the term $\sum_{ij} C_{ij}^{-1} N_i N_j$ dominates in this limit. The leading coefficients of both $a$ and $c$ are
\begin{equation} \label{a_from_I6}
a,c \sim  \frac{27}{8}(g-1) \sum_{i,j} C_{ij}^{-1} N_i N_j.
\end{equation}
As discussed at length in \cite{ct2015}, the role of the Cartan matrix in the holographic limit is like that of a discrete second derivative, so that (in the conventions of this paper)
\begin{equation}
\sum_{i,j} C_{ij}^{-1} N_i N_j \sim \frac{1}{(9\pi)^4}\int -\alpha \ddot{\alpha} dz.
\end{equation}
This identification leads to a leading-order match between \eqref{a_from_I6} and the holographic $a$-central charge in \eqref{AFPRT_a}, which we can verify for a given quiver.

As an example, consider a balanced linear quiver with the same rank $r$ for all of $(P-1)$ gauge groups, and a pair of additional $SU(r)$ flavour groups, one coupled to each of the first and last gauge nodes. In this case \eqref{a_from_I6} gives
\begin{equation}\label{anomaly_match_example}
\sum_{i,j}C_{ij}^{-1}=\frac{1}{12}P(P-1)(P+1),  \qquad a \sim \frac{9}{32}(g-1) r^2 P^3,
\end{equation}
again with only the leading-order term shown.
The function $\alpha(z)$ associated to the holgraphic dual of this quiver is 
\begin{equation}
    \alpha(z)=\frac{27\pi^2}{2} \begin{cases} 
    - r z^3+ 3r(P-1) z & 0 \leq z \leq 1 \\
      -3r z^2 +3r P z-r  & 1\leq z\leq P-1 \\
      r(z-P)^3 +3r(P-1)(P-z) & P-1\leq z \leq P 
   \end{cases},
\end{equation}
which satisfies \eqref{alpha_eq} for $\frac{F_0}{2\pi}$ taking values $\{r,0,-r\}$ on the various intervals, and satisfies the boundary conditions $\alpha(0)=\alpha(P)=0$. Jumps in $F_0$ at $z=1$ and $z=P-1$ are sourced by D8 flavour branes.
Computing the holographic central charge using eq.\eqref{AFPRT_a},
\begin{equation}
   a= \frac{V_\Sigma}{(6\pi)^5} \int -\alpha \ddot{\alpha} dz
   =\frac{9}{32}(g-1)\left( r^2 P^3+ \mathcal{O}(P) \right).
\end{equation}
This is a leading-order match to the field theory result \eqref{anomaly_match_example} obtained from reducing the anomaly polynomial on $\Sigma$.

We conclude this appendix by commenting on the prospect for a quiver description of the 4d SCFTs in question. As noted in the main text, a heuristic proposal for such a description was made in \cite{Merrikin:2022yho}. There the proposed quiver consisted of the same basic content of vector and hyper multiplets as in the 6d quiver, but with a multiplicity proportional to volume of the Riemann surface. However, this proposal fails to reproduce the leading order $a$ anomaly derived here from both holographic and field-theoretic approaches.

In fact we can go further. In that proposal, the 4d fields contributing to the anomaly polynomial were the fermions of the vector and hypermultiplets, with the former of R-charge $q_v=1$ and the latter $q_h=-1/2$. A natural generalization of that proposal would be to consider these same fermion species, and ask if there is any possible integer multiplicity of each that reproduces the anomaly polynomial in \eqref{I6}. In fact there is not. Furthermore, one can show that there is no rational R-charge assignment for $q_h$ which could reproduce \eqref{I6} (again assuming integer multiplicities of both hyper and vector multiplets). These observations make it appear even less likely that these theories admit a quiver description.

\section{Gaiotto-Maldacena background Page charges}
\label{app:GM}

In this appendix we provide more details for the calculation of the Page charges in the GM background. For more details see \cite{Macpherson:2024frt}.

Following the discussion in Section \ref{5c}, we Fourier expand the Rank function and $\dot{V}$ as
\begin{align}
{\cal R}= \sum_{n=1}^{\infty} {\cal R}_n \sin\left(\frac{n\pi}{P}\eta\right), \qquad \dot{V}&=\frac{\pi}{P}\sum_{n=1}^{\infty}{\cal R}_n\sigma\sin\bigg(\frac{n\pi}{P}\eta\bigg) K_1\bigg(\frac{n\pi}{P}\sigma\bigg).
\end{align}
In some limits of the geometry, it is useful to have another parameterisation for $\dot{V}$ \cite{Reid-Edwards:2010vpm},
\begin{align}
\dot{V}
&=\frac{1}{2}\sum_{m=-\infty}^{\infty}\sum_{k=1}^Pb_k\left(\sqrt{\sigma^2+(\eta-2m P+k)^2}-\sqrt{\sigma^2+(\eta-2m P-k)^2}\right).\label{eq:alternative}
\end{align}
In the $k$'th cell, with $\eta\in[k,k+1]$ one has 
\beq
B^{k}_2=2\kappa\left(-(\eta-k)+ \frac{1}{4}\dot{V}f_5f_6\right) \text{vol}(\text{S}^2),
\eeq
hence there is a large gauge transformation $B_2\to B_2+2\kappa ~ k \text{vol}(\text{S}^2)$ as we move through each sector while moving in the $\eta$ axis.\\

Now we consider the behaviour of the metric near special points which there is possibility of singularities or D-branes. All of the calculation are performed in the $r\to\infty$ limit. In this limit, the 5d subspace in eq. \ref{eq:N=2} denoted by $ds_5^2$, asymptotes to AdS$_5$ and the fibered sphere $\tilde{S}^2$ can be re-written as a round sphere $S^2$ by absorption of the constant factors in the metric fibration. 
First, we will focus on the boundary $\sigma\to \infty$. Using the asymptotics relation $x\to \infty$, $K_0(x)\to\sqrt{\pi} (2x)^{-\frac{1}{2}}e^{-x}$ and the fact that the leading term in \eqref{eqn:potential} is the $n=1$ term, one has 
\beq
V=-{\cal R}_1 e^{-\frac{\pi}{P}\sigma}\sqrt{\frac{P}{2\sigma}}\sin\left( \frac{2\pi}{P} \eta\right) + \cdots,
\eeq
leading to 
\begin{align}
ds^2&=\kappa \bigg[4\sigma\bigg(ds^2(\text{AdS}_5)+d\chi^2\bigg)+\frac{2P}{\pi}\bigg(d\left(\frac{\pi}{P}\sigma\right)^2+ d\left(\frac{\pi}{P}\eta\right)^2+ \sin^2\left(\frac{\pi}{P}\eta\right)ds^2(\text{S}^2)\bigg)\bigg],
\nonumber \\
e^{-\Phi}&=\frac{{\cal R}_1\pi^2}{2 P^{\frac{3}{2}}\sqrt{\kappa}}e^{-\frac{\pi}{P}\sigma}\left(\frac{\pi}{P}\sigma\right)^{-\frac{1}{2}},~~~~H_3=-\frac{4\kappa P}{\pi}\sin^2\left(\frac{\pi}{P}\eta\right) d\left(\frac{\pi}{P}\eta\right)\wedge \text{vol}(\text{S}^2).
\end{align}
The RR fluxes are zero at this order and the subspace $(\frac{\pi}{P}\eta,\text{S}^2)$  now forms a unit radius 3-sphere. One can show that by a change of coordinate $ \tilde{r}= e^{-\frac{\pi}{P}\sigma}(\frac{\pi}{P}\sigma)^{-\frac{1}{2}}$  the metric reduces to the near horizon limit of a stack of spherically symmetric NS5 branes in flat space. By  choosing $2\kappa=\pi$  we find  the appropriately quantized charges of the NS5 branes,
\beq
Q_{\text{NS5}}=-\frac{1}{(2\pi)^2}\int_{S^3} H_3=P.
\eeq
By similar analysis, the limits for $\eta= 0, P$ for $\sigma$ far from its bounds shows that the solution is regular except at $\sigma=0$ which requires more detailed study.

 For $(\sigma=0, \eta= k)$, in the range $0<k<P$, we use the coordinate change $(\eta=k- r\cos\alpha,\sigma=r\sin\alpha)$ for small $r$ and find that 
\beq
\dot{V}=N_k,~~~V''=\frac{b_k}{2r },~~~~\dot{V}'=\frac{b_k}{2}(1+\cos\alpha)+N_{k+1}-N_{k},\label{eq:VsattheD6s}
\eeq
leading to the asymptotic form of
\begin{align}
\frac{ds^2}{2\kappa\sqrt{N_k}}&= \frac{1}{\sqrt{\frac{b_k}{r}}}\bigg(4ds^2(\text{AdS}_5)+ ds^2(\text{S}^2)\bigg)+ \frac{\sqrt{\frac{b_k}{r}}}{N_k}\bigg(dr^2+ r^2 ds^2(\tilde{\text{S}}^2)\bigg),~~~e^{-\Phi}=\left(\frac{N_kb_k^3}{2^6\kappa^2r^3 }\right)^{\frac{1}{4}}.\nn
\end{align}
This matches with the near horizon limit of a stack of D6 branes wrapping AdS$_5\times$S$^2$ with $\tilde{\text{S}}^2$ spanned by $(\alpha,\chi)$. The RR fluxes to leading order are
\beq
B_2=0,~~~C_1=\left(\frac{b_k}{2}(1+\cos\alpha)+N_{k+1}-N_k\right)d\chi,~~~C_3=-2\kappa N_k d\chi\wedge \text{vol}(\text{S}^2),
\eeq
The Page charge of D6 branes is quantised and one gets
\beq
F_2=-\frac{1}{2}b_k \text{vol}(\tilde{\text{S}}^2)~~~\Rightarrow~~~Q^k_{D6}=-\frac{1}{2\pi}\int_{\tilde{S}^2}F_2= b_k=2N_k-N_{k-1}-N_{k+1}.
\eeq
The solution is regular everywhere else and has a stack of source NS5 branes at $\sigma=\infty$ and stack of D6 branes at $(\sigma=0,\eta=k)$ for $k=1,...P-1$. 

We can also define a quantised Page charge for D4 branes sitting at $\sigma=0$. At the loci of the D6 branes, near $\eta=k$, with the help of \eqref{eq:VsattheD6s} one can integrate on $(\chi,\text{S}^2)$ and the semi circul defined by $(\eta=k- r\cos\alpha,\sigma=r\sin\alpha)$, with $r$ small and $0\leq \alpha\leq \pi$. 
Performing the integral one finds
\beq
Q^k_{D4}=-\frac{1}{(2\pi)^3}\int_{\text{S}^2\times \tilde{\text{S}}^2}\hat F_4=N_k-N_{k-1},\label{eq:GMF4pagecharge}
\eeq
which are interpreted as colour branes. The $\hat{F}_4$ is locally given by $dC_3$, or $\hat{F}_4=F_4-B_2 \wedge F_2$.
The total charge of D6 and D4 branes reads
\beq
Q_{D6}=\sum_{k=1}^{P-1}Q^k_{D6}=N_{P-1}+N_1,~~~~
Q_{D4}=\sum_{k=1}^{P-1}Q^k_{D4}=N_{P-1},
\eeq
 This total amount of charge of D4 branes quoted above includes the 'true' colour D4 present in the background, but also the charge of  four-brane induced on the D6 and NS branes. 
 If one is interested only in the 'true' D4 charge in the interval $[k,k+1]$, excluding the charge of  four-brane induced on the D6 and NS branes, there are $N_k$ of them. 

\bibliographystyle{JHEP}
\bibliography{main.bib}

\end{document}